\documentclass[11pt]{article}
\usepackage[dvips]{graphicx}
\usepackage{amsmath,amsfonts}
\usepackage{amssymb}
\usepackage{epsfig,slashed}
\usepackage{cite}
\usepackage{color,comment}
\usepackage[hmargin=2.5cm,vmargin=3.5cm]{geometry}

\newcommand{\be}{\begin{equation}}
\newcommand{\ee}{\end{equation}}
\newcommand{\bea}{\begin{eqnarray}}
\newcommand{\eea}{\end{eqnarray}}
\newcommand{\mbb}{\mathbb}
\newcommand{\ti}{\times}
\newcommand{\half}{\frac{1}{2}}
\newcommand{\mc}{\mathcal}

\newcommand{\beqa}{\begin{eqnarray}}
\newcommand{\eeqa}{\end{eqnarray}}

\newcommand{\thba}[3]{\theta[\!\!\begin{array}{c}{\phantom{}\vspace{-.5mm}\scriptstyle#1}%
                        \\[-1.6mm]{\scriptstyle #2}\end{array}\!\!]( { #3} )}

\newcommand{\im}{\mathrm{Im\;}}
\newcommand{\re}{\mathrm{Re\;}}

\def\bra{\langle}
\def\ket{\rangle}

\def\ap{\alpha^{\prime}}

\def\beq{\begin{equation}}
\def\eeq{\end{equation}}
\def\mr#1{\mathrm{#1}}

\newcommand{\vt}{\theta_1}

\def\ov{\overline}

\newcommand{\tab}[4]{\theta\left[ \begin{array}{c}#1 \\ #2 \end{array} \right] (#3,#4)}

\newcommand{\nn}{\nonumber}

\begin{document}

\title{}
\author{}
\date{}
\thispagestyle{empty}

\begin{flushright}
\vspace{-3cm}
{\small CPHT-RR056.0810 \\
\small DESY 10-135 \\
\small OUTP-10/21P  }
\end{flushright}
\vspace{1cm}

\begin{center}
{\bf\LARGE
Anomaly Mediation in Superstring Theory}

\vspace{1.5cm}

{\bf Joseph P. Conlon\;}$^{1,2}$
{\bf,\hspace{.2cm} Mark Goodsell\;}$^{3}$
{\bf\hspace{.1cm} and\hspace{.2cm} Eran Palti\;}$^{4}$
\vspace{1cm}

{\it
$^1$ Rudolf Peierls Center for Theoretical Physics, 1 Keble Road, \\
Oxford, OX1 3NP, United Kingdom\\
$^2$ Balliol College, Oxford, OX1 3BJ, United Kingdom \\
$^3$ Deutsches Elektronen-Synchrotron DESY, Notkestrasse 85, D-22603 Hamburg, Germany.\\
$^4$ Centre de Physique Th´eorique, Ecole Polytechnique, CNRS, 91128 Palaiseau, France. \\
\vspace{3mm}
j.conlon1@physics.ox.ac.uk, mark.goodsell@desy.de, eran.palti@cpht.polytechnique.fr
}

\vspace{1cm}

{\bf Abstract}
\end{center}
\vspace{-.5cm}

We study anomaly mediated supersymmetry breaking in type IIB string
 theory and use our results to test the supergravity formula for anomaly mediated
gaugino masses. We compute 1-loop gaugino masses for models of D3-branes
on orbifold singularities with 3-form fluxes by calculating the annulus
 correlator of 3-form flux and two gauginos in the zero momentum limit. 
 Consistent with supergravity expectations we find both anomalous and running contributions to 1-loop gaugino masses.
For background Neveu-Schwarz H-flux we find an exact match with the supergravity formula. For Ramond-Ramond flux
there is an off-shell ambiguity that precludes a full matching. The anomaly mediated
gaugino masses, while determined by the infrared spectrum, arise from an explicit sum over
UV open string winding modes.
We also calculate brane-to-brane tree-level gravity mediated gaugino masses and show that there are two contributions coming from the dilaton and from the twisted modes, which are suppressed by the full $T^6$ volume and the untwisted $T^2$ volume respectively.

\clearpage

\tableofcontents

%\vspace{3cm}
%\clearpage

%%%%%%%%%%%%%%%%%%%%%%%%%%%%%%%%%%%%%%%%%%%%%%%%%%%%%%%%%%%%%%%%%%%%%%%%%%%%%%%%%%%%%%%%%%%%%%
\section{Introduction}
%%%%%%%%%%%%%%%%%%%%%%%%%%%%%%%%%%%%%%%%%%%%%%%%%%%%%%%%%%%%%%%%%%%%%%%%%%%%%%%%%%%%%%%%%%%%%%

Supersymmetry is both one of the most promising phenomenological ideas for Beyond-the-Standard-Model physics and
also an apparently crucial component of consistent ultraviolet physics. The structure of supersymmetry breaking is
an area where experimental results within the next few years could plausibly give information as to the
nature of Planck scale physics, in particular through gravity mediation.
This makes a thorough understanding of supersymmetry breaking essential, both to understand observations from the Large Hadron Collider
and to connect these observations with theories of fundamental physics.

Once supersymmetry is broken in some sector it will be mediated to the observable sector thereby inducing soft masses. Within a 4-dimensional $\mc{N}=1$ supergravity framework the Lagrangian is specified by three functions, the holomorphic superpotential $W$, the holomorphic gauge kinetic function $f$ and the real K\"ahler potential $K$. The soft terms can be expressed in terms of the above functions and
the tree-level formulae for gravity mediation are well known, being found in e.g. \cite{hepph9707209}. For example the tree-level gaugino masses are given by
\be
M_{1/2}^{\mr{tree}} = \frac{1}{2\re{f}} F^i \partial_i f \;,
\ee
with $F^i$ the F-terms for the supersymmetry breaking fields.

Tree-level expressions for soft terms are however often insufficient. This is particularly the case if the tree-level soft terms vanish.
While this will not occur for generic supergravity theories it does occur for certain special cases,
for example no-scale or sequestered supergravities. However these are the cases relevant for string compactifications, where
the effective supergravity often takes a no-scale structure. In such cases the form of loop contributions to soft masses is crucial.

Loop corrections to soft masses are closely related to loop corrections to physical couplings, as is known in field theory through techniques of
analyticity in superspace \cite{hepph9706540,hepph9803290}. At 1-loop gauge couplings are modified not just by conventional running but also by anomalous contributions that are only generated at the loop level. In supergravity these anomalies arise from (for example) the Konishi or K\"ahler-Weyl anomalies associated to non-canonical normalisation for matter or metric kinetic terms. For the physical
gauge couplings these are captured by the Kaplunovsky-Louis formula \cite{hepth9303040, hepth9402005}
\bea
\label{KL}
g_a^{-2}(\Phi, \bar{\Phi}, \mu) & = & \hbox{Re}(f_a(\Phi)) + \frac{\left( \sum_r n_r T_a(r) - 3T_a(G)\right)}{8 \pi^2}
\ln \left( \frac{M_P}{\mu}\right) + \frac{T(G)}{8 \pi^2} \ln g_a^{-2} (\Phi, \bar{\Phi}, \mu)
\nonumber \\
& & + \frac{(\sum_r n_r T_a(r) - T(G))}{16 \pi^2} \hat{K}(\Phi, \bar{\Phi})
 - \sum_r \frac{T_a(r)}{8 \pi^2} \ln \det Z^r(\Phi, \bar{\Phi}, \mu).
\eea
Here $a$ is a gauge group index, $\mu$ is the running scale and the matter kinetic matrix $Z$ is only for fields charged under the gauge group.\footnote{The $T_R$ and $T_G$ stand for the usual quantities appearing in the beta functions and $d_R$ is the dimension of the representation.}
One would expect that anomalies in the gauge couplings also appear as anomalous contributions to gaugino masses.
Anomalous gaugino masses appear implicitly in \cite{hepth9303040}, where the gaugino masses are said to be given by
\be
\label{dundee}
M_a = \frac12 F^i \partial_i \ln g_a^{-2},
\ee
with $g_a^{-2}$ given by (\ref{KL}).

One further source of anomalous loop-generated soft masses was proposed by Randall and Sundrum and by Giudice, Luty, Murayama and Rattazzi
in \cite{hepth9810155, hepph9810442}. These are associated to the conformal compensator of supergravity (or F-terms associated to the supergravity multiplet) and give a contribution to gaugino masses of
\be
M_a = g_a^2 \frac{b_a}{16 \pi^2} m_{3/2} = - \frac{g_a^2}{16 \pi^2} (3 T(G) - T(R)) m_{3/2}.
\label{hibs}
\ee
This contribution
is often called `anomaly mediation'. Within general supergravity theories
the physics of this was further studied in \cite{hepth9911029, hepth0004170, Dine:2007me, ChoiNilles}.
In \cite{hepth9911029} an expression was given for the 1-loop anomaly-induced gaugino masses in a general supergravity theory:
\be
m_{1/2} = -\frac{g^2}{16\pi^2}\left[\left(3T_G-T_R\right)m_{3/2} - \left(T_G-T_R\right)K_iF^i - \frac{2T_R}{d_R}F^i\partial_i\left(\mathrm{ln\;det\;}Z \right) \right]\;. \label{hearts}
\ee
This includes both the term (\ref{hibs}) and also some of the terms of (\ref{dundee}).

Let us establish our conventions on nomenclature.
The gaugino mass contribution of (\ref{hibs}) is often simply called `anomaly mediation' and much of the phenomenological
analysis of anomaly mediation makes use purely of this term.
However the other terms in (\ref{hearts}) are also generated by anomalies
and are of similar importance to (\ref{hibs}), with a characteristic
magnitude of the gravitino mass suppressed by a loop factor.
These terms are phenomenologically just as important as
the term given in (\ref{hibs}) and there is no real reason to ignore them.
We shall use `anomaly mediation' to refer to all terms that contribute at 1-loop to gaugino masses
and have magnitudes that are characteristically given by the gravitino mass suppressed by a loop factor.

The Kaplunovsky-Louis expression (\ref{KL}) has been studied extensively in the context of string theory, in particular in the context
of gauge threshold corrections. However the same is not true of the expression (\ref{hearts}).
In fact there has been only a very limited direct study of anomaly-mediated gaugino masses in string theory.
This is surprising. First, gravity-mediated supersymmetry breaking is one of the most plausible loci for
string theory and phenomenology to intersect. Secondly, anomaly mediation is associated to the UV physics
of supergravity and superstring theory is the only known UV consistent completion of supergravity. Therefore studying how anomaly mediation
arises within superstring theory is of crucial importance to having a complete picture of the phenomenon. Finally, string theory has the useful feature that it is calculationally decoupled from supergravity. The covariant Neveu-Schwarz-Ramond formalism of the superstring is in some sense a dual calculational formalism to supergravity, as the spacetime symmetries are not manifest on the worldsheet. The calculations therefore provide a check of the supergravity results in a technically independent fashion.

The need for an understanding of anomaly mediation in string theory is further amplified by some previous studies
in the literature. Indeed, as far as we are aware, the only direct study of anomaly mediation in string theory prior to this work was performed in \cite{hepth0507244, hepth0509048} in which it was argued that anomaly mediation does not occur in perturbative string theory.\footnote{There are of course many more papers which study the properties and phenomenology of anomaly-mediated gaugino masses in the effective supergravities that descend from string theory.} More recently, arguments were also
presented in \cite{08010578, 09122950} that the `superconformal anomaly' term in (\ref{hearts}) directly proportional to $m_{3/2}$
is absent.

In this paper we perform the first detailed study of the expression (\ref{hearts}) in a string theory context.
The overall aim of the paper is to compute 1-loop gaugino masses for gauge theories living on D-branes in a supersymmetry breaking flux background.
Our setup is that of D3 branes on a $\mathbb{Z}_4$ orbifold singularity as studied in \cite{09014350}. We study annulus diagrams with two gaugino vertex operators on the boundary. The flux background can be accounted for by the insertion of the appropriate vertex operators in the bulk corresponding to 3-form fluxes (NSNS or RR).
As is well known, the worldsheet NSR theory cannot be consistently formulated in the presence of background RR fluxes (the RR vertex operators cannot
 be exponentiated into the worldsheet action) and even an NSNS flux background is difficult to work with. However the
flux vertex operators are well-defined and we can probe the 1-loop gaugino masses in a fluxed background by using the CFT appropriate to
a fluxless background and inserting the flux vertex operator. In a similar way, we can probe supersymmetry breaking effects in a supersymmetric background. This same approach has been used in \cite{08071666} to study the tree-level gaugino masses on D3 branes that are induced by fluxes.
At a technical level, the study of 1-loop gaugino masses then requires the computation of the correlators $\langle \lambda \lambda H_3 \rangle$ and $\langle \lambda \lambda F_3 \rangle$ on the annulus.

There is a non-zero amount of technical machinery that goes into this computation. In section \ref{sec:cftbuildblock}
we review the basic CFT building blocks that are needed for the computation. This section establishes notation and
convention, as well as reviewing bosonic and fermionic correlators on the torus and annulus. Although most of this is standard there
are some results which we have not otherwise found in the literature, for example the twisted bosonic correlators.
In section \ref{sec:themodel} we introduce the model that we study throughout this paper, fractional D3 branes on the $\mbb{C}^3/\mbb{Z}_4$ orbifold.

In section \ref{sec:gmgm} we study an example of tree-level gravity mediated soft terms: the annulus correlator of an open string superpotential with two gaugini, $\langle \phi_1 \phi_2 \phi_3 \lambda \lambda \rangle$. The `susy breaking sector' of $\phi_1 \phi_2 \phi_3$ and the
gaugini $\lambda \lambda$ are on separate brane stacks. This is a toy model of the communication of susy breaking
from a distant `susy breaking brane' to a visible sector. Although this is an annulus diagram it turns out it should actually
be regarded as tree-level supersymmetry breaking, mediated by the closed strings exchanged in the open string ultraviolet limit.
The field theory expectations can then be found via tree-level supergravity. This amplitude (an annulus 5-point amplitude) is technically
quite similar to the flux computations that are our ultimate focus. We use it as a test both of the CFT formalism and also to show
that susy breaking masses can be accurately found by computations performed in a supersymmetric background. We find agreement
between the string computations and the supergravity expectations.

In section \ref{sec:anommedgaugmass} we come to our main focus, the test of the formula (\ref{hearts}) and the
study of 1-loop gaugino masses in string theory. String theory makes no reference to K\"ahler potentials or superpotentials
and so one could never `derive' the formula (\ref{hearts}) from a string computation. Instead we study the supergravity predictions
of (\ref{hearts}) for 1-loop gaugino masses from background NS-NS and RR 3-form fluxes, and compare with the results of the string computations.

In appendix \ref{sec:thresholds} we study 1-loop gauge threshold corrections for gauginos. This is a new calculation but its
main purpose is simply to serve as a check on our formalism. In appendix \ref{sec:coupcon} we present an analysis of the string coupling for the two-gaugini three-boson tree-level gravity mediation amplitude to show that although it is an annulus amplitude it still corresponds to
tree-level physics. In appendix \ref{sec:nsapp} we present an important check on the anomaly mediated mass induced by NS H-flux by calculating a 4-point amplitude where one of the gaugini is decomposed into a scalar and a fermion. In appendix \ref{sec:rrapp} we present the details of the RR-flux calculation. Finally in appendix \ref{sec:useiden} we collect some useful mathematical expressions.

\subsection*{Summary of results regarding anomaly mediation}

As the calculations are technically involved we present here a brief summary of our methodology and key results regarding the anomaly mediation formula (\ref{hearts}).

Firstly we find that the formula (\ref{hearts}) is incomplete and actually
needs an additional term corresponding to the supersymmetrisation of the NSVZ term in the gauge couplings:
\be
\label{kilmarnock}
m_{1/2} = -\frac{g^2}{16\pi^2}\left[\left(3T_G-T_R\right)m_{3/2} - \left(T_G-T_R\right)K_iF^i - \frac{2T_R}{d_R}F^i\partial_i\left(\mathrm{ln\;det\;}Z \right) + 2T_G F^I \partial_I \ln \left( \frac{1}{g_0^2}  \right) \right]\;.
\ee
This term has previously appeared in the expression (\ref{dundee}) (from \cite{hepth9303040})
and also in field theory studies of higher-order contributions to gaugino masses \cite{hepph9803290}, which is related to earlier
work of \cite{ShifmanVainshtein91}. In the string computation the presence of this term can be discerned almost trivially from the requirement for the group factors $T_G$ and $T_R$ to match that in the Chan-Paton trace. It is an interesting test of (\ref{kilmarnock}) that
the relative coefficients of the last three terms of (\ref{kilmarnock}) must take the form they do to
match the Chan-Paton traces.

We pay particular attention to verifying the presence of the term (\ref{hibs}) which is the first term of (\ref{kilmarnock}). This is most accurately probed by considering only NSNS flux. It turns out that in this case there is a cancellation between the K\"ahler, Konishi and NSVZ anomalies (the last three terms of (\ref{kilmarnock})). A non-zero anomalous mass term in the string computation can then occur if and only if the term (\ref{hibs}) is present. We do find such an anomalous mass term, with a coefficient that is proportional to the beta function of the gauge theory, exactly as predicted. Further, the supergravity formula also predicts the relative sign and magnitude of this term to the 1-loop running mass of the gauginos which is probed by the same string calculation and indeed we find an exact match also for this. Therefore the string calculation provides two independent checks on the validity of the formula (\ref{kilmarnock}).

The anomaly mediated mass term arises as follows in the string calculation. The annulus amplitude $\langle \lambda \lambda H_3 \rangle$ gives the 1-loop gaugino mass and so includes both the anomaly mediated contribution and the running mass. We find that up to unimportant overall factors the amplitude takes the form
\be
{\cal A} \sim \left( 3T_G-T_R\right) m_{3/2} \int_0^{\infty} \frac{dt}{t} \left( -Z(t) + t \frac{d}{dt} Z(t)\right) \;. \label{nsnspreview}
\ee
The parameter $t$ is the annulus modulus, where $t \to 0$ is the open string UV or the closed string IR (which is a long cylinder), and $t \to \infty$ is the open string IR and closed string UV (which is a thin annulus). $Z(t)$ is the internal $N=2$ classical partition function which is a sum over winding modes in the untwisted torus of the orbifold
\bea
Z(t) &=& {\cal Z}_{\mr{Int,Cl}}(t) = \sum_i \sum_{n,m} \delta_i e^{-\frac{t}{4\pi\alpha'}|\Delta X_i(m,n)|^2} \;, \nn \\
\Delta X_i(m,n) &\equiv& 2\pi \sqrt{\frac{\im{T}}{\im{U}}} \left( n + U m + X_i \right) \;. \label{n2partfun}
\eea
Here $n$ and $m$ sum over the winding modes, $T$ and $U$ are the torus (along the untwisted direction) K\"ahler and complex-structure moduli. We have added a sum over $i$ which stands for the different brane stacks present in the construction so that $X_i$ denote the separation between the gaugino brane stack and another stack at the other end of the cylinder. We denote the gaugino brane stack $i=0$ and the sum over all the states includes the $X_{i=0}=0$ states and also the states which are strings stretching between the stack and other ones required for tadpole cancellation. The factor $\delta_i$ accounts for the Chan-Paton traces so that tadpole cancellation implies $Z\left(t \to 0\right) \to 0$.

The amplitude can be integrated exactly and gives
\be
\int_0^{\infty} \frac{dt}{t} \left( -Z(t) + t \frac{d}{dt} Z(t)\right) = \left[ Z\left(\infty\right) - Z\left(0\right) \right] - \int_0^{\infty} \frac{dt}{t} Z(t) \;. \label{anomrunfinalprev}
\ee
The first term is the anomaly mediated mass and the second term is the running mass, with the relative factors
exactly matching the supergravity prediction. We concentrate here on the first term. Tadpole cancellation implies that in the UV $t \to 0$
limit the partition function vanishes which guarantees finiteness and so
\be
Z(0) = 0 \;.
\ee
In the IR $t \to \infty$ only the massless modes can contribute which gives
\be
Z(\infty) = 1 \;.
\ee
It is worth looking a little closer at the anomaly mediated contribution
\bea
\label{firsttt}
Z(0) - Z(\infty) & = & \underset{t \to 0}{\mr{lim}} \sum_{\Delta X_i \neq 0} \sum_{n,m} e^{-\frac{t}{4\pi\alpha'}|\Delta X_i(m,n)|^2} \;  \\
\label{secondtt}
& \equiv & \int_{0}^{\infty} dt \sum_i \sum_{n,m} \delta_i \frac{| \Delta X_i(m,n) |^2}{4 \pi \alpha'}
e^{-\frac{t}{4\pi\alpha'}|\Delta X_i(m,n)|^2}.
\eea
We see that although the UV part $Z(0)$ vanishes leaving just a contribution from the IR piece $Z(\infty)$, we can think of this as
 the contribution from all the heavy modes not present in the IR limit i.e. $Z(0)=0$ only if we also include the IR modes.
 Another way to see this is to note that in (\ref{secondtt}) all contributing terms must have $\Delta X_i \neq 0$ and so arise from
 a heavy string mode with non-zero winding. Furthermore since values of $t$
 larger than $|\Delta X_i|^{-2}$ give only an exponentially suppressed contribution, we see the
  dominant contribution arise from values of the loop parameter $t \sim |\Delta X_i|^{-2}$.
  In that sense the anomaly mediated contribution comes purely from the heavy (open string) modes in the theory that lead to the finite UV completion. Of course in a closed string picture these are related to light IR degrees of freedom.

We conclude here the brief summary and refer to the main text for more details. We do note though that for the RR flux the string computation is substantially more complicated and we are only able to recreate some of the expected results.
While we do obtain both anomalous and running masses there is an ambiguity in the off-shell
continuation that precludes extracting full numerical results.
We also note that in both the RR and NSNS cases there is a technical issue with picture changing both gaugini,
for which we cannot obtain any
anomalous mass term. We discuss these issues in much more detail in the text.

To follow the calculations it is essential to read section \ref{sec:cftbuildblock} and the mathematical appendix \ref{sec:useiden}.
A reader purely interested in anomaly mediation and 1-loop susy breaking could skip section \ref{sec:gmgm} and just focus on \ref{sec:anommedgaugmass}. If uninterested in the details of the string computations we suggest reading simply the sections on supergravity predictions and then the overall result of the string computation.

%%%%%%%%%%%%%%%%%%%%%%%%%%%%%%%%%%%%%%%%%%%%%%%%%%%%%%%%%%%%%%%%%%%%%%%%%%%%%%%%%%%%%%%%%%%%%%
\section{CFT Building Blocks}
\label{sec:cftbuildblock}
%%%%%%%%%%%%%%%%%%%%%%%%%%%%%%%%%%%%%%%%%%%%%%%%%%%%%%%%%%%%%%%%%%%%%%%%%%%%%%%%%%%%%%%%%%%%%%

The computation of the amplitudes requires the evaluation of various CFT correlators between world-sheet fields that are introduced through the vertex operators. In this section we collate the relevant correlators and also other miscellaneous
CFT results that are used. Some of the results presented, such as the twisted bosonic correlators,
are new to our knowledge while others can be found within the list of useful references \cite{Polchinski,Friedan:1985ge,Atick:1986rs,Atick:1986ns,Atick:1987gy,0412206,Abel:2005qn,Bianchi:2006nf,Benakli:2008ub, Olly}. All the amplitudes evaluated in this paper are cylinder (annulus) amplitudes and so all correlators are on this topology. We therefore begin with a brief description of this geometry before describing the relevant correlators.

The cylinder has a single real modulus $t$ and is parameterised by a complex coordinate $z$. The circles at each end of the cylinder are positioned at $\re{(z)}=0,\half$ and are parameterised by $0\leq \im{z}\leq \frac{t}{2}$. The long cylinder limit is given by $t\rightarrow 0$ and corresponds to the open string UV and the closed string IR. The long strip limit is $t \rightarrow \infty$ and gives the open string IR and closed string UV. There is a single conformal Killing vector corresponding to translations parallel to the boundary.

The target space coordinates are the real worldsheet bosons $x^{M}\left(z,\bar{z}\right)$ where $M=0,...,9$. We further decompose $x^{M}=\left\{x^{\mu},x^m\right\}$ with $\mu=0,..,3$ denoting external directions and $m=4,..,9$ denoting internal directions. It will also be useful to group the directions into complex pairs and we define
\be
X^{i} = x^{2i-2} + i x^{2i-1} \;,
\ee
where $i=1,...,5$. To save on clatter we usually drop the indices on the coordinates unless needed and denote
\be
X = x_1 + ix_2 \;.
\ee

There are two basic boundary conditions that can be imposed at each end of the cylinder
\be
\hbox{Neumann: } \partial_n X (z, \bar{z}) \equiv \half(\partial + \bar{\partial}) X(z, \bar{z}) = 0,
\ee
\be
\hbox{Dirichlet: } \partial_t X (z, \bar{z}) \equiv \half(\partial - \bar{\partial}) X(z, \bar{z}) = 0.
\ee
We have also defined the normal and tangential derivatives. In principle we can consider different boundary conditions at each end of the annulus but since we are only studying D3 branes we restrict either to NN or DD boundary conditions. Henceforth to save on clatter we denote the coordinate dependence $X(z)$ without implying holomorphic properties.

The cylinder can be obtained from the torus by quotienting under the identification $z \to 1 - \bar{z}$, with boundaries at $z = 1 - \bar{z}$. This is useful for relating bosonic ($X(z,\bar{z})$) correlators on the torus to those on the cylinder. The method of images
can then be used to obtain the cylinder correlators by starting with torus correlators and adding an image field at $1- \bar{z}$ for any field at $z$. The sign of the image correlator is positive for Neumann boundary conditions and negative for Dirichlet boundary conditions. The torus modular parameter $\tau$ is related to the cylinder modulus by $\tau=\frac{it}{2}$.

%%%%%%%%%%%%%%%%%%%%%%%%%%%%%%%%%%%%%%%%%%%%%%%%%%%%%%%%%%%%%%%%%%%%%%%%%%%%%%%%%%%%%%%%%%%%%%
\subsection{Vertex operators}
\label{sec:vertexop}
%%%%%%%%%%%%%%%%%%%%%%%%%%%%%%%%%%%%%%%%%%%%%%%%%%%%%%%%%%%%%%%%%%%%%%%%%%%%%%%%%%%%%%%%%%%%%%

The amplitudes are calculated by inserting the vertex operators of
appropriate pictures into the partition function integral. In this section we briefly summarise the expressions for the vertex operators. We also note that we are always calculating cylinder amplitudes for which the ghost charge should be zero so that the sum of all the vertex operator `pictures' should vanish.

The bosonic vertex operator for a four-dimensional scalar $\phi$ is given in the $(-1)$ picture as
\be
{\cal V}_{\phi}^{-1} \left(z\right) = t^a e^{-\phi} \psi^i e^{ik \cdot x} \left(z\right)\;.
\ee
Here $z$ denotes the point on the worldsheet at which the vertex operator is inserted (which we integrate over). The scalar
Chan-Paton wavefunction is denoted $t^a$ and the field $\phi$ is the ghost from bosonising the $(\beta, \gamma)$ CFT.
The field $\psi^i$ can be bosonised in terms of free fields $H_i$ so that
\be
\label{psiboson}
\psi^i = e^{iH_i(z)}\;.
\ee
Here $i$ labels the complex direction.
Note that this bosonisation is purely local as the $\psi^i$ correlators depend on the spin structure and so cannot be
globally bosonised. However these
amplitudes (which we give in section \ref{sec:fermcorr} below) are fixed uniquely in terms of this local bosonisation.
For economy of notation we typically suppress the CP index and wavefunction so that for a four-dimensional scalar
we have the (-1)-picture vertex operator
\be
{\cal V}_{\phi}^{-1}\left(z\right) = e^{-\phi} \psi^i e^{ik \cdot x} \left(z\right)\;.
\ee
The four-dimensional gauge field vertex operator is given by
\be
{\cal V}_{A}^{-1} \left(z\right) = A^a e^{-\phi} \epsilon_{\mu} \psi^{\mu} e^{ik \cdot x} \left(z\right)\;.
\ee
Here again $\psi^{\mu}$ can be bosonised with H-charge of $\pm 1$ and $\epsilon_{\mu}$ is the polarisation vector of the gauge boson which satisfies $\epsilon \cdot k = 0$.

The fermion vertex operator in the $-\half$ picture is given by
\be
{\cal V}_{\lambda}^{-\half} \left(z\right) = \lambda^a e^{-\frac{\phi}{2}} S_{10} e^{ik \cdot x} \left(z\right)\;.
\ee
Here $S_{10}$ is the ten-dimensional spin field which can be locally bosonised to
\be
\label{spinbos}
S_{10} = \prod_{i=1}^5 e^{iq^iH_i} \;,
\ee
where the H-charges $q_i$ are given by the spin $\pm \half$ of the complex direction components of the spinor.

To bring the amplitude into the appropriate zero ghost charge picture we can change pictures following the
prescription of Friedan, Martinec and Shenker \cite{Friedan:1985ge} using
\be
{\cal V}^{i+1}\left(z\right) = \underset{{z\rightarrow w}}{\mr{lim}} e^{\phi(z, \bar{z})} \mr{T_F} \left(z\right) {\cal V}^{i}\left(w\right) \;,
\ee
where we have the picture changing operator
\be
\mr{T_F}\left(w\right) = \frac12 \left( \psi_i \partial \overline{X}^i\left(w\right)  + \overline{\psi}_i\partial X^i\left(w\right)  \right) \;.
\ee
In practice the picture changing is evaluated using the operator product expansions (OPE)
\bea
e^{iaH\left(w\right)}e^{ibH\left(z\right)} &=& \left(w-z\right)^{ab}e^{i\left(a+b\right)H\left(z\right)} + ... \;, \label{opespin}\\
e^{ia\phi\left(w\right)}e^{ib\phi\left(z\right)} &=& \left(w-z\right)^{-ab}e^{i\left(a+b\right)\phi\left(z\right)} + ... \;, \label{opeghost} \\
\partial X\left(w\right)e^{ikX(z)} &=& -\frac{i\alpha'}{2}k^+\left(w-z\right)^{-1} e^{ikX(z)} + ... \;, \label{opeboson} \\
\partial \overline{X}\left(w\right)e^{ikX(z)} &=& -\frac{i\alpha'}{2}k^-\left(w-z\right)^{-1} e^{ikX(z)} + ... \;,
\eea
where the ellipses denote less divergent terms. Recall that the $H_i$ are free fields and so
only OPEs with the same direction are non-vanishing.
We have also introduced the notation of complex momenta $k^{\pm}=k^1\pm ik^2$ and defined
\be
\label{kXcomplex}
kX\left(z\right) \equiv \half \left(k^+ \cdot \overline{X}\left(z\right) + k^- \cdot X\left(z\right)\right)\;,
\ee
so that in complex notation we can write
\be
k\cdot x\left(z\right) = k_i X^i\left(z\right) \;.
\ee

%%%%%%%%%%%%%%%%%%%%%%%%%%%%%%%%%%%%%%%%%%%%%%%%%%%%%%%%%%%%%%%%%%%%%%%%%%%%%%%%%%%%%%%%%%%%%%
\subsection{Bosonic Correlators}
\label{sec:boscorr}
%%%%%%%%%%%%%%%%%%%%%%%%%%%%%%%%%%%%%%%%%%%%%%%%%%%%%%%%%%%%%%%%%%%%%%%%%%%%%%%%%%%%%%%%%%%%%%

We first evaluate the bosonic correlators, namely those involving the worldsheet bosons $X(z,\bar{z})$.
 Since the bosons are free worldsheet fields, for a correlator to be non-vanishing it must involve the same complex directions. Therefore such a correlator can be labeled by the associated direction: correlators involving $X^{1,2}$ are labeled external, while $X^{3,4,5}$ are internal.

 Internal correlators are either twisted or untwisted. Twisted correlators involve directions in which an orbifold twist acts
 whereas untwisted directions have no orbifold action. Twisted correlators have no zero modes (as these are projected out by the orbifold) and
 so there is only a quantum contribution to the correlator coming from the path integral over the string oscillator modes. In addition to the quantum contribution untwisted correlators may also have a classical contribution coming the zero mode solutions. For the case of D3 branes studied
 here, this is associated to winding modes in the compact space.

 For external modes with Neumann boundary conditions, the classical contribution instead comes from momentum modes whereas the quantum correlator
is the same as for internal directions except with Neumann boundary conditions.
 We now proceed to calculate the correlators according to the preceding classification.

%%%%%%%%%%%%%%%%%%%%%%%%%%%%%%%%%%%%%%%%%%%%%%%%%%%%%%%%%%%%%%%%%%%%%%%%%%%%%%%%%%%%%%%%%%%%%%
\subsubsection{Internal untwisted quantum correlators}
%%%%%%%%%%%%%%%%%%%%%%%%%%%%%%%%%%%%%%%%%%%%%%%%%%%%%%%%%%%%%%%%%%%%%%%%%%%%%%%%%%%%%%%%%%%%%%

The quantum bosonic correlator on the cylinder can be derived from that on the covering torus (denoted by a subscript ${\cal T}$) which reads
\be
\langle X(z) \overline{X}(w) \rangle_{\mr{{\cal T},Qu}} = -\alpha' \log |\theta_1 (z-w)|^2 + \frac{2\pi\alpha'}{\im{\tau}} \left(\im (z-w)\right)^2\;.
\label{torusuntwicorr}
\ee
Here $\tau$ is the torus modular parameter.
For comparison with say \cite{Polchinski} note that $X$ here is a complexified coordinate.
As only correlators involving the same directions are non-vanishing
\be
\langle X(z) X(w) \rangle_{\mr{{\cal T},Qu}} = 0 \;,
\ee
as the two real directions give equal contributions of opposite sign.
From (\ref{torusuntwicorr}) one can obtain correlators on the cylinder (denoted by a subscript ${\cal A}$) through use
of the method of images.
\be
\langle X(z) \overline{X}(w) \rangle_{\mr{{\cal A}}} =  \half \left[ \langle X(z) \overline{X}(w) \rangle_{\mr{{\cal T}}} \, \pm \, \langle X(1-\bar{z}) \overline{X}(w) \rangle_{\mr{{\cal T}}} \pm \langle X(z) \overline{X}(1 - \bar{w})\rangle_{\mr{{\cal T}}} + \langle X(1-\bar{z}) \overline{X}(1-\bar{w}) \rangle_{\mr{{\cal T}}} \right],
\ee
where the plus sign applies for Neumann boundary conditions and the minus sign applies for Dirichlet boundary conditions. We can write the Neumann and Dirichlet correlator explicitly as
\bea
\label{neumannxx}
\bra X (z) \ov{X} (w) \ket_{\mr{{\cal A},Qu}}^{\mr{N}} &=&  -\alpha' \left(
\log \left|\theta_1 (z - w) \right|^2 + \log \left| {\theta_1 (\ov{z} + w)}\right|^2\right) + \frac{8\pi\alpha'}{t} \left(\im (z-w)\right)^2 \;, \\
\label{dirichletxx}
\bra X (z) \ov{X} (w) \ket_{\mr{{\cal A},Qu}}^{\mr{D}} &=&  -\alpha' \left(
\log \left| \theta_1 (z - w) \right|^2 - \log \left| {\theta_1 (\ov{z} + w)}\right|^2 \right)\;.
\eea
Here we have used the relation $\tau=\frac{it}{2}$ for the modular parameters of the cylinder and the covering torus.
The Dirichlet correlator has no zero mode since the string center of mass is fixed, whereas for Neumann boundary conditions the string can take any position.
Note also that when restricted to the boundary the Dirichlet correlator vanishes
\be
\left.\bra X (z) \ov{X} (w) \ket_{\mr{{\cal A},Qu}}^{\mr{D}}\right|_{\mr{Boundary}} = 0\;.
\ee

Vertex operator computations with the bosonic fields normally involve not the bare fields but rather their derivatives.
For operators on the boundary, under Neumann boundary conditions the vertex operators involve tangential derivatives $\partial_t X$ whereas for Dirichlet boundary conditions vertex operators involve normal derivatives $\partial_n X$. The relevant boundary correlators are
\bea
\label{Neumanntt}
\bra \partial_t X(z) \partial_t \ov{X}(w) \ket_{\mr{{\cal A},Qu}}^{N} &=&
-\frac{\alpha'}{2} \left( \partial_z \partial_w \log \theta_1 (z-w)+ \mr{c.c.} \right) + \frac{4 \pi \alpha'}{t} \;, \\
\label{Dirictt}
\bra \partial_n X(z) \partial_n \ov{X}(w) \ket_{\mr{{\cal A},Qu}}^{D} &=&
-\frac{\alpha'}{2} \left( \partial_z \partial_w \log \theta_1 (z-w) + \mr{c.c.} \right) \;.
\eea
It is important to note that the Neumann correlator (\ref{Neumanntt}) is obtained as the derivative of
a function periodic under $z \to z + \frac{it}{2}$, and is therefore exact on integration around the boundary
of the annulus:
\be
\int_{0}^{\frac{it}{2}} dz\; \bra \partial_t X(z) \partial_t \ov{X}(w) \ket_{\mr{{\cal A},Qu}}^{N} = 0.
\ee
As (\ref{Neumanntt}) and (\ref{Dirictt}) differ only by the zero mode term this also implies that
\be
\label{dirdercorrint}
\int_{0}^{\frac{it}{2}} dz\; \bra \partial_n X(z) \partial_n \ov{X}(w) \ket_{\mr{{\cal A},Qu}}^{D} = -2\pi i \alpha'.
\ee

We will also encounter correlators involving a single derivative but with some of the operators being closed string vertex operators inserted in the bulk of the cylinder. If we denote boundary positions by $z_i$ and bulk position with $w$ (for left moving and $\bar{w}$ for right moving) we have
\be
\begin{array}{ll}
\left<\partial X(z_1)\overline{X}(z_2)\right>^N_{Qu} = -2\alpha' \frac{\theta_1'(z_1-z_2)}{\theta_1(z_1-z_2)} \;,
& \left<\partial X(z_1)\overline{X}(z_2)\right>^D_{Qu} = 0 \;, \\
\left<\partial X(z_1)\overline{X}(w)\right>^N_{Qu} = -\alpha' \left[\frac{\theta_1'(z_1-w)}{\theta_1(z_1-w)} + \frac{\theta_1'(z_1+\bar{w})}{\theta_1(z_1+\bar{w})} \right] \;,
& \left<\partial X(z_1)\overline{X}(w)\right>^D_{Qu} = -\alpha' \left[\frac{\theta_1'(z_1-w)}{\theta_1(z_1-w)} - \frac{\theta_1'(z_1+\bar{w})}{\theta_1(z_1+\bar{w})} \right] \;, \\
\left<\partial X(w)\overline{X}(z_1)\right>^N_{Qu} = -2\alpha' \frac{\theta_1'(w-z_1)}{\theta_1(w-z_1)} \;,
& \left<\partial X(w)\overline{X}(z_1)\right>^D_{Qu} = 0 \;,  \\
\left<\partial X(w)\overline{X}(-\bar{w})\right>^N_{Qu} = -\alpha' \frac{\theta_1'(w+\bar{w})}{\theta_1(w+\bar{w})} \;,
& \left<\partial X(w)\overline{X}(-\bar{w})\right>^D_{Qu} = -\alpha' \frac{\theta_1'(w+\bar{w})}{\theta_1(w+\bar{w})} \;. \label{corrfac}
\end{array}
\ee

%%%%%%%%%%%%%%%%%%%%%%%%%%%%%%%%%%%%%%%%%%%%%%%%%%%%%%%%%%%%%%%%%%%%%%%%%%%%%%%%%%%%%%%%%%%%%%
\subsubsection{Internal twisted quantum correlators}
%%%%%%%%%%%%%%%%%%%%%%%%%%%%%%%%%%%%%%%%%%%%%%%%%%%%%%%%%%%%%%%%%%%%%%%%%%%%%%%%%%%%%%%%%%%%%%

In some cases we also require bosonic correlators along directions
that are twisted by the orbifold action and satisfy
\be
X(z + \tau) = e^{-2\pi i\theta} X(z)\;.
\ee
Here $\tau$ can be the modulus parameter of the covering torus or the associated cylinder. To evaluate the corresponding correlator we use the covering torus and then map it to the cylinder through the method of images. We denote the correlator on the torus as
\be
G^{\mr{\cal T}}_{\theta}\left(z-w\right) \equiv \bra \partial X (z) \partial \ov{X} (w) \ket_{\mr{{\cal T},Qu}} \;.
\ee
The correlator satisfies the following boundary conditions
\bea
G^{\mr{\cal T}}_{\theta}\left(z-w+\tau\right) &=& e^{-2\pi i\theta}G^{\mr{\cal T}}_{\theta}\left(z-w\right) \;, \nn \\
G^{\mr{\cal T}}_{\theta}\left(z-w+1\right) &=& G^{\mr{\cal T}}_{\theta}\left(z-w\right) \;.
\eea
We also know that since it involves two derivatives acting on the coordinate correlator it satisfies the operator product expansion (OPE)
\be
\underset{{z\rightarrow w}}{\mr{lim}}\;G^{\mr{\cal T}}_{\theta}\left(z-w\right) \sim \frac{\alpha^\prime}{\left(z-w\right)^2} + \alpha^\prime \bra T(0) \ket \;. \label{twqucoope}
\ee
where $T(0)$ is the stress-energy tensor for the compact dimensions. There is no zero mode because of the twist $\theta$. These conditions are enough to determine the correlator exactly. First we use the fact that any meromorphic function on a complex torus is given by a ratio of translated theta functions, the particular ratio can be determined using the transformation property
\be
\label{inttwiscorrthetatrans}
\theta_1\left(z + \theta + \tau\right) =  e^{-\pi i\tau}e^{-2\pi i\left( z +\theta \right)}\theta_1\left(z+\theta\right)\;.
\ee
This gives\footnote{Note that although the correlator $G^{\mr{\cal T}}_{\theta}\left(z-w\right)$ is not a periodic function on the torus it does have the same phase transformation as (\ref{twistcorr}) and so the two can still be identified.}
\be
\label{twistcorr}
G^{\mr{\cal T}}_{\theta}\left(z-w\right) \sim \ap \left(\frac{\theta_1^\prime (0)}{\theta_1 \left(z-w\right)}\right)^2 \bigg[ \frac{\theta_1 \left(z-w + \theta - U\right) \theta_1 \left(z-w + U\right)}{\theta_1 \left(\theta-U\right) \theta_1 \left(U\right)} \bigg] \;.
\ee
Here and henceforth in this section $\sim$ denotes up to a constant.
$U$ is defined as the solution to
\beq
\label{Ucondition}
\frac{\theta_1^\prime \left(\theta-U\right)}{\theta_1 \left(\theta-U\right)} + \frac{\theta_1^\prime (U)}{\theta_1 (U)} = 0 \;.
\eeq
This follows from the requirement that (\ref{twistcorr}) matches the
OPE (\ref{twqucoope}) with no single poles. It is manifest that (\ref{Ucondition}) has solutions: the left hand side is a periodic function of $U$ on the torus with two poles, and since any meromorphic function on the torus has as many zeros as poles, it must also have two zeros.\footnote{We can check the result for the twisted correlator (\ref{twistcorr}) by studying the $\theta \rightarrow 0$ limit which should lead to the untwisted correlator (\ref{torusuntwicorr}) without the zero mode. Expanding (\ref{Ucondition}) for small $\theta$ to first order gives the constraint
\beq
\label{eqUU}
\frac{\theta_1^{\prime\prime} (U)}{\theta_1 (U)} -  \bigg(\frac{\theta_1^\prime (U)}{\theta_1 (U)}\bigg)^2 = 0.
\eeq
This equation can be written as
\be
\partial_z \partial_w \log \theta_1 (z-w)|_{\pm U} = 0 \;,
\ee
which implies that the untwisted correlator (\ref{torusuntwicorr}) has zeros at $z-w=\pm U$. These are also the zeros of (\ref{twistcorr}) in the $\theta \rightarrow 0$ limit and so since the two functions share zeros and poles and have the same periodicity they can be identified.}

There is a very useful way to write the twisted correlator (\ref{twistcorr}):
\be
\label{twiscorrderv}
G^{\mr{\cal T}}_{\theta}\left(z-w\right) \sim -\ap \partial_z \bigg[ \frac{\theta_1 (z - w + \theta)}{\theta_1 (\theta)} \frac{\theta_1^\prime (0)}{\theta_1 (z-w)}\bigg] \;.
\ee
To see this, note that (\ref{twiscorrderv}) has the correct double pole at $z-w = 0$,
no single pole $(z-w)^{-1}$ (the two terms cancel upon expanding), and zeros when
\beq
\frac{\theta_1^\prime (z-w + \theta)}{\theta_1 (z-w + \theta)} - \frac{\theta_1^\prime (z-w)}{\theta_1 (z-w)}=0 \;.
\eeq
Using (\ref{Ucondition}) we see that this is solved by $z-w + \theta = U$ and $z-w = -U$. We therefore see that (\ref{twiscorrderv}) has the same periodicity, zeros and poles as (\ref{twistcorr}) and so the two must be identified.

To complete the check that this is indeed the correlator, we use the fact that \cite{Polchinski}
\beq
\frac{1}{4\pi}\re(\bra T(0) \ket) = \partial_t \log Z (t)\;.
\eeq
Here $Z_{tw}(t)$ is the twisted bosonic partition function which is given by
\be
Z_{tw}(t) = \frac{\eta(it/2)}{\vt(\theta)} \;.
\ee
From expanding (\ref{twiscorrderv}) we find
\begin{align}
G^{\mr{\cal T}}_{\theta}\left(z-w\right) \sim& \frac{\ap}{(z-w)^2}   - \ap \bigg[ \frac{1}{2} \frac{\theta_1^{\prime\prime}(\theta)}{\theta_1(\theta)} - \frac{1}{6} \frac{\theta_1^{\prime\prime\prime}(0)}{\theta_1^\prime(0)}\bigg]\;.
\end{align}
For one compact twisted complex dimension the derivative of the bosonic partition function is
\begin{align}
\partial_t \log Z_{tw} = \frac{1}{8\pi} \bigg[\frac{1}{3} \frac{\theta_1^{\prime\prime\prime}(0)}{\theta_1^\prime(0)} - \frac{\theta_1^{\prime\prime}(\theta)}{\theta_1(\theta)}\bigg]\;.
\end{align}
Comparing the two above we find complete agreement.

Unlike the untwisted case the expression (\ref{twiscorrderv}) is not a total derivative as it stands
since the part in the bracket is not periodic under $z \rightarrow z+\tau$, and so (\ref{twiscorrderv})
does not integrate to zero around the annulus. However when we use the twisted amplitudes
in section \ref{sec:gravmedstringamp} we will see that there is
an additional phase factor from the Chan-Paton matrices, and once this is
taken into account (\ref{twiscorrderv}) does indeed integrate to zero.

Finally we use the method of images to construct the Neumann and Dirichlet correlators on the cylinder which gives
\bea
\bra \partial_t X(z) \partial_t \ov{X}(w) \ket_{\mr{{\cal A},Qu},\theta}^{N} = \bra \partial_n X(z) \partial_n \ov{X}(w) \ket_{\mr{{\cal A},Qu},\theta}^{D} = \half \left(G^{\mr{\cal T}}_{\theta}\left(z-w\right) + \mr{c.c.} \right) \;.
\eea

%%%%%%%%%%%%%%%%%%%%%%%%%%%%%%%%%%%%%%%%%%%%%%%%%%%%%%%%%%%%%%%%%%%%%%%%%%%%%%%%%%%%%%%%%%%%%%
\subsubsection{Internal classical correlators}
\label{sec:intclascorr}
%%%%%%%%%%%%%%%%%%%%%%%%%%%%%%%%%%%%%%%%%%%%%%%%%%%%%%%%%%%%%%%%%%%%%%%%%%%%%%%%%%%%%%%%%%%%%%

There is also a classical contribution to the untwisted Dirichlet correlator $\bra \partial_n X(z) \partial_n \overline{X}(w) \ket^{D}_{\mr{{\cal A},Cl}}$ which comes from the winding modes along the compact internal directions.\footnote{There are no open string twisted winding modes and since we are considering only space-filling $D3$ branes there are no internal Kaluza-Klein modes.} We first of all evaluate the partition function
of the winding modes. This exists as part of the identity correlator $\langle 1 \rangle$ and contributes as an overall
factor even for the quantum correlator.

In general we can write the change in the target space coordinates as we go around a path $C$ on the string worldsheet as
\be
\Delta X = \int_{C} dX = \int_{C} \left[ dz \partial X + d\bar{z} \overline{\partial} X \right] \;.
\ee
We can split the path $C$ into a component that comes from integrating around the circle on the boundary of the cylinder $A \equiv [0,it/2]$ and a component coming from integrating from one end of the cylinder to the other $B\equiv [0,1/2]$. The former contribution can be seen to vanish using the Dirichlet boundary conditions
\be
\int_A dX = \int_0^{t/2} i dy \bigg[ \partial X - \ov{\partial} X \bigg] = 2\int_0^{t/2} i dy \;\partial_t X = 0 \;.
\ee
The full contribution comes from the $B$ path which gives
\be
\Delta X = \int_B dX = \int_0^{1/2} dx \bigg[ \partial X + \overline{\partial} X \bigg] \;. \label{windBpath}
\ee
Since the classical contribution is just linear on the worldsheet we can use (\ref{windBpath}) to write\footnote{The relation $\partial X = \overline{\partial} X$ comes from the fact that winding modes couple with opposite signs to left moving and right moving sectors.}
\be
\Delta X = \partial X = \overline{\partial} X\;. \label{linearclasspath}
\ee
For a rectangular torus with diameters $2 \pi R_1$ and $2 \pi R_2$ we can write
\beq
\Delta X = ( 2\pi n R_1 + x_1) + i (2\pi m R_2 + x_2) \;,
\eeq
where $x_1 + i x_2$ is the location of a D3-brane on the other boundary or is zero for branes at the same singularity. The winding numbers $n$ and $m$ are integers. For a complex torus this generalises to
\beq
\Delta X = 2\pi \sqrt{\frac{\im{T}}{\im{U}}} \left( n + U m + X_0 \right) \;,
\eeq
where $T = i R_1 R_2 \sin \alpha$ and $U = \frac{R_2}{R_1} e^{i\alpha}$ are the K\"ahler and complex structure moduli. The complex D3 position is $X_0 \equiv \frac{1}{2\pi} \sqrt{\frac{\im{U}}{\im{T}}} \left(x_1 + i x_2\right)$.

Let us first evaluate the partition function.
The worldsheet action is given by
\be
\label{bosonwsact}
S = \frac{1}{2\pi\alpha'} \int d^2z \half\left( \partial X \overline{\partial}\overline{X} + \overline{\partial}X \partial\overline{X} \right)  \;.
\ee
Evaluating this over the cylinder we have $\int d^2z=2 \int_0^{\half}dx \int_0^{\frac{t}{2}} dy$ so that from the classical contribution we obtain
\be
S = \frac{1}{2\pi\alpha'} \frac{t}{2} \left|\Delta X\right|^2 \;.
\ee
Therefore the classical part of the partition function associated to the internal directions is given by the sum over the winding modes
\be
{\cal Z}_{\mr{Int,Cl}} = \sum_{n,m} e^{-\frac{t}{4\pi\alpha'}|\Delta X(m,n)|^2} \;.
\ee
We can now evaluate the correlator
\bea
\bra \partial_n X(z) \partial_n \overline{X}(w) \ket^{D}_{\mr{{\cal A},Cl}} & = & \bra \left|\Delta X\right|^2 \ket \; \nonumber \\
& = & \sum_{n,m} | \Delta X|^2 e^{-\frac{t}{4\pi\alpha'}|\Delta X(m,n)|^2}.
\eea
We note this can be written as
\be
\label{classicalnormalderivative}
\bra \partial_n X(z) \partial_n \overline{X}(w) \ket^{D}_{\mr{{\cal A},Cl}} = -4\pi\alpha' \partial_t {\cal Z}_{\mr{Int,Cl}} \;.
\ee
This implies the full Dirichlet correlator (classical plus quantum parts) can be written as
\be
\bra \partial_n X(z) \partial_n \overline{X}(w) \ket^{D}_{\mr{{\cal A},full}} = -4\pi\alpha' (\frac{1}{t} + \partial_t) {\cal Z}_{\mr{Int,Cl}} \;.
\ee

%%%%%%%%%%%%%%%%%%%%%%%%%%%%%%%%%%%%%%%%%%%%%%%%%%%%%%%%%%%%%%%%%%%%%%%%%%%%%%%%%%%%%%%%%%%%%%
\subsubsection{Momentum exponential correlators and pole structures}
\label{sec:momenexpopole}
%%%%%%%%%%%%%%%%%%%%%%%%%%%%%%%%%%%%%%%%%%%%%%%%%%%%%%%%%%%%%%%%%%%%%%%%%%%%%%%%%%%%%%%%%%%%%%

We also encounter correlators involving exponentials $e^{ikX}$. These are most easily calculated using real coordinates $x^{M}$ and momenta $k^M$. The relevant correlator
\be
\bra \prod_i e^{ik_i \cdot x\left(z,\bar{z}\right)} \ket \;,
\ee
is evaluated by contracting the scalars using the real forms\footnote{These are simply related to the complex versions by a factor of $\half$.} of the cylinder correlators (\ref{neumannxx}) and (\ref{dirichletxx}).
In general this is given by
\be
\prod_{i<j} e^{- k_i \cdot k_j \mc{G}(z_i - z_j)},
\ee
where $\mc{G}(z_i - z_j)$ is the bosonic correlator.

However, we also provide more explicit expressions in the case we
only require
the Neumann correlator in the limit $z_i\rightarrow z_j$, where we can drop the zero mode piece of (\ref{neumannxx}).  This is given by
\be
\bra \prod_i e^{ik_i \cdot x\left(z_i,\bar{z}_i\right)} \ket^{N}_{{\cal A}} = \prod_{i<j} \left| \frac{\theta_1\left(z_{ij}\right)}{\theta_1'(0)} \right|^{\alpha' k_i k_j} \;. \label{expcorrreal}
\ee
We may also write (\ref{expcorrreal}) in complex co-ordinates and momenta as
\be
\bra \prod_i e^{ik_iX\left(z_j\right)} \ket^{N}_{{\cal A}} = \prod_{i<j} \left| \frac{\theta_1\left(z_{ij}\right)}{\theta_1'(0)} \right|^{\frac{\alpha'}{2}\left(k^+_i k^-_j + k^-_i k^+_j \right)} \;. \label{momexpcorrcom}
\ee
where we recall that the complex notation $k_iX\left(z\right)$ is defined in (\ref{kXcomplex}).

Another correlator that we require is
\be
\bra \partial X(w) \prod_i e^{ik_iX\left(z_j\right)} \ket^{N}_{{\cal A}} = -i\alpha' \prod_{i<j} k_j^+ \frac{\theta'_1\left(w-z_j\right)}{\theta_1\left(w-z_j\right)} \left| \frac{\theta_1\left(z_{ij}\right)}{\theta_1'(0)} \right|^{\frac{\alpha'}{2}\left(k^+_i k^-_j + k^-_i k^+_j \right)} \;, \label{momexpdercorr}
\ee
which can be deduced by acting on (\ref{expcorrreal}) with a derivative.

At this point we discuss a principle which greatly simplifies our calculations. The important point is that since we are probing
non-derivative terms in the action we do not need to know the full amplitude but rather only its zero momentum limit $k_i\rightarrow 0$. Given this it seems naively that bosonic correlators such as (\ref{momexpdercorr}) vanish. However it is also possible to generate a pole in the amplitude which when combined with the correlator (\ref{expcorrreal}) can generate inverse powers of momenta that cancel against the momenta in the amplitude leaving a result that is non-vanishing in the zero momentum limit. To see this consider the amplitude factor
\be
\label{polecancmom}
{\cal A} \supset \underset{{k_1\cdot k_2\rightarrow 0}}{\mr{lim}} \left[ \left(k_1 \cdot k_2\right) \int dz_1 \left|\frac{\theta_1\left(z_1-z_2\right)}{\theta_1'\left(0\right)}\right|^{k_1\cdot k_2} \left(\frac{\theta'\left(0\right)}{\theta_1\left(z_1-z_2\right)}\right)\right] = \frac{\left(k_1 \cdot k_2\right)}{\left(k_1 \cdot k_2\right)} = 1\;,
\ee
where we have used
\be
\frac{\theta_1\left(z\right)}{\theta_1'\left(0\right)} = z + {\cal O}\left(z^3\right) \;.
\ee
The pole at $z_1=z_2$ has canceled the vanishing momentum prefactor. In practice this means that
evaluating certain amplitudes can simply amount to analysing their pole structure.

%%%%%%%%%%%%%%%%%%%%%%%%%%%%%%%%%%%%%%%%%%%%%%%%%%%%%%%%%%%%%%%%%%%%%%%%%%%%%%%%%%%%%%%%%%%%%%
\subsection{Fermionic and Ghost Correlators}
\label{sec:fermcorr}
%%%%%%%%%%%%%%%%%%%%%%%%%%%%%%%%%%%%%%%%%%%%%%%%%%%%%%%%%%%%%%%%%%%%%%%%%%%%%%%%%%%%%%%%%%%%%%

The amplitudes also involve correlators of spin fields, which after bosonisation as in (\ref{spinbos}) correspond
 to correlators of $H$ fields.
This includes the case of the $\psi$ correlators which are spin fields with $\pm 1$ H charge.
The correlators depend on the spin structure, denoted by indices $\left(\alpha\beta\right)=\left\{\left(00\right),\left(10\right),\left(01\right),\left(11\right)\right\}$, and read
\be
\label{hchargecorr}
\bra \prod_i e^{ia_iH\left(z_i\right)} \ket = K_{\alpha\beta} \left[\prod_{i<j} \left(\frac{\theta_1\left(z_{ij}\right)}{\theta'_1\left(0\right)}\right)^{a_ia_j}  \right]\theta_{\alpha\beta}\left(\sum_i a_i z_i + \theta_I\right) \;,
\ee
where $\theta_I$ is the orbifold twist in torus $I$. The constants $K_{\alpha\beta}$ are determined for each amplitude by the factorisation limit. This amounts to taking the limit $z_i \rightarrow z_j$ for all $i,j$ so that the amplitude factorises to the field theory amplitude times the string partition function. The spin structure is then matched to that of the partition function.
Note that using (\ref{opespin}) we deduce that only correlators where the total $H$-charge is zero are non-vanishing. This is known as $H$-charge conservation. These correlators were derived by Atick and Sen by considering their OPEs with the stress tensor, giving a set
of differential equations that can be solved to obtain the correlator. The details can be found in
\cite{Atick:1986ns, Atick:1986rs, 0412206}.

The ghost correlators can be found by the same method \cite{Atick:1986ns, Atick:1986rs}. The resulting correlators are very similar to the fermionic
correlators except with signs and powers reversed,
\be
\label{ghostchargecorr}
\bra \prod_i e^{ia_i\phi\left(z_i\right)} \ket = K_{\alpha\beta} \left[\prod_{i<j} \left(\frac{\theta_1\left(z_{ij}\right)}{\theta'_1\left(0\right)}\right)^{-a_ia_j}  \right]\theta_{\alpha\beta}^{-1}\left(-\sum_i a_i z_i \right) \;.
\ee
Again, the factors $K_{\alpha\beta}$ are determined by factorisation onto the partition function limit.

%%%%%%%%%%%%%%%%%%%%%%%%%%%%%%%%%%%%%%%%%%%%%%%%%%%%%%%%%%%%%%%%%%%%%%%%%%%%%%%%%%%%%%%%%%%%%%
\subsection{Partition functions}
\label{sec:partfuncpre}
%%%%%%%%%%%%%%%%%%%%%%%%%%%%%%%%%%%%%%%%%%%%%%%%%%%%%%%%%%%%%%%%%%%%%%%%%%%%%%%%%%%%%%%%%%%%%%

In the $2,3,4$ spin structures - those involving $\theta_{00}, \theta_{01},$ and  $\theta_{10}$ - the partition functions for the non-compact dimensions are given as follows
\begin{align}
\mathrm{Bosonic} :& \frac{1}{\eta^4 (it)} \frac{1}{(4\pi^2 \ap t)^2}, \nonumber \\
\mathrm{Fermionic} :& \bigg(\frac{\theta_\nu (0)}{\eta (it)} \bigg)^2, \nonumber \\
bc\ \mathrm{ghosts} :& \eta^2 (it), \nonumber \\
\beta \gamma \ \mathrm{ghosts} :& \frac{\eta(it)}{\theta_\nu (0)}, \nonumber \\
\mathrm{Total} :& \frac{\theta_\nu (0)}{\eta^3 (it)}  \frac{1}{(4\pi^2 \ap t)^2}.
\end{align}
For the 1 spin structure, which involves $\theta_{11}$, the above expressions must be changed, and they become
\begin{align}
\mathrm{Bosonic} :& \frac{1}{\eta^4 (it)} \frac{1}{(4\pi^2 \ap t)^2}, \nonumber \\
\mathrm{Fermionic} :& \bigg(\eta^4 (it) \bigg)^2, \nonumber \\
bc\ \mathrm{ghosts} :& \eta^2 (it), \nonumber \\
\beta \gamma \ \mathrm{ghosts} :& \frac{1}{\eta^2 (it)}, \nonumber \\
\mathrm{Total} :& \frac{1}{(4\pi^2 \ap t)^2}.
\end{align}
which assumes that the zero modes in the fermionic sector are saturated.
If this is not the case that the partition function vanishes due to integrating over the fermionic zero modes.
Note that we require no additional insertions for the $\beta \gamma$ ghosts; their zero modes must be explicitly excluded.
In practice however the effect of the fermionic and ghost partition functions are already incorporated into the
correlators (\ref{hchargecorr}) and (\ref{ghostchargecorr}).

The partition function for one compact torus $I$  with twist $\theta_I \ne 0$ is\footnote{For the partition function derivation see
\cite{09014350,09061920} for example}
\beq
Z_{I} = (-2\sin \pi \theta_I) \frac{\theta_\nu (\theta_I)}{\theta_1 (\theta_I)},
\eeq
while for an untwisted torus of area $T_2$ and complex structure $U= U_1 + i U_2$ it is
\beq
Z_{I} =   Z(t) \times \left\{ \begin{array}{cc} \frac{\theta_\nu (0)}{\eta^3 (it/2)} & \nu = 2,3,4 \\  1 & \nu = 1 \end{array} \right. , \eeq
where
\beq
Z(t) \equiv  \sum_{n,m=-\infty}^{\infty} \exp [ -t \frac{\pi T_2}{\ap U_2} | n+ U m |^2 ] .
\label{CompactUntwistedPF}
\eeq
This assumes that both ends of the string are attached to the same brane stack, hence there is a zero mode as $t\rightarrow \infty$. If one end is on a stack displaced from the first by a (complex) displacement $z$, then we should modify $| n + Um|^2 \rightarrow |n + Um + \frac{z}{2\pi} \sqrt{\frac{U_2}{T_2}}|^2$ and there is no such zero mode.
In the following we shall define $Z(t)$ to be equal to $1$ when there is no $N=2$ sector in the amplitude.

%%%%%%%%%%%%%%%%%%%%%%%%%%%%%%%%%%%%%%%%%%%%%%%%%%%%%%%%%%%%%%%%%%%%%%%%%%%%%%%%%%%%%%%%%%%%%%%%%%%%%%%%%%%%%%%%%%%%%%%%%%%%%%%%
\section{The Model}
\label{sec:themodel}
%%%%%%%%%%%%%%%%%%%%%%%%%%%%%%%%%%%%%%%%%%%%%%%%%%%%%%%%%%%%%%%%%%%%%%%%%%%%%%%%%%%%%%%%%%%%%%%%%%%%%%%%%%%%%%%%%%%%%%%%%%%%%%%%

Although many of our results for the scattering amplitudes hold in an arbitrary D3-brane setting the complete evaluation requires an embedding in an explicit model. To this end we now introduce the model that we will use throughout this paper. The model is a $\mbb{Z}_4$ toroidal orbifold with fractional D3 branes at fixed point singularities. Before describing this model in detail we note that our expressions only rely on certain elements of the model. In particular, results concerning threshold corrections of tree-level quantities only rely on the local behaviour of the model near the singularity. This is in the spirit of \cite{09014350,09061920} where the threshold correction can be extracted as the coefficient of the appropriate logarithmic divergence. For the purpose of such calculations we could equally well work in a non-compact setting $\mbb{C}^3/\mbb{Z}_4$.

The expressions for tree-level and anomaly-mediated quantities rely on the use of $N=2$ winding modes. Therefore in this case the compact global completion of the model is important and in particular it is important that all the $N=2$ tadpoles are canceled, which requires the
 third torus to be compactified. We will choose our distribution of D3-branes so that this is the case.
As we do not introduce orientifolds the global $N=4$ tadpole remains uncancelled. One way to deal with this is to not compactify the first two complex
tori, as we are never sensitive to the global structure of these tori. Alternatively, we can note that as we are studying
effects that depend on the $\beta$-functions of a gauge theory, and $\mc{N}=4$ sectors do not involve running couplings,
our calculations are not sensitive to $\mc{N}=4$ tadpoles and so remain unaffected. Therefore, although strictly the model is incomplete, it is sufficient for our purposes.

%%%%%%%%%%%%%%%%%%%%%%%%%%%%%%%%%%%%%%%%%%%%%%%%%%%%%%%%%%%%%%%%%%%%%%%%%%%%%%%%%%%%%%%%%%%%%%%%%%%%%%%%%%%%%%%%%%%%%%%%%%%%%%%%
\subsection{Local aspects}
%%%%%%%%%%%%%%%%%%%%%%%%%%%%%%%%%%%%%%%%%%%%%%%%%%%%%%%%%%%%%%%%%%%%%%%%%%%%%%%%%%%%%%%%%%%%%%%%%%%%%%%%%%%%%%%%%%%%%%%%%%%%%%%%

We begin by describing the local properties of the model near a $\mbb{C}^3/\mbb{Z}_4$ singularity. This model has been previously studied in \cite{09014350} and orientifolded versions of it have been analysed in \cite{09061920}. The advantages of this model are that despite being very simple it still has chiral matter with running gauge couplings.
\begin{figure}
\begin{center}
\epsfig{file=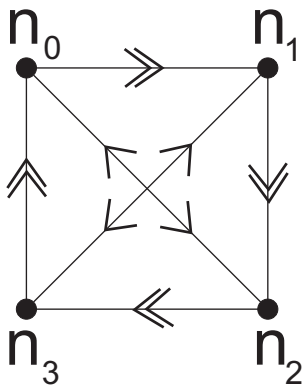,height=5cm}
\caption{The $\mathbb{Z}_4$ quiver.}
\label{z4quiver}\end{center}\end{figure}
Locally the orbifold is $\mbb{C}^3/\mbb{Z}_4$, with the orbifold action $\theta^i$ given by $\theta:(z_1, z_2, z_3) \to (e^{2 \pi i/4} z_1,
e^{2 \pi i/4} z_2, e^{-2 \pi i/2} z_3)$. The orbifold twist vector is then $\frac14(a_1, a_2, a_3) = \frac{1}{4}(1,1,-2)$. The non-Abelian part of the
gauge group is $SU(n_0) \ti SU(n_1) \ti SU(n_2) \ti SU(n_3)$ and the spectrum is
\be
\sum_{i=0}^{3} \sum_{r=1}^3 \left(n_i,\bar{n}_{i+a_r} \right) \;,
\ee
where $\left(n_i,\bar{n}_{i+a_r} \right)$ denotes matter in the bi-fundamental representation of $SU\left(n_i\right)\times SU\left(n_{i+a_r}\right)$.
The quiver diagram for the model is shown in figure \ref{z4quiver}. The superpotential is given by
\be
\label{superP}
W = \sum_{i=0}^3 \sum_{r,s,t=1}^3 \epsilon_{rst} \mathrm{Tr}\left( \Phi^r_{i,i+a_r}\Phi^s_{i+a_r,i+a_r+a_s}\Phi^t_{i+a_r+a_s,i} \right) \;,
\ee
where we define
\be
\Phi^r_{i,i+a_r}=\left(n_i,\bar{n}_{i+a_r}\right) \;.
\ee
The indices $r,s,t$ denote the plane that the bosonic field corresponds to (in terms of vertex operators this is equivalent
to the plane in which the boson has non-zero H charge). Local tadpole cancellation (equivalently cancellation of non-Abelian anomalies) requires
\be
n_0=n_2\;,\;n_1=n_3\;,
\ee
and after imposing these the $\beta$ functions for the local gauge groups are given by
\be
\beta_{n_0} = \beta_{n_2} = -\beta_{n_1} = -\beta_{n_3} = \frac{1}{16\pi^2}(2n_1-2n_0) \;.
\ee
The Chan-Paton realisation of the orbifold twist is given for $N=1$ and $N=2$ sectors by
\bea
\Theta_{N=1} & = & \mathrm{diag\;}\left(1_{n_0},i_{n_1},-1_{n_2},-i_{n_3}\right)\;, \\
\Theta_{N=2} & = & \mathrm{diag\;}\left(1_{n_0},-1_{n_1},1_{n_2},-1_{n_3}\right)\;,
\eea
where $1_{n_i}$ corresponds to the unit $n_i\times n_i$ matrix. The embedding of the CP factors $\lambda_{n_in_j}$ of the gauginos and $\Phi^r$s into the full CP matrix of the singularity are given by
\bea
G^{1,2} &=& \mathrm{diag}\left(\lambda_{n_0n_0},\lambda_{n_1n_1},\lambda_{n_2n_2},\lambda_{n_3n_3}\right) \;, \nn \\
\Phi^{1,2} &=&
\left( \begin{array}{cccc}
0 & \lambda_{n_0n_1} & 0 &  0 \\
0 & 0 & \lambda_{n_1n_2} & 0 \\
0 & 0 & 0 & \lambda_{n_2n_3} \\
\lambda_{n_3n_0} & 0 & 0 &  0
 \end{array}\right) \;, \nn \\
\Phi^{3} &=&
\left( \begin{array}{cccc}
0 & 0 & \lambda_{n_0n_2}  &  0 \\
0 & 0 & 0 & \lambda_{n_1n_3} \\
\lambda_{n_2n_0} & 0 & 0 & 0 \\
0 & \lambda_{n_3n_1} & 0 &  0
 \end{array}\right) \;.
\eea
Note that the matrices satisfy the following
\be
\label{thetphicomm}
\Phi^i \Theta = \Theta \Phi^i e^{2\pi i \theta^i} \;.
\ee

%%%%%%%%%%%%%%%%%%%%%%%%%%%%%%%%%%%%%%%%%%%%%%%%%%%%%%%%%%%%%%%%%%%%%%%%%%%%%%%%%%%%%%%%%%%%%%%%%%%%%%%%%%%%%%%%%%%%%%%%%%%%%%%%
\subsection{Global aspects}
%%%%%%%%%%%%%%%%%%%%%%%%%%%%%%%%%%%%%%%%%%%%%%%%%%%%%%%%%%%%%%%%%%%%%%%%%%%%%%%%%%%%%%%%%%%%%%%%%%%%%%%%%%%%%%%%%%%%%%%%%%%%%%%%

\begin{figure}
\label{z4orbifold}
\begin{center}
\epsfig{file=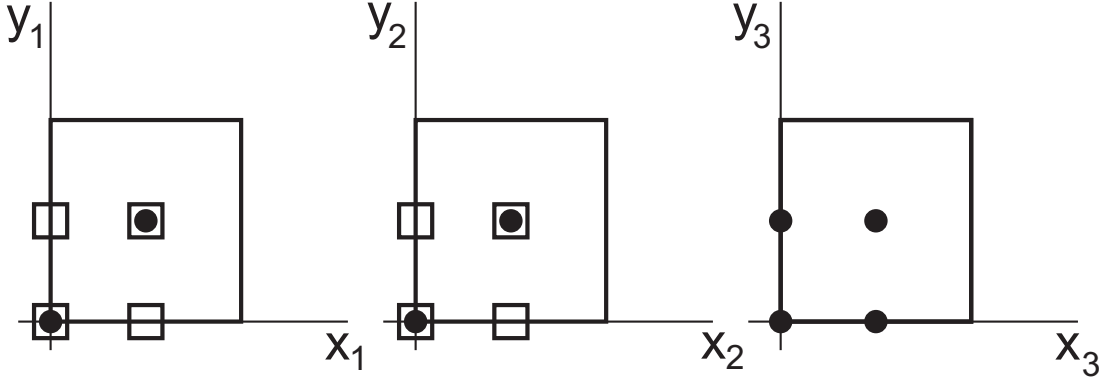,height=5cm}
\caption{The $T^6/\mbb{Z}_4$ orbifold. Dark circles correspond to $\theta$ fixed points and hollow squares correspond to
$\theta^2$ fixed points.}
\end{center}
\end{figure}

We now turn to more global aspects of the model. Although the particular compact completion will not affect our general results it is useful to have a concrete realisation in mind and so we will use the $T^6/\mbb{Z}_4$ orbifold of \cite{09014350} shown in figure \ref{z4orbifold}. We will introduce an additional brane stack to cancel twisted tadpoles.
As a compact space this orbifold has $h^{1,1} = 31, h^{2,1} =7$. The 31 elements of $h^{1,1}$ decomposes as
5 untwisted 2-cycles, 16 $\theta^1$ twisted cycles stuck at the 16 $\mbb{Z}_4$ fixed points, 6 $\theta^2$ twisted cycles stuck
at $\mbb{Z}_4$ invariant combinations of $\theta^2$ fixed points, and 4 $\theta^2$ twisted cycles at $\mbb{Z}_4$ fixed points and
propagating across the third $T^2$.

The primary use of this explicit global completion is to give an explicit cancellation of the $N=2$ twisted tadpoles. Recall that tracing over the $N=2$ sector gives a factor of $n_0-n_1+n_2-n_3$.  Therefore we place a single stack of fractional branes at the origin $(0,0,0)$ (point A) of multiplicity $(n^A_0, n^A_1, n^A_2, n^A_3) = (N, M, N, M)$ and a stack of fractional branes on the (0,0,i/2) (point B) of multiplicity $(n^B_0, n^B_1, n^B_2, n^B_3) = (M, N, M, N)$. This configuration cancels the $N=2$ tadpoles and importantly for our purposes implies that the classical partition function vanishes once all the winding modes are taken into account. Explicitly, for this configuration, from section \ref{sec:intclascorr} we have that the classical partition function of the $N=2$ winding modes takes the form
\be
Z(t) = \sum_{m,n} \left( e^{-\left(m^2+n^2\right)R^2t} - e^{-\left(m^2+(n+1/2)^2\right)R^2t} \right) \;,
\ee
where the first term comes from $AA$ strings and the second from $AB$ strings.
Using Poisson resummation (\ref{poissonsimple}) we can write this as
\be
Z(t) = \left(\frac{\pi}{R^2t}\right)\sum_{m,n} e^{-\frac{m^2+n^2}{R^2t}} \left( 1 - e^{n\pi i}\right) \;. \label{parttrans}
\ee
The coefficient vanishes for $(m, n) = (0,0)$, while all other terms vanish in the limit
 $t \to 0$ which is the open string UV or closed string IR. An important property which we will use throughout
 is that $Z(\infty)=1$ and $Z(0)=0$. Note also that only the details of the third torus has entered into the computation.

In section \ref{sec:gmgm} we also require the winding mode partition function for the case of a double trace operator. This differs from the previous case in that there is no trace over an open end of the string which means it is not connected to tadpole cancellation. The primary difference is then the absence of the last factor of (\ref{parttrans}) so that in the closed string channel the $n=0$ mode can also contribute.

%%%%%%%%%%%%%%%%%%%%%%%%%%%%%%%%%%%%%%%%%%%%%%%%%%%%%%%%%%%%%%%%%%%%%%%%%%%%%%%%%%%%%%%%%%%%%%%%%%%%%%%%%%%%%%%%%%%%%%%%%%%%%%%%
\subsection{K\"ahler potential}
%%%%%%%%%%%%%%%%%%%%%%%%%%%%%%%%%%%%%%%%%%%%%%%%%%%%%%%%%%%%%%%%%%%%%%%%%%%%%%%%%%%%%%%%%%%%%%%%%%%%%%%%%%%%%%%%%%%%%%%%%%%%%%%%

In order to compare the string calculations to the supergravity formulae we also need the K\"ahler potential for the matter and moduli fields.
The inheritance property of orbifold models implies the tree-level K\"ahler potential for the matter fields is
inherited from that for position moduli of D3 branes. The K\"ahler potential for the motion of D3 branes on our orbifold can therefore be read off from the results of \cite{0508171, 0508043} which gives
\be
\label{bhkk}
K = - \ln (S + \bar{S}) - \sum_{I=1}^3 \ln \left( (T_I + \bar{T}_I)(U_I + \bar{U}_I) - \frac{1}{6}(\Phi_I + \bar{\Phi}_I)^2\right) \;.
\ee
Here $\Phi_I$ are the position loci of the D-branes, $T_I$ are the K\"ahler moduli and $U_I$ the complex structure moduli.
This implies a matter metric of the from
\be
K_{\Phi_I\bar{\Phi}_{\bar{I}}} \equiv Z_I = \frac{1}{(T_I + \bar{T}_I)(U_I + \bar{U}_I)} \;.\label{matmoduli}
\ee
We note this is consistent with expectations since starting from the holomorphic Yukawa coupling (\ref{superP}), which we note always has one contribution from each direction, and computing the physical Yukawa couplings $\hat{Y}_{\alpha\beta\gamma}$ using the standard supergravity formula
\be
\hat{Y}_{\alpha\beta\gamma} = e^{K/2} \frac{Y_{\alpha\beta\gamma}}{\sqrt{Z_{\alpha} Z_{\beta} Z_{\gamma}}},
\ee
we obtain
\be
\hat{Y}_{\alpha \beta \gamma} = \frac{1}{(S + \bar{S})^{\half}} Y_{\alpha \beta \gamma} = g_{YM} Y_{\alpha \beta \gamma}.
\ee
This is precisely the correct behaviour of the N=4 SYM Yukawa couplings, in particular the factor of  $g_s^{\half} = g_{YM}$.
We can then use the K\"ahler potential (\ref{bhkk}) to determine the supergravity predictions for gaugino masses.

%%%%%%%%%%%%%%%%%%%%%%%%%%%%%%%%%%%%%%%%%%%%%%%%%%%%%%%%%%%%%%%%%%%%%%%%%%%%%%%%%%%%%%%%%%
\section{Gravity mediated gaugino masses}
\label{sec:gmgm}
%%%%%%%%%%%%%%%%%%%%%%%%%%%%%%%%%%%%%%%%%%%%%%%%%%%%%%%%%%%%%%%%%%%%%%%%%%%%%%%%%%%%%%%%%%

The primary motivation for this work is of course to study terms induced by supersymmetry breaking and in particular through anomaly mediation. As discussed in the introduction one way to approach this, for example as in \cite{hepth0507244,hepth0509048}, is to start from a background with broken supersymmetry and directly calculate the gaugino mass term in that background. However such an approach is limited by the
fact that there are very few backgrounds with no supersymmetry for which a CFT description is available.

In this paper we take a different approach. We consider a supersymmetric setting but instead
 compute correlators such that once one of the fields involved develops a background expectation value supersymmetry is broken.
 This is analogous to the computations of tree-level flux supersymmetry breaking in \cite{08071666} or for computations
 of supersymmetry breaking in gauge mediation \cite{hepph9706540}.
 In section \ref{sec:anommedgaugmass} we use this method to obtain the
 anomaly mediated gaugino masses associated
 to a vev for flux.
 However initially we would like to test this method in a
 better understood setting: gaugino masses induced by tree-level gravity mediation.

%%%%%%%%%%%%%%%%%%%%%%%%%%%%%%%%%%%%%%%%%%%%%%%%%%%%%%%%%%%%%%%%%%%%%%%%%%%%%%%%%%%%%%%%%%
\subsection{The supergravity prediction}
%%%%%%%%%%%%%%%%%%%%%%%%%%%%%%%%%%%%%%%%%%%%%%%%%%%%%%%%%%%%%%%%%%%%%%%%%%%%%%%%%%%%%%%%%%

We consider two stacks of branes, and study the communication of
 non-zero susy breaking on one stack to generate gaugino masses on the other stack.
The tree-level gravity induced gaugino mass is given by the formula
\be
m_{1/2}^{\mr{tree}} = \frac{1}{2\re{f}} F^i \partial_i f \;.
\ee
Here $f$ is the holomorphic gauge kinetic function of the associated gauge group which in the case of
bulk D3-branes at tree-level takes the universal form
\be
f = S \;,
\ee
with $S$ the dilaton superfield. In the case that the D3 branes are fractional branes living on orbifold singularities
the gauge kinetic function can take the form
\be
f = S + \sum_i \lambda_i M_i,
\ee
where $M_i$ are twisted fields associated to the singularity.

We define
\be
K_i = \partial_i K \;,\;\; F^i = e^{K/2}K^{i\bar{j}}\left(\partial_{\bar{j}}\bar{W}+K_{\bar{j}}\bar{W}\right) \;,\;\; \bar{m}_{3/2}=e^{K/2}\bar{W} \;.
\ee
where $m_{3/2}$ is denoted the gravitino mass and is given by the scalar components of the superfields in the superpotential.
Provided the dilaton does not enter the superpotential, we can use (\ref{bhkk}) and (\ref{superP}) to obtain
\be
\label{dilatonfterm}
F^S = - (S + \bar{S}) \bar{m}_{3/2} \;.
\ee
Putting all this together gives the dilaton-induced gaugino masses
\be
m_{1/2}^{\mr{tree},S} = - \bar{m}_{3/2}\;.
\ee
There is also a gaugino mass induced via the twisted fields. Since the $N=1$ twisted modes have a canonical K\"ahler potential they also have vanishing F-terms at the singularity where they have a vanishing vev. The $N=2$ twisted modes can have non-vanishing F-terms and therefore contribute to the gaugino masses.
\footnote{The $N=2$ chiral superfields that appear in the gauge kinetic function come from reducing $B_2+iC_2$ on the collapsed 2-cycle. In order to determine their F-terms we require knowledge of how they appear in the K\"ahler potential - knowledge which we do not yet have. However we still expect them to lead to a non-vanishing F-term. Further we expect that this F-term comes solely from the piece $e^{K/2}K^{M_i\bar{\Phi}_i}\partial_{\bar{\Phi}_i}\bar{W}$ while the piece $e^{K/2}K^{M_i\bar{X}_i}K_{\bar{X}_i}\bar{W}$ vanishes. Here the $\Phi_i$ are the matter fields appearing in the superpotential and $X_i$ denote any superfields that appear in the K\"ahler potential. If this conjecture is true then we note that there is no contribution to the tree-level gaugino masses in the absence of a $\Phi_i$ vev and in particular for the flux case. Now we discuss why we expect this F-term behaviour. Imagine separating the brane stacks of the gauginos and the bosons in the first 2 tori. This means that in the partition function there are no zero modes and the first modes are the `winding' modes stretching between the two branes. However these are now winding along a twisted direction which means that when we sum over the orbifold actions in the partition function their contribution sums to zero. Hence the resulting amplitude vanishes. This can only be matched by the supergravity formula if the F-term contribution is as described above. In that case since the $N=2$ twisted blow-up mode only couples to the bosons on its own singularity it does not couple to the bosons on the other singularity and so $K^{M\Phi}$ vanishes and the resulting F-term vanishes.
\label{n2foot}}

This implies an interaction of the form
\be
\label{gaugino3bosonfull}
\mr{Tr}\left(\lambda \lambda\right) \sum_{i=0}^3 \sum_{r,s,t=1}^3 \epsilon_{rst} \mathrm{Tr}\left( \phi^r_{i,i+a_r}\phi^s_{i+a_r,i+a_r+a_s}\phi^t_{i+a_r+a_s,i} \right) \;,
\ee
where $\phi$ denotes the scalar component of the superfield $\Phi$. The full expression (\ref{gaugino3bosonfull}) includes the particular term
\be
\label{gaugino3bosonpart}
{\cal L}_{0} \equiv \mr{Tr}\left(\lambda_{n_3n_3}\lambda_{n_3n_3}\right) \mr{Tr}\left(\phi^1_{n_0n_1}\phi^2_{n_1n_2}\phi^3_{n_2n_0} \right)\;.
\ee
These terms are essentially the correlator of the field theory superpotential with two gauginos. The fields are chosen so that the gauginos and bosons
are not charged under any common gauge group. Furthermore, the gauginos and bosons can be placed on separate stacks, implying the absence
of any massless open string modes connecting the two stacks.

Our aim is to recreate the interaction (\ref{gaugino3bosonpart}) by studying the zero momentum limit of a string scattering amplitude. Note that since the gauginos and bosons are not charged under any common gauge groups this interaction must be induced by gravity. Note also that since this is a double trace operator, within the global picture of the orbifold we can place the gauginos and bosons on different singularities. For now we do not specify this and return to these possibilities later.

%%%%%%%%%%%%%%%%%%%%%%%%%%%%%%%%%%%%%%%%%%%%%%%%%%%%%%%%%%%%%%%%%%%%%%%%%%%%%%%%%%%%%%%%%%
\subsection{The string amplitude}
\label{sec:gravmedstringamp}
%%%%%%%%%%%%%%%%%%%%%%%%%%%%%%%%%%%%%%%%%%%%%%%%%%%%%%%%%%%%%%%%%%%%%%%%%%%%%%%%%%%%%%%%%%

The relevant string scattering amplitude involves two gauginos and three bosons at vanishing momentum. Although the operator is tree-level the appropriate topology is not the disc but rather the cylinder. This can be heuristically seen in two ways.
First, since this is a double trace operator we require two boundaries to have any non-zero answer.
Second, since it is gravity mediated it is tree-level in the closed string channel which is 1-loop in the open string channel. The latter reason is a little subtle since it depends on whether the vertex operators themselves are open or closed string but since in this case they are all open string operators the reasoning is valid. There is also a more formal way to determine the topology which is by counting powers of the string coupling. This is carried out in appendix \ref{sec:coupcon} where it is shown that indeed the cylinder reproduces the correct dilaton dependence
for a tree level amplitude.

To specify the vertex operators in the amplitude we require the spin structure or H-charge of the gauginos and bosons. These are given by
\bea
g_1^{-1/2}(z_1) &=& \half(+,+,+,+,+) \;, \nn \\
g_2^{-1/2}(z_2) &=& \half(-,-,+,+,+) \;, \nn \\
b_1^{-1}(z_3) &=& \half(0,0,--,0,0) \;, \nn \\
b_2^{-1}(z_4) &=& \half(0,0,0,--,0) \;, \nn \\
b_3^{-1}(z_5) &=& \half(0,0,0,0,--) \;,
\eea
where as usual the $z_i$ denote the insertion points on the worldsheet and the superscripts give the ghost charges. In the canonical picture as in section \ref{sec:vertexop} the vertex operators give a total ghost charge of -4 which means that we require 4 picture changing operators to reach the correct picture on the cylinder. In order to conserve H-charge the 4 PCOs split into 2 pairs of opposite H-charge. We choose the PCOs to act on the second gaugino and on the 3 bosons. Each PCO can act in an internal or external direction and we decompose the calculation into parts according to this action.

\subsection*{Amplitudes with no internal PCOs}

We first consider the case where all picture changing operators act on the external coordinates. There are many combinatorial
options here. We only do one case in detail but shall then explain how the other cases promote this to a Lorentz-invariant
structure and state the overall result.

The example we do corresponds to the picture-changed H-charges
\bea
g^{-1/2}_1(z_1) &=& \half(+,+,+,+,+) \;, \nn \\
g^{+1/2}_2(z_2) &=& \half(-,---,+,+,+) \;, \nn \\
b^0_1(z_3) &=& \half(0,++,--,0,0) \;, \nn \\
b^0_2(z_4) &=& \half(++,0,0,--,0) \;, \nn \\
b^0_3(z_5) &=& \half(--,0,0,0,--) \;, \label{gravmedpcoexthcharge}
\eea
The superscripts give the picture-changed ghost charges of the operators. We now proceed to evaluate the correlators coming from the vertex operators and PCOs using the results of section \ref{sec:cftbuildblock}. The spin structure dependent part from fermion and ghost fields gives
\bea
&&\sum \eta_{\alpha \beta} \frac{\thba{\alpha}{\beta}{\frac{z_1 - z_2}{2} + z_4 - z_5} \thba{\alpha}{\beta}{\frac{z_1 - 3z_2}{2} + z_3}
}{\thba{\alpha}{\beta}{\frac{z_1 - z_2}{2}}} \nn \\
&&\times \thba{\alpha}{\beta}{\frac{z_1 + z_2}{2} - z_3 + \theta_1} \thba{\alpha}{\beta}{\frac{z_1 + z_2}{2} - z_4 + \theta_2}
\thba{\alpha}{\beta}{\frac{z_1 + z_2}{2} - z_5 + \theta_3}\;. \label{gravspin}
\eea
Here the $\theta_i$ are the orbifold twisting angles. The spin structure independent part (for fermion and ghost fields) gives
\be
\frac{\eta^3}{\theta_1(z_1 - z_5)} \frac{\eta^3}{\theta_1(z_4 - z_5)}
\frac{\eta^6}{\theta_1(z_2 - z_3)^2} \frac{\eta^3}{\theta_1(z_2 - z_4)} \;. \label{gravnonspin}
\ee
Now these correlators are supplemented by the bosonic correlators and are multiplied by a momentum factor coming form the picture changing of $k_2^{2+} k_3^{2-} k_4^{1-} k_5^{1+}$. The other external picture changing possibilities promote this to the Lorentz-invariant
structure $(k_2 \cdot k_3)(k_4 \cdot k_5)$. As discussed in section \ref{sec:momenexpopole}, we should cancel these factors with poles in the amplitude. Since the gauginos and bosons are on opposite ends of the cylinder it is not possible to bring them together which means it is not possible to generate poles from functions such as $\theta_1\left(z_1-z_5\right)$. Therefore from (\ref{gravspin}) and (\ref{gravnonspin}) we see that it is not possible to cancel the momentum factors by poles alone. However we shall see that in the UV limit of the amplitude the momentum factors will still be canceled leaving a finite answer. We begin by using the pole from $\theta_1(z_4 - z_5)$ to cancel the factor of $(k_4 \cdot k_5)$. In that case we set $z_4=z_5$ and remove the term $(k_4 \cdot k_5) \eta^3/\theta_1(z_4 - z_5)$. The resulting combined amplitude, after using the Riemann summation formula (\ref{rieiden1}) reads
\be
-2 \left(k_2 \cdot k_3\right) \eta^{12} \frac{\theta_1\left(-z_2+z_5+\theta_1\right)\theta_1\left(-z_2+z_3+\theta_2\right)\theta_1\left(-z_2+z_3+\theta_3\right)}{\theta_1\left(-z_2+z_5\right)\theta_1\left(-z_2+z_3\right)\theta_1\left(-z_2+z_3\right)} \;. \label{ampton4}
\ee
To extract a non-vanishing contribution from such an amplitude let us consider the $N=4$ sector first and we return to the other possibilities later. In that case we can set $\theta_1=\theta_2=\theta_3=0$.
We also need to include the extra factor coming from the bosonic partition function of $\eta^{-12}$ and also
$Z(t)$ from the winding string partition function, where $Z(t)$ is the trace over winding states stretching between the
two stacks of branes. There is also a factor $\langle \prod_i e^{ik_i \cdot X(z_i)} \rangle$ from the bosonic correlators.
The overall amplitude is then given by
\be
\label{vader}
{\cal A} = A_0 (k_2 \cdot k_3)\int \frac{dt}{t} \frac{1}{(t/2)^2} \int dz_1 dz_2 dz_3 dz_5
\langle \prod_i e^{ik_i \cdot X (z_i)}\rangle  Z(t)\;.
\ee
Here $A_0$ is an unimportant constant normalisation factor.
Let us temporarily neglect the $\langle \prod_i e^{ik \cdot X} \rangle$ terms to study the divergence
 structure. After performing the $z$ integrals we would be left with
\be
\label{chewie}
\mc{A} \sim (k_2 \cdot k_3) \int dt \, t \, Z(t).
\ee
Naively this appears finite in the limit that $t \to 0$.
However, as discussed in section \ref{sec:themodel}, for a compact space the winding mode partition function, in the case of a double trace operator, behaves as (recall that we are in the $N=4$ sector so all the tori are untwisted)
\be
Z(t) \sim \prod_i \left( \sum_{n_i,m_i} e^{-((\Delta R)^2 + (n_i^2  + m_i^2 ) R_i^2) t} \right)
\to_{t \to 0} \frac{1}{R_1^2 R_2^2 R_3^2 t^3} \sum_{n,m} e^{-(n_i^2 + m_i^2)/(R_i^2 t)}
\ee
using the Poisson resummation formula (\ref{poisson}).
In the limit that $t \to 0$ the amplitude therefore behaves as
\be
k_2 \cdot k_3 \int \frac{1}{R^6} \frac{dt}{t^2}
\ee
giving an ultraviolet divergence as $t \to 0$ which is the open string UV.
However there are also the factors of
\be
\langle \prod_i e^{i k \cdot X} \rangle = \prod_{i < j} e^{-k_i \cdot k_j  \mc{G}(z_i - z_j)}
\ee
which need to be taken into account.
From section \ref{sec:cftbuildblock} we have that
\be
\mc{G}(z_i - z_j) = -2\alpha' \ln |\theta_1\left(z_i-z_j\right)|^2 + \frac{8 \pi \alpha'}{t} \left(\hbox{Im}(z_i - z_j)\right)^2 \;,
\ee
where recall that the correlator can only be in the external Neumann directions because of the vanishing internal momentum.
Using the transformation property
\be
\theta_{1}(z,\tau) = i(-i \tau)^{-1/2} e^{-\pi i z^2/\tau} \theta_{1}(z/\tau, -1/\tau) \;,
\ee
we have that in the limit $t \to 0$
\be
\label{luke}
\theta_{1}(z,t) \to \sqrt{\frac{2}{t}} \;e^{-2\pi z^2/t} e^{-\pi/2t} \left(e^{2\pi z/t}-e^{-2\pi z/t} \right)\;.
\ee
Now we are interested in studying the leading contribution in this limit to the bosonic correlators. There are two distinct possibilities: either the two operators are on the same end of the cylinder, in which case $z=ixt/2$ or on opposite ends of the cylinder which gives $z=1/2+ixt/2$. Here $x$ is some constant fraction which gives the relative separation between the operators on the boundary. Now for $x \neq 1$ we have that as $t \to 0$
\be
\mc{G}(z_i - z_j) \to \left\{ \begin{array}{cl} \frac{\pi\alpha'}{t} & \hbox{ for operators on same end} \\
2\alpha'\ln t & \hbox{ for operators on opposite ends } \end{array} \right.
\ee
As a result the dominant contributions to $\langle e^{i k \cdot X} \rangle$ come from operators on the same end of the cylinder,
and we obtain
\be
\langle \prod_i e^{i k_i \cdot X(z_i)} \rangle \sim e^{- \pi \alpha' \frac{(k_1 + k_2)^2}{t}} \equiv e^{- \pi \alpha'\frac{(k_3 + k_4 + k_5)^2}{t}}\;.
\ee
The divergence then takes the form
\be
\frac{k_2 \cdot k_3}{R^6} \int \frac{dt}{t^2}  e^{-\pi \alpha'\frac{k_1 \cdot k_2 }{t}} \to \frac{1}{R^6}\frac{k_2 \cdot k_3}{k_1 \cdot k_2} \;.
\ee
The fact that the zero behaves as $(k_2 \cdot k_3)$ is simply an artifact of which terms we evaluated in the PCO
(as there is a symmetry in $k_3, k_4, k_5$). It is easy to see that other choices of picture changing give identical results with $k_3 \to k_4, k_5$.
Summing all contributions the pole structure then becomes
\be
\frac{k_2 \cdot (k_3 + k_4 + k_5)}{k_1 \cdot k_2} \frac{1}{R^6} = \frac{k_2 \cdot (-k_1 -k_2)}{k_1 \cdot k_2} \frac{1}{R^6}
\to \frac{1}{R^6}
\ee
in the on-shell limit.

This $R^{-6}$ behaviour of the induced gaugino mass is fully in accord with supergravity expectations, as it corresponds to mediation
via the propagation of the dilaton, which is suppressed by $\mc{V}^{-1} = R^{-6}$ as the dilaton probes the entirety of the compact space.

Finally we can return to the $N=1$ and $N=2$ contributions to (\ref{ampton4}). From the results above we see that for operators on opposite ends of the cylinder we have that as $t \to 0$, $\theta_1(z,t) \to \sqrt{2/t}$. Therefore the extra contributions of $t^{3/2}$ and $t$ for $N=1$ and $N=2$ sectors respectively compared to the $N=4$ sector mean that the UV divergence is regulated and so the amplitude vanishes due to the momentum factors.

\begin{figure}
\begin{center}
\epsfig{file=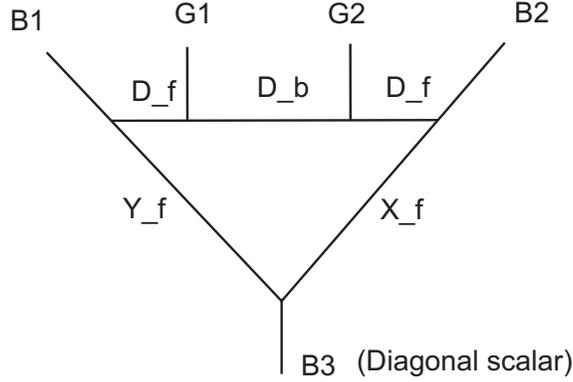,height=5cm}
\caption{The field theory diagram giving the infrared divergence. B denotes a bosonic field and G a gaugino.}
\label{pentagon}\end{center}\end{figure}

There is also a (potential) infrared divergence in (\ref{chewie})
as $t \to \infty$. This divergence is present only in the case that the bosons and the gauginos are on the same stack, so that
$Z(t) \to 1$ as $t \to \infty$.
This divergence is likewise regulated by the exponentials from the $\langle e^{i k \cdot X} \rangle$ correlator.
This regulator gives schematically
\be
\int t dt e^{-k_i \cdot k_j t} \sim \frac{1}{(k_i \cdot k_j)^2} \;,
\ee
giving a double pole in momentum. The overall infrared divergence is then given by a single pole, behaving as
$(k_i \cdot k_j)^{-1}$. This divergence is not associated to the induced soft gaugino masses but rather can be understood in field theory as arising from the pentagon diagram of figure \ref{pentagon}.
In the limit that all external momenta vanish the pentagon loop integral diverges as
\be
\int d^4 k \frac{1}{k^6} \sim \frac{1}{\mu^2} \;,
\ee
consistent with the pole found in $(k_i \cdot k_j)^{-1}$. However the integral over the vertex operator coordinates is less
straightforward as in the large $t$ limit
\be
\theta_{1}(z,t) = -2 e^{-\pi t/8} \sin (\pi z) \;,
\ee
and so the integral over $z$ does not decouple.

\subsection*{Amplitude with internal PCOs}

We now consider the case where at least one PCO is internal, which by H-charge conservation implies that a conjugate pair of PCOs must be internal.

First consider the case where this pair of PCOS acts on a pair of bosons. Since in any particular internal direction only one boson has H-charge it must be that for one boson the PCO contractions must involve the derivative term $\partial X^i$ rather than the $\psi^i$.
In the external direction case this could contract with the external momentum to generate a picture-changed operator.
However this is not possible for the internal directions since we have D3 branes and there are no internal momentum modes.
This term must therefore vanish.

The only other possibility is to have one external pair of PCOs acting on bosons and one internal pair of PCOs acting on a boson and a gaugino.
Let us take this latter pair for now as $g_2(z_2)$ and $b_3(z_5)$ and remember
that we subsequently need to sum over cyclic permutations of the bosons. In this case we have the picture changed H-charges
\bea
\label{someintspinstru}
g_1(z_1) &=& \half(+,+,+,+,+) \;, \nn \\
g_2(z_2) &=& \half(-,-,+,+,-) \;, \nn \\
b_1(z_3) &=& \half(++,0,--,0,0) \;, \nn \\
b_2(z_4) &=& \half(--,0,0,--,0) \;, \nn \\
b_3(z_5) &=& \half(0,0,0,0,0) \;,
\eea
As before there are similar cases that complete the above amplitude into a Lorentz covariant structure.

The spin structure dependent part is
\bea
& & \thba{\alpha}{\beta}{\frac{z_1 - z_2}{2} + z_3 - z_4}\thba{\alpha}{\beta}{\frac{z_1 + z_2}{2} - z_3 + \theta_1}
\thba{\alpha}{\beta}{\frac{z_1 + z_2}{2} - z_4 + \theta_2}\thba{\alpha}{\beta}{\frac{z_1 - z_2}{2} + \theta_3} \nn \\
& = & 2 \theta_1(z_1 - z_4) \theta_1(\theta_1) \theta_1(z_3 - z_4 + \theta_2) \theta_1(-z_2 + z_3 + \theta_3).
\eea
This clearly vanishes for the $\mc{N}=4$ sector due to the presence of the $\theta_1(\theta_1)$ term. The spin structure
independent part is
\be
\frac{\eta^3}{\theta_1(z_1 - z_4)} \frac{\eta^3}{\theta_1(z_2 - z_3)} \frac{\eta^3}{\theta_1(z_3 - z_4)}
\ee
This gives for the combined spin and ghost systems
\be
\label{aleph}
2 \frac{\theta_1(z_3 - z_4 + \theta_2)}{\theta_1(z_3 - z_4)} \frac{\theta_1(-z_2 + z_3 + \theta_3)}{\theta_1(z_2 - z_3)}
\theta_1(\theta_1) \eta^9.
\ee
It will turn out that the only relevant case is the $N=2$ sector where $\theta_3 = 0$. To see this, let us
consider the contributions of the bosonic fields in the case that the $X^5$ direction is twisted.

\subsubsection*{Twisted $N=1$ and $N=2$ sectors}

The $N=1$ and $N=2$ sectors have two types of contributions according to whether the direction is twisted or untwisted. All the $N=1$ directions are twisted and there is a single $N=2$ untwisted direction. We return to the latter case in the next section and consider first the twisted contributions which we show vanishes. The analysis applies to any twisted direction but for explicitness we denote the direction $X^5$.

First note that since $b_3\left(z_5\right)$ has no H-charge, no ghost charge and has vanishing internal momentum $z_5$ does not appear in the amplitude apart from in the correlator
\be
\label{3bostwistint}
{\cal A}_{N=1,2} \sim \int dz_5 \left<\partial_n X^{5}(z_2)\partial_n \overline{X}^{5}(z_5) \right> \;.
\ee
Since the direction is twisted there is no classical contribution from winding modes and there is only the quantum contribution which is given by (\ref{twiscorrderv}). Since (\ref{twiscorrderv}) is a total derivative we can perform the $z_5$ integral exactly.
However before doing this explicitly we should note an important contribution coming from the Chan-Paton factors which read
\be
\mr{Tr}\left(\Phi^1\Phi^2\Phi^3 \Theta \right) \;,
\ee
where $\Phi^i$ denote the CP factors of the three bosons and $\Theta$ is the CP
factor of the twisting angle of $X^5$, see section \ref{sec:themodel}. The important point is that as we
integrate the position of $b_3$ around the boundary circle we have to commute $\Phi^3$ and $\Theta$ which, using (\ref{thetphicomm}), gives an extra factor of $e^{2\pi i\theta_3}$ compared to the pure bosonic correlator.
Taking this factor into account (\ref{3bostwistint}) gives
\be
{\cal A}_{N=1,2} \sim \left[ \left(\frac{\theta_1 ( z_2 + \theta_3)}{\theta_1 (\theta_3)} \frac{\theta_1^\prime (0)}{\theta_1 (z_2)}\right)
 - e^{2\pi i\theta_3}\left( \frac{\theta_1 ( z_2 -\frac{it}{2} + \theta_3)}{\theta_1 (\theta_3)} \frac{\theta_1^\prime (0)}{\theta_1 (z_2 -\frac{it}{2})}\right)\right] = 0 \;,
\ee
where we take the bosons to be on the boundary at $\re{z}=0$ and used the transformation property (\ref{inttwiscorrthetatrans}).
The CP phase essentially cancels the phase from the twisted bosonic correlator, and as a result
all such twisted contributions integrate to zero and vanish.

\subsubsection*{Untwisted $N=2$ sectors}

The upshot of this is that the only contributions can come from the $\mc{N}=2$ sector with $\theta_3 = 0$.
This implies that the bosonic oscillator partition function takes the form
$$
\frac{1}{\theta_1(\theta_1) \theta_1(\theta_2) \eta^6}
$$
and so (\ref{aleph}) becomes
\be
-2 \frac{\eta^3 \theta_1(z_3 - z_4 + \theta_1)}{\theta_1(z_3 - z_4) \theta_1(\theta_1)}.
\ee
Summing over all choices of PCOs now completes the amplitude into a Lorentz covariant structure, giving for
the full amplitude
\bea
{\cal A}^{\mr{Untwisted}}_{N=2} &=& A_0 \mr{Tr}\left(g_1g_2\right)\mr{Tr}\left(\Phi_1\Phi_2\Phi_3\right)\int_0^{\infty} \frac{dt}{t} \frac{1}{\left(it/2\right)^2} \int dz_1dz_2dz_3dz_4dz_5 \nn \\
&\times& \left[ -8\sin^2\left(\pi\theta\right) \frac{\theta_1\left(z_{34}-\theta\right)}{\theta_1\left(\theta\right)}\frac{\theta'_1\left(0\right)}{\theta_1\left(z_{34}\right)}
\left<\prod_i e^{ik_i \cdot X(z_i)} \partial_n X^5\left(z_2\right)\partial_n \overline{X}^5\left(z_5\right) \right> \right] \nn \\
 &\times& \left(-\frac{1}{16}\right) \left( k_3^{1+}k_4^{1-} + k_3^{1-}k_4^{1+} + k_3^{2+}k_4^{2-} + k_3^{2-}k_4^{2+} \right) \;.
\eea
Here $A_0$ is a normalisation. The traces are over the Chap-Paton indices. We then have the integrals over the cylinder modulus, having already included the external partition function to give the inverse powers of $t$, and the positions of the vertex operators. The part in the square brackets comes from the bosonic and spin correlators as in section \ref{sec:cftbuildblock}. Here $\theta=\frac12$ denotes the $N=2$ twisting angle on the first and second internal directions. The factor of $\frac{1}{16}$ comes from the 4 PCOs. The momentum factors come from contracting the derivatives in the PCOs with the exponentials.
 Finally we are left with the derivative correlator which is in an internal direction and so is given by the Dirichlet correlator.

 There is a pole as $z_3 \to z_4$ which cancels the factor of $k_3 \cdot k_4$ and effectively puts $z_3 = z_4$. Extracting this pole using (\ref{polecancmom}) and (\ref{dirdercorrint}) the amplitude further simplifies to
\be
{\cal A}^{\mr{Untwisted}}_{N=2} = -A_0 \mr{Tr}\left(g_1g_2\right)\mr{Tr}\left(\Phi_1\Phi_2\Phi_3\right)\int_0^{\infty} \frac{dt}{t^3}
\int dz_1 dz_2 dz_3 dz_4 \left< \prod_i e^{i k \cdot X(z_i)} \partial_n X^5\left(z_2\right)\partial_n \overline{X}^5\left(z_5\right) \right>.
\ee
If we temporarily neglect the $e^{ik \cdot X}$ factors then this amplitude simplifies to
\bea
{\cal A}^{\mr{Untwisted}}_{N=2} & = & - A_0 \mr{Tr}\left(g_1g_2\right)\mr{Tr}\left(\Phi_1\Phi_2\Phi_3\right) \pi \alpha' \int_0^{\infty} dt
\left({\cal Z}_{\mr{Int,Cl}}+t\partial_t{\cal Z}_{\mr{Int,Cl}} \right) \;. \label{gravmedn2untwis} \\
& = & -A_0 \mr{Tr}\left(g_1g_2\right)\mr{Tr}\left(\Phi_1\Phi_2\Phi_3\right) \pi \alpha' \int_0^{\infty} dt \partial_t \left( t{\cal Z}_{\mr{Int,Cl}} \right)  \nonumber \\
& = & -A_0 \mr{Tr}\left(g_1g_2\right)\mr{Tr}\left(\Phi_1\Phi_2\Phi_3\right) \pi \alpha' \left[  t{\cal Z}_{\mr{Int,Cl}} \right]^{t=\infty}_{t=0} \;.
\eea
In the case that $Z_{Int,Cl} = 1 + \ldots $ (i.e. when the gauginos and bosons are on the same brane stack) there is an infrared
 divergence as $t \to \infty$. This divergence is similar to the one previously encountered for the case of external picture-changing.
 It is regulated by the exponentials and gives a pole of order $\frac{1}{k_i \cdot k_j}$ and its field theory origin is presumably
 the pentagon diagram of figure \ref{pentagon}. For the case of separate brane stacks the loop of the pentagon involves massive
 fields and the $t \to \infty$ divergence is absent consistent with this interpretation.

In the ultraviolet limit $t \to 0$ we obtain a finite answer by Poisson resummation of the classical partition function:
\begin{align}
A=& - \frac{1}{8\pi^3} \mathrm{Tr[CP]}  \bigg[ t \sum_{n,m} \exp [ - \frac{\pi t}{\alpha'} \frac{T_2}{U_2} \bigg| n + U m + z \bigg|^2 \bigg]^\infty_0 \nonumber \\
=& - \frac{1}{8\pi^3} \mathrm{Tr[CP]} \frac{\alpha^\prime}{T_2} \sum_{n',m'} \exp \bigg[ -\frac{\pi \alpha^\prime}{t T_2 U_2}  | n' - \ov{U} m'|^2 - \frac{2\pi i}{U_2} \Im \bigg( z (n' - m' \ov{U} )\bigg) \bigg] \bigg|_{t=0} \nonumber \\
=& - \frac{1}{8\pi^3} \frac{\alpha'}{T_2}\mathrm{Tr[CP]}. \label{result1}
\end{align}
This term is associated to generation of the gaugino mass by propagation of the $\mc{N}=2$ twisted mode in the third toroidal direction.
We note that this contribution scales only as $\sim \frac{1}{T_2}$, where $T_2$ is the area of the third torus. This is substantially
larger than the contribution due to the dilaton, which scaled as $\sim \frac{1}{\mc{V}}$, associated to the propagation of the dilaton
in the entirety of the compact space. This is consistent as the twisted $\mc{N}=2$ mode can only propagate along a single torus and so
has a wavefunction that is much less diluted than that of the dilaton.

In summary, the study of the three boson amplitude $\langle \phi_1 \phi_2 \phi_3 \lambda \lambda \rangle$ is fully consistent with
field theory expectations. In the ultraviolet limit $t \to 0$
the string amplitudes give two sources of gaugino masses, one associated to mediation by the dilaton and one associated to
mediation by the $\mc{N}=2$ twisted mode. The latter gives a much larger contribution than the former, consistent with the localisation
of its wavefunction onto a single torus. In the case that gauginos and bosons are on coincident brane stacks
the amplitudes also diverge in the infrared limit, which is understood as a field theory effect associated to a divergent
pentagon loop diagram.

%%%%%%%%%%%%%%%%%%%%%%%%%%%%%%%%%%%%%%%%%%%%%%%%%%%%%%%%%%%%%%%%%%%%%%%%%%%%%%%%%%%%%%%%%%
\section{Anomaly mediated gaugino masses}
\label{sec:anommedgaugmass}
%%%%%%%%%%%%%%%%%%%%%%%%%%%%%%%%%%%%%%%%%%%%%%%%%%%%%%%%%%%%%%%%%%%%%%%%%%%%%%%%%%%%%%%%%%

In the last section we showed that string amplitudes can recreate the supersymmetry breaking effects induced by tree-level gravity mediation. In this section we apply the same principle to anomaly mediation. For the case of the 5-pt open string amplitude, the cylinder was the lowest topology
for which a non-zero amplitude was possible, and so represented the tree level amplitude. For the case of 3-form fluxes, the
lowest possible topology is the disk with a flux vertex operator in the interior and 2 gaugini on the boundary (studied in \cite{08071666}).
To study anomaly mediated contributions
 the appropriate amplitude is then the cylinder with closed string vertex operators in the bulk and two open string gaugino vertex operators on the boundary. It is well known that in type IIB string theory turning on background flux can break supersymmetry.
 We can therefore study the anomaly mediated gaugino masses induced by a non-supersymmetric background flux by studying the cylinder amplitude with gaugini on the boundary and a flux vertex operator in the interior.
This case is of particular interest due to the role of flux compactifications in moduli stabilisation.

We begin the section with a discussion regarding the predictions of anomaly mediation and pick out the particular aspects that we wish to test with a string calculation. We then perform the string calculation for NSNS and RR flux separately.

%%%%%%%%%%%%%%%%%%%%%%%%%%%%%%%%%%%%%%%%%%%%%%%%%%%%%%%%%%%%%%%%%%%%%%%%%%%%%%%%%%%%%%%%%%%%%%%%%%%%%%%%%%%%%%%%%%%%%%%%%%%%%%%%
\subsection{Anomaly mediation in supergravity}
\label{sec:anommedsuper}
%%%%%%%%%%%%%%%%%%%%%%%%%%%%%%%%%%%%%%%%%%%%%%%%%%%%%%%%%%%%%%%%%%%%%%%%%%%%%%%%%%%%%%%%%%%%%%%%%%%%%%%%%%%%%%%%%%%%%%%%%%%%%%%%

In \cite{hepth9303040,hepth9402005} gaugino masses were calculated for general supersymmetry breaking effects by studying Weyl, K\"ahler and Konishi anomalies in supergravity. However in \cite{hepth9810155,hepph9810442} a new contribution to the gaugino masses (also to other soft terms) was proposed which is often labeled anomaly mediation. This contribution was implemented into the full supergravity framework in \cite{hepth9911029} where the expression for the gaugino masses was given as
\be
m_{1/2} = -\frac{g^2}{16\pi^2}\left[\left(3T_G-T_R\right)m_{3/2} - \left(T_G-T_R\right)K_iF^i - \frac{2T_R}{d_R}F^i\partial_i\left(\mathrm{ln\;det\;}Z \right) \right]\;. \label{hearts2}
\ee
It is the first term proportional to $m_{3/2}$ that was pointed out in \cite{hepth9810155,hepph9810442}. Here we have
\be
K_i = \partial_i K \;,\;\; F^i = e^{K/2}K^{i\bar{j}}\left(\partial_{\bar{j}}\bar{W}+K_{\bar{j}}\bar{W}\right) \;,\;\; m_{3/2}=e^{K/2}\bar{W} \;.
\ee
An important point is that the kinetic matrix $Z$ should only be for fields charged under the gauge group of the gaugino. The $T_R$ and $T_G$ stand for the usual quantities appearing in the beta functions and $d_R$ is the dimension of the representation.

We shall in fact see that the formula (\ref{hearts2}) cannot be correct as it does not incorporate the effect of the NSVZ term
in the 1PI gauge couplings. We therefore modify (\ref{hearts2}) to
\be
m_{1/2} = -\frac{g^2}{16\pi^2}\left[\left(3T_G-T_R\right)m_{3/2} - \left(T_G-T_R\right)K_iF^i - \frac{2T_R}{d_R}F^i\partial_i\left(\mathrm{ln\;det\;}Z \right) + 2T_G F^I \partial_I \ln \left( \frac{1}{g_0^2}  \right) \right]\;. \label{gmassanom2}
\ee
This last term is proportional to the tree-level gaugino masses and so is expected to be always subleading. Nevertheless it is present and is important in reconciling the supergravity prediction with the string calculation. We use `anomaly mediation' to refer to the full content of
equation (\ref{gmassanom2}).

We wish to check (\ref{gmassanom2}) through a string calculation. First we should extract the predictions of (\ref{gmassanom2}) for the case of background flux. Recall that the three-form flux and superpotential are given by
\be
G = F - i S H, \qquad W = \int G_3 \wedge \Omega \;.
\ee
Using (\ref{bhkk}) and (\ref{matmoduli}) for the K\"ahler potential and matter metrics
the anomaly mediation formula (\ref{gmassanom2}) gives
\be
m_{1/2} = -\frac{g^2}{16\pi^2}\left[ -\left(T_G-T_R\right)\left(K_SF^S + K_UF^U\right) - \frac23 T_R K_U F^U - 2T_GK_SF^S \right] \;.
\ee
There is a subtlety here due to the presence of the blow-up modes. Although they do not affect the first 3 terms of (\ref{gmassanom2}) they could affect the last term since they appear in the gauge kinetic function. However, as discussed in footnote \ref{n2foot}, we do not expect an F-term to be induced for them by the flux.  It is simple to show that
\be
K_SF^S = \int G \wedge \overline{\Omega} \;.
\ee
Also using
\be
\partial_U \Omega = -\left(\partial_U K\right) \Omega + \chi^{(2,1)} \;,
\ee
which simply follows from $K \sim \ln{\int \Omega\wedge \overline\Omega}$, we find
\be
K_U F^U = -9 \left(U+\bar{U}\right) \int \overline{G} \wedge \overline{\chi}^{(1,2)} \;.
\ee
We therefore obtain the following field theory expectations for flux-induced gaugino masses
\bea
G^{(0,3)} (F^T \neq 0) &\rightarrow& m_{\lambda} = 0 \;, \\
G^{(1,2)} (F^U \neq 0) &\rightarrow& m_{\lambda} \sim \left (3T_G - T_R\right) m_{3/2} = \beta m_{3/2}= \mathrm{Tr} \Theta_{N=2} m_{3/2}\;, \\
G^{(2,1)} (\hbox{unbroken susy}) &\rightarrow& m_{\lambda} = 0  \;, \\
G^{(3,0)} (F^S \neq 0) &\rightarrow& m_{\lambda} \sim \left (3T_G-T_R\right) m_{3/2} =  \beta m_{3/2}= \mathrm{Tr} \Theta_{N=2} m_{3/2} \;. \label{fluxfterms}
\eea
Here the indices on the flux refer to the complex type. $\beta$ refers to the beta function of the gauge group of the gauginos and $\mathrm{Tr} \Theta_{N=2}$ is the trace over the $N=2$ CP indices. Note that this provides the first non-trivial check on the validity of the formula (\ref{gmassanom2}): the group factors combine to give the $N=2$ CP trace which arises from the end of the cylinder which does not have the gaugino vertex operators and so, apart from the orbifold twist operator, is empty. Tracing over an empty end of the string gives no contribution from $N=1$ sectors (in the absence of orientifolds) since this is the tadpole condition. This just leaves the $N=2$ and $N=4$ sectors, the latter of which we shall show vanishes.

The purpose of the string calculation is to study (\ref{gmassanom2}) and how it arises in string theory.
Some parts of (\ref{gmassanom2}) are closely related to the analogous anomalous contributions to gauge couplings.
One part for which this is not the case is the first term of (\ref{gmassanom2}).
So one of our focuses in the string computation will be to
extract particular predictions coming from this term and check them in a string calculation. We also find that rather than working with the complex flux $G$ we find it simpler to work with the NSNS and RR pieces separately. Further we only study the case where $G$ has $(0,3)$ and $(3,0)$ components, which implies $H$ and $F$ are composed of $(3,0)$ and $(0,3)$ components. So the complex-structure moduli F-terms always vanish. Therefore we would like to extract the coupling of the first term in (\ref{gmassanom2}) to this type of NSNS and RR flux.

First recall that we do not expect a string computation to be sensitive to the $c_0$ part of the complex 3-form flux $G$ as this would correspond to an amplitude with two closed string insertions (the 3-form flux and the RR 0-form). So for practical purposes
\be
G_3 = F_3 - i S H_3 \to F_3 - \frac{i}{g_s} H_3.
\ee
Let us add a factor of $\delta_{AM}$ to the first term of (\ref{gmassanom2})
\be
m_{1/2} = -\frac{g^2}{16\pi^2}\left[ \delta_{AM} \left(3T_G-T_R\right)m_{3/2} - \left(T_G-T_R\right)K_iF^i - \frac{2T_R}{d_R}F^i\partial_i\left(\mathrm{ln\;det\;}Z \right) + 2T_G F^I \partial_I \ln \left( \frac{1}{g_0^2}  \right) \right]\;, \label{gmassanom2dam}
\ee
so that $\delta_{AM} = 1(0)$ for the cases where this term is present (absent).
Now consider the case where we only turn on NSNS flux. In this case we have
\be
W = -i S \int H_3 \wedge \Omega, \qquad \bar{W} = i \bar{S} \int H \wedge \bar{\Omega}.
\ee
We also have the no-scale structure $K_T F^T = 3 \bar{m}_{3/2}$ where we abuse notation and let $T$ run over the blow-up moduli as well. The dilaton F-term reads
\be
F^S  = e^{K/2} K^{S \bar{S}} (\partial_{\bar{S}} \bar{W} + (\partial_{\bar{S}} K) \bar{W}) = (S + \bar{S})\bar{m}_{3/2},
\ee
and so $K_S F^S = - \bar{m}_{3/2}$. The tree-level gaugino mass is therefore
\be
M_{\lambda, tree} = \frac{F^S}{2 \hbox{Re}(S)} =  \bar{m}_{3/2}.
\ee
In this case we have at one loop
\begin{enumerate}
\item
Running gauge couplings
\be
\frac{1}{g^2(\mu)} = \frac{1}{g_{tree}^2} \left( 1 + \frac{g^2 b_a}{16 \pi^2} \ln \left( \frac{M_W^2}{\mu^2} \right) \right) \;.
\ee
\item
Running gaugino masses (using field theory running of gaugino masses)
\be
M_{\lambda, running} = \bar{m}_{3/2} \left( 1 - \left( \frac{g^2 b_a}{16 \pi^2} \right) \ln \left( \frac{M_W^2}{\mu^2} \right) \right) \;. \label{nsanompredrun}
\ee
\item
Anomaly-induced masses
\be
M_{\lambda,anomaly} = \frac{g^2 b_a}{16 \pi^2} \delta_{AM}  \bar{m}_{3/2} \;. \label{nsanompredanom}
\ee
\end{enumerate}
Note that here $b_a$ refers to the beta function of the gaugino gauge group which is given by the CP trace over the $N=2$ sector (\ref{fluxfterms}). From (\ref{nsanompredrun}) and (\ref{nsanompredanom}) we gain two sharp predictions that depend on $\delta_{AM}$ which we can test. The first is that there should be a mass term of the form (\ref{nsanompredanom}). If this is present, $\delta_{AM}=1$. If it is not, $\delta_{AM}=0$. It is a very clean test. The second prediction is this mass term should be of equal magnitude and opposite sign to the running mass term. It is these two predictions that we calculate and verify in section \ref{sec:nsnsflux}.

We can repeat a similar analysis for RR flux although the predictions turn out to be not as clean as the NSNS case. Suppose we only have RR flux. Then
\be
W = \int F_3 \wedge \Omega, \qquad \bar{W} = \int F_3 \wedge \bar{\Omega}
\ee
Then
\be
K_T F^T = 3 \bar{m}_{3/2} \;, K_S F^S = \bar{m}_{3/2}\;.
\ee
At tree-level we have using the standard supergravity formulae
\be
M_{\lambda, tree} = \frac{F^S}{2 \hbox{Re}(S)} = - \bar{m}_{3/2}.
\ee
At one-loop we therefore have
\begin{enumerate}
\item
Running gauge couplings
\be
\frac{1}{g^2(\mu)} = \frac{1}{g_{tree}^2} \left( 1 + \frac{g^2 b_a}{16 \pi^2} \ln \left( \frac{M_W^2}{\mu^2} \right) \right).
\ee
\item
Running gaugino masses (using field theory running of gaugino masses)
\be
M_{\lambda, running} = -\bar{m}_{3/2} \left( 1 - \left( \frac{g^2 b_a}{16 \pi^2} \right) \ln \left( \frac{M_W^2}{\mu^2} \right) \right).
\ee
\item
Anomaly-induced masses
\be
M_{\lambda,anomaly} = \frac{g^2 b_a}{16 \pi^2} (\delta_{AM} - 2)  \bar{m}_{3/2}.
\ee
\end{enumerate}
So we can probe the existence of the $\delta_{AM}$ term from the relative magnitude of the running gaugino mass
and the anomaly-induced gaugino mass. We study this in section \ref{sec:rrflux}.

For the case of RR flux the existence of an anomalous mass term is not in itself a check on the parameter $\delta_{AM}$, as we instead
need the relative sign and magnitude of the running and anomalous mass terms.

%%%%%%%%%%%%%%%%%%%%%%%%%%%%%%%%%%%%%%%%%%%%%%%%%%%%%%%%%%%%%%%%%%%%%%%%%%%%%%%%%%%%%%%%%%%%%%%%%%%%%%%%%%%%%%%%%%%%%%%%%%%%%%%%
\subsection{NSNS flux}
\label{sec:nsnsflux}
%%%%%%%%%%%%%%%%%%%%%%%%%%%%%%%%%%%%%%%%%%%%%%%%%%%%%%%%%%%%%%%%%%%%%%%%%%%%%%%%%%%%%%%%%%%%%%%%%%%%%%%%%%%%%%%%%%%%%%%%%%%%%%%%

We seek to compute a correlator $\bra \lambda \lambda H \ket$ which involves two gaugino vertex operators on the boundary and one flux vertex operator in the bulk of the cylinder.  The basic vertex operator for an NSNS potential is \cite{Polchinski}
\be
\mc{V}_B^{(-1,-1)} = e^{-\phi - \tilde{\phi}} B_{jk} \psi^j \tilde{\psi}^k e^{i k \cdot X(z, \bar{z})}. \label{bvertexop}
\ee
We require a constant field strength profile, $H = H_{345} (dX^3 \wedge dX^4 \wedge dX^5 + d \bar{X}^3 \wedge d \bar{X}^4 \wedge d \bar{X}^5)$. This can be achieved by taking
\be
B = H_{345} \left( X^5 dX^3 \wedge dX^4 + \bar{X}^5 d \bar{X}^3 \wedge d \bar{X}^4 \right). \label{hflux}
\ee
This profile is sufficient to generate the required field strength. Although permissible, we do not need
to consider terms of the form $B_{45} \sim X^3, B_{53} \sim X^4$. For convenience we recall the gaugino vertex operator
\be
{\cal V}_{\lambda}^{-\half} \left(z\right) = e^{-\frac{\phi}{2}} S_{10} e^{ik \cdot x} \left(z\right)\;.
\ee
In their canonical pictures the H-charges of the gauginos are given by
\begin{align}
g_1^{-1/2}(z_1) =& \half(+,+,+,+,+) \;, \nn \\
g_2^{-1/2}(z_2) =& \half(-,-,+,+,+) \;.
\end{align}
For an H-charge conserving amplitude we require the anti-holomorphic part of the flux $H$ in (\ref{hflux}) so that the bosonic operators in (\ref{bvertexop}) have negative H-charges. The extra H-charge changes required for neutrality is induced by the 3 PCO insertions.
The full anti-holomorphic part of the vertex operator is then
\be
\mc{V}_B^{(-1,-1)} = H_{345} e^{-\phi(w) - \tilde{\phi}(\bar{w})} \overline{X}^5(w, \bar{w})
\left( \overline{\psi^3}(w) \overline{\tilde{\psi}^4}(\bar{w}) - \overline{\psi^4}(w) \overline{\tilde{\psi}^3}(\bar{w}) \right) e^{i k \cdot X(w, \bar{w})}.
\ee
We have introduced the left-moving and right-moving vertex operator insertion positions on the cylinder $w$ and $\bar{w}$.

For the Annulus amplitude we need to picture change to the $(0,0)$ picture. When we picture change $\mc{V}_B^{-1,-1}$ to $\mc{V}_B^{0,-1}$, we get
\bea
\mc{V}_B^{(0,-1)} & = & \half H_{345}e^{-\tilde{\phi}(\bar{w})} e^{i k \cdot X(w)} \left[
-\alpha' \overline{\psi^5}(w) \left( \overline{\psi^3}(w) \overline{\tilde{\psi}^4}(\bar{w}) - \overline{\psi^4}(w) \overline{\tilde{\psi}^3}(\bar{w}) \right)   \right. \nn \\
 &+& \overline{X}^5(w) \left( \partial\overline{X}^3(w) -\frac{i\alpha'}{2} \left(k\cdot\psi\right)\overline{\psi^3}(w) \right) \overline{\tilde{\psi}^4}(\bar{w}) \nn \\
 &-& \left.\overline{X}^5(w) \left( \partial\overline{X}^4(w) -\frac{i\alpha'}{2} \left(k\cdot\psi\right)\overline{\psi^4}(w) \right) \overline{\tilde{\psi}^3}(\bar{w}) \right]\;.
\eea
We have introduced the notation $k\cdot\psi=\left(k^+\cdot \overline{\psi} + k^-\cdot \psi \right)$. For the $(0,0)$ picture we therefore obtain
\bea
\mc{V}_B^{(0,0)} & = & \frac14 H_{345} e^{i k \cdot X}\left[ -\alpha'\overline{\psi^5} \left( \overline{\psi^3}\left(\overline{\partial}\overline{\tilde{X}^4}  - \frac{i\alpha'}{2}\left(k\cdot\tilde{\psi}\right)\overline{\tilde{\psi}^4} \right)
 - \overline{\psi^4}\left(\overline{\partial}\overline{\tilde{X}^3}  - \frac{i\alpha'}{2}\left(k\cdot\tilde{\psi}\right)\overline{\tilde{\psi}^3} \right)  \right) \right. \nn \\
 &-& \alpha' \overline{\tilde{\psi}^5} \left( \left(\partial \overline{X}^3 - \frac{i\alpha'}{2}\left(k\cdot\psi\right)\overline{\psi^3} \right) \overline{\tilde{\psi}^4} - \left(\partial \overline{X}^4 - \frac{i\alpha'}{2}\left(k\cdot\psi\right)\overline{\psi^4} \right) \overline{\tilde{\psi}^3} \right)  \nn \\
&+& \overline{X}^5 \left( \left(\partial \overline{X}^3 - \frac{i\alpha'}{2}\left(k\cdot\psi\right)\overline{\psi^3} \right) \left(\partial \overline{\tilde{X}}^4 - \frac{i\alpha'}{2}\left(k\cdot\tilde{\psi}\right)\overline{\tilde{\psi}^4} \right) \right.\nn \\
&-& \left. \left. \left(\partial \overline{X}^4 - \frac{i\alpha'}{2}\left(k\cdot\psi\right)\overline{\psi^4} \right) \left(\partial \overline{\tilde{X}}^3 - \frac{i\alpha'}{2}\left(k\cdot\tilde{\psi}\right)\overline{\tilde{\psi}^3} \right) \right) \right] \;. \label{vb1}
\eea
The computation is on an annulus with Dirichlet boundary conditions on the boundary for internal directions
and Neumann boundary conditions for external directions. We can deal with these by replacing
 $\tilde{\psi}(\bar{w})$ by $-\psi(- \bar{w})$ (Dirichlet) and
 $\tilde{\psi}(\bar{w})$ by $\psi(- \bar{w})$ (Neumann).
Also the amplitude computation requires us to picture change a single gaugino and the structure of the internal gaugino
H-charges implies the $(0,0)$ picture B vertex operator must contain at least two internal
$\overline{\psi}$, $\overline{\tilde{\psi}}$ fields to be able to give a non-vanishing amplitude. Applying these two constraints gives the effective $(0,0)$ picture flux operator as
\bea
\label{dbdb}
\mc{V}_B^{(0,0)}\left(w,-\bar{w}\right) &=& \frac14 H_{345} e^{i k \cdot X} \left[
- \alpha' \overline{\psi^5}(w) \left( -\overline{\psi^3}(w) \left( \overline{\partial} \overline{X^4} (-\bar{w}) - \frac{i \alpha'}{2} \left(k\cdot\psi(-\bar{w})\right) \overline{\psi^4}(-\bar{w}) \right) \right. \right. \nn \\
&+&  \left. \overline{\psi^4}(w) \left( \overline{\partial} \overline{X^3}(-\bar{w}) - \frac{i\alpha'}{2} \left(k \cdot \psi(-\bar{w})\right)
\overline{\psi^3}(-\bar{w}) \right) \right) \nn \\
 &-& \alpha' \overline{\psi^5}(-\bar{w}) \left( \left( \partial \overline{X^3} (w) - \frac{i \alpha'}{2} \left(k \cdot \psi(w)\right) \overline{\psi^3}(w) \right) \overline{\psi^4}(-\bar{w}) \right. \nonumber \\
&-& \left.\left( \partial \overline{X^4}(w) - \frac{i \alpha'}{2} \left(k \cdot \psi(w)\right) \overline{\psi^4}(w) \right) \overline{\psi^3}(-\bar{w}) \right)  \\
&+& \left. \left(\frac{\alpha'}{2}\right)^2 \overline{X^5}(w, \bar{w}) \left(k \cdot \psi(w)\right)\left(k \cdot \psi(-\bar{w})\right) \left(
\overline{\psi^4}(w)\overline{\psi^3}(-\bar{w}) - \overline{\psi^3}(w)\overline{\psi^4}(-\bar{w}) \right) \right] \; . \nn
\eea
There are three different kind of terms in the vertex operator (\ref{dbdb}) that can give a non-zero amplitude.

The first case are terms of the form
\be
\left(k \cdot \psi (w)\right) \overline{\psi^4}(w) \left(k \cdot \psi(-\bar{w})\right) \overline{\psi^3}(-\bar{w})\;.
\ee
A non-zero amplitude requires picture changing of the gaugino in the internal direction, with the resulting H-charges being (for example)
\begin{align}
g_1^{-1/2}(z_1) =& \half(+,+,+,+,+) \;, \nn \\
g_2^{+1/2}(z_2) =& \half(-,-,+,+,-) \;, \nn \\
\psi (w) =& (+1,0,-1,0,0) \;, \nn \\
\psi (-\bar{w}) =& (-1,0,0,-1,0) \;.
\end{align}
This carries a coefficient $k_3^{1-} k_3^{1+}$ from the external PCOs. There are three similar terms with different external PCOs which complete the coefficient of this term into $k_3 \cdot k_3$. The spin-structure dependent part gives
\be
2 \theta_1(z_1 + \bar{w})\theta_1(-z_2 + w) \theta_1(w + \bar{w} - \theta) \theta_1(\theta) k_3 \cdot k_3.
\ee
Here $\theta$ is the orbifold twist along the first internal direction and $-\theta$ is the twist along the second. The last direction is untwisted since we know the $N=1$ sectors vanish from the CP trace. Note that setting $\theta=0$ which corresponds to the $N=4$ sector gives a vanishing contribution. The spin-structure independent part of the fermionic amplitude is
\be
\theta_1(z_1 + \bar{w})^{-1} \theta_1(z_2 - w)^{-1} \theta_1(w + \bar{w})^{-1},
\ee
and so overall we get (including the bosonic part)
\bea
{\cal A} &=& A_0 \int \frac{dt}{t} \frac{1}{(t/2)^2} \int dz_1 dz_2 d^2 w \frac{\theta_1(w + \bar{w} - \theta) \theta_1(\theta)}
{\theta_1(w + \bar{w}) \theta_1(\theta) \theta_1(-\theta)} k_3 \cdot k_3
\bra \partial_n \overline{X}^{5}(z_2) X^5(w,-\bar{w}) \ket \;. \label{nsnsamp1}
\eea
Here $A_0$ is some overall normalisation factor which does not play a role in the calculation. There are two non-vanishing contributions from this expression associated to the classical and quantum correlator pieces. The quantum correlator is given in (\ref{corrfac}) which we recall here for convenience
\be
\bra \partial_n \overline{X}^{5}(z_2) X^5(w,-\bar{w}) \ket_{\mr{{\cal A},Qu}}^{\mr{D}} = -\alpha'\left[ \frac{w + \bar{w}}{(z_2 - w)(z_2 + \bar{w})} \right]\;. \label{nsnsqcor}
\ee
We see that this induces a pole when $w \to z_2$ which takes the form
\be
-\alpha' k_3 \cdot k_3 \int d^2 w \frac{1}{(z_2 - w)(z_2 + \bar{w})} = - 2\pi \alpha' \int \frac{dr}{r}\;,
\ee
where have introduced $z_2-w=re^{i\theta}$ and integrated over the $\theta=\{0,\pi\}$ region (note $d^2w=2rdrd\theta$).
Note that $w \to z_2$ also implies $w \to -\bar{w}$ which gives a pole in (\ref{nsnsamp1}) from the $\theta_1(w+\bar{w})^{-1}$ factor that cancels against the zero in (\ref{nsnsqcor}). In this limit the exponential terms $\bra e^{ik_2 \cdot X(z_2)} e^{i k_3 \cdot X(w, -\bar{w})} \ket$ regularise the divergence, giving a pole in $2k_2 \cdot k_3 + k_3 \cdot k_3$. So overall this term
gives (for the vertex operator integrals)
\be
\label{yiq}
{\cal A}_1 = A_0 \int \frac{dt}{t} \frac{1}{(t/2)^2} \frac{-2\pi \alpha' k_3 \cdot k_3}{2 k_2 \cdot k_3 + k_3 \cdot k_3}\int dz_1 dz_2 \;,
\ee
where $k_i\cdot k_j = \half \left(k_i^+\cdot k^-_j + k_i^-\cdot k^+_j \right)$.

The second contribution comes from the classical correlator which from (\ref{linearclasspath}) reads
\be
\bra \partial_n X^{\bar{5}}(z_2) X^5(w,-\bar{w}) \ket_{\mr{{\cal A},Qu}}^{\mr{D}} = \Delta \bar{X}^5(z_2) X^5(w,-\bar{w}) \;.
\ee
The pole comes from $w + \bar{w} \to 0$ which can occur on either boundary of the annulus.  Here $\theta_1(w + \bar{w})
\sim (w + \bar{w}) \theta_1^{'}(0)$.
Note that even though this involves the closed string fields a single pole is sufficient
here, as the divergence occurs as we take the one-dimensional limit of moving the fields to the boundary. In fact there are two possible
ways of doing this, one by taking $\hbox{Re}(w) \to 0$ and the other by taking $\hbox{Re}(w) \to \half$.
There are then poles on either end of the annulus: $\hbox{Re}(w) \to 0$ (when $w + \bar{w} > 0$)
and when $\hbox{Re}(w) \to \half$ (when $w + \bar{w} < 0$). The resulting poles look like (note: $d^2w = 2 dw_1dw_2$ with $w_1$ and $w_2$ being the real and imaginary parts of $w$)
\bea
\int_0^{t/2} dw_2 \int_0^{1/2} \frac{dw_1}{w_1} \left|2w_1\right|^{k_3\cdot k_3}&\to& \frac{1}{k_3 \cdot k_3} \;\;(\hbox{near $w_1=0$}) \;, \nn \\
  &\to& \frac{-1}{k_3 \cdot k_3} (\hbox{near $w_1=1/2$}) \;.
\eea
We then obtain for the vertex operator integral
\be
\int dz_1 dz_2 dw_2 \bra \partial_n \overline{X}^{5} (z_2) \left( X^5(iw_2) - X^5 ( 1/2 + iw_2 ) \right) \ket.
\ee
Now we can write
\be
(X^5(iw_2) - X^5(\half + iw_2)) = -\Delta X^5(iw_2)   \;.
\ee
So we then obtain (\ref{classicalnormalderivative})
\be
-\int dz_1 dz_2 dw_2 \, \left|\Delta X \right|^2 = 2\pi \alpha' \int dz_1 dz_2 \, \, t\frac{d}{dt} Z(t) \;,
\label{df}
\ee
Here $Z(t)$ is the internal partition function summing over all the winding modes. Combining (\ref{yiq}) with (\ref{df}), we have overall
\bea
{\cal A}_1+{\cal A}_2 &=& A_0 \int \frac{dt}{t} \frac{2\pi\alpha'}{(t/2)^2} \left[ \frac{-k_3 \cdot k_3}{2 k_2 \cdot k_3 + k_3 \cdot k_3} \int dz_1 dz_2 Z(t) + \int dz_1 dz_2 t \frac{d}{dt} Z(t) \right] \nn \\
&=& 2\pi\alpha' A_0 \int \frac{dt}{t}  \left[ \frac{-k_3 \cdot k_3}{2 k_2 \cdot k_3 + k_3 \cdot k_3} Z(t) + t \frac{d}{dt} Z(t) \right] \;. \label{nsnsamp12}
\eea

The second set of terms from (\ref{dbdb}) that could contribute have the form
\be
\overline{\psi^5(w)} \overline{\psi^3(w)} \bar{\partial} \overline{X^4}(- \bar{w}) \;.
\ee
The form of the H-charges also require the gaugino to be
picture changed in the internal dimensions.
The H-charge structure of these is given by
\begin{align}
g_1^{-1/2}(z_1) =& \half(+,+,+,+,+) \;, \nn \\
g_2^{+1/2}(z_2) =& \half(-,-,+,-,+) \;, \nn \\
\psi (w) =& (0,0,-1,0,-1) \;, \nn \\
\psi (-\bar{w}) =& (0,0,0,0,0) \;.
\end{align}
The spin structure dependent parts give
\be
2 \theta_1(z_1 - z_3) \theta_1(\theta) \theta_1(-z_2 + z_3 - \theta) \theta_1(0).
\ee
and so these terms vanish automatically.\footnote{If we had allowed $B_{45}$ and $B_{53}$ to be non-zero, then these
terms could give a contribution. However as we only have $B_{34}$ non vanishing,
there is no term of the form $\overline{\psi^3}\overline{\psi^4} \bar{\partial} \overline{X^5}$, in which case all terms of this form vanish after the spin structure summation.}

The final possible case corresponds to terms in (\ref{dbdb}) of the schematic form
\be
\overline{\psi^5}(w) \overline{\psi^3}(w) \left(k \cdot \psi(-\bar{w})\right) \overline{\psi^4}(-\bar{w})\;.
\ee
In this case picture changing has to be in the external directions and there are two basic sub-options.
The first sub-option has the following H-charges:
\begin{align}
g_1^{-1/2}(z_1) =& \half(+,+,+,+,+) \;, \nn \\
g_2^{+1/2}(z_2) =& \half(-3,-,+,+,+) \;, \nn \\
\psi (w) =& (0,0,-1,0,-1) \;, \nn \\
\psi (-\bar{w}) =& (+1,0,0,-1,0) \;.
\end{align}
with a coefficient of $k_2^{1+} k_3^{1-}$.
%The factor of $2 \alpha'$ occurs as this is an open string picture changing and so has an extra factor of 4 compared to closed string
%picture changing \cite{Polchinski}.
The spin structure dependent parts give
\be
2 \theta_1(z_1 - w) \theta_1(-z_2 - \bar{w} + \theta) \theta_1(-z_2 + w - \theta) \theta_1(-z_2 - \bar{w}) \;.
\ee
The spin structure independent parts give
\be
\theta_1(z_1 - w)^{-1} \theta_1(z_2 - w)^{-1} \theta_1(z_2 + \bar{w})^{-2} \;.
\ee
The fermionic modes combine into
\be
-2 k_2^{1+} k_3^{1-} \frac{\theta_1(-z_2 - \bar{w} + \theta) \theta_1(-z_2 + w - \theta)}{\theta_1(\theta) \theta_1(-\theta) \theta_1(z_2 - w) \theta_1(z_2 + \bar{w})} \;. \label{ir1}
\ee
There are no bosonic contractions to be done. These have a clear pole as $w \to z_2$ giving\footnote{Note again that for an $N=4$ sector there is no pole and the contribution vanishes. The $\theta_1(\theta)$ factor in the denominator come from the bosonic sector and are not present in the $N=4$ case.}
\be
{\cal A}_3 = 2\pi\alpha' A_0 \frac{(-k_2^{1+}k_3^{1-} - k_2^{2+} k_3^{2-})}{2 k_2 \cdot k_3 + k_3 \cdot k_3} \int \frac{dt}{t} Z(t) \;. \label{nsnsa3}
\ee
We have included the contribution from the similar H-charge configuration along the second external direction.

The second sub-option (and final case overall) can have contractions between the PCO derivative piece and the second gaugino exponential piece. The H-charges are
\begin{align}
g_1^{-1/2}(z_1) =& \half(+,+,+,+,+) \;, \nn \\
PCO(u) =& (+1,0,0,0,0) \;, \nn \\
g_2^{+1/2}(z_2) =& \half(-,-,+,+,+) \;, \nn \\
\psi (w) =& (0,0,-1,0,-1) \;, \nn \\
\psi (-\bar{w}) =& (-1,0,0,-1,0) \;.
\end{align}
The spin structure dependent term gives
\be
2 \theta_1\left(z_1 + \half u + \half z_2 - w +\bar{w}\right) \theta_1\left(\half u -\half z_2 + \theta\right) \theta_1\left(\half u -\half z_2 + w +\bar{w} - \theta\right) \theta_1\left(\half u - \half z_2\right) \;.
\ee
We require to take the limit $u \to z_2$ in which case the following vanishes linearly which cancels against the pole obtained in contracting
$\partial X(u)$ with $e^{i k_2 \cdot X(z_2)}$. Therefore we have a momentum factor multiplying the amplitude of $k_2^{1-}k_3^{1+}$. The spin structure dependent part then gives
\be
- k_2^{1-}k_3^{1+} \theta_1\left(z_1 + z_2 - w +\bar{w}\right) \theta_1\left(\theta\right) \theta_1\left(w +\bar{w} - \theta\right)\;.
\ee
The spin-structure independent part gives
\be
\theta_1(z_1 - z_2) \theta_1(z_1 - w)^{-1} \theta_1(z_1 + \bar{w})^{-1} \theta_1(z_2 + \bar{w})^{-1} \theta_1(z_2 - w)^{-1} \;,
\ee
so that overall we obtain, adding the bosonic piece,
\be
-k_2^{1-} k_3^{1+} \frac{\theta_1\left(z_1 + z_2 - w +\bar{w}\right)\theta_1\left(w +\bar{w} - \theta\right)
\theta_1(z_1 - z_2) }{
\theta_1(z_1 - w)\theta_1(z_1 + \bar{w}) \theta_1(z_2 + \bar{w})\theta_1(z_2 - w)\theta_1(-\theta)} \;. \label{ir2}
\ee
This has a poles as $w \to z_1$ and as $w \to z_2$ (which also imply $w+\bar{w} \to 0$) giving a factor
\be
\pi\alpha' A_0 \left( \frac{-k_2^{1-} k_3^{1+}}{2 k_2 \cdot k_3 + k_3 \cdot k_3} +  \frac{k_2^{1-} k_3^{1+}}{2 k_1 \cdot k_3 + k_3 \cdot k_3} \right) \int \frac{dt}{t} Z(t) \;.
\ee
After using $2 k_2 \cdot k_3 + k_3 \cdot k_3 = - \left( 2 k_1 \cdot k_3 + k_3 \cdot k_3\right)$, this combines with a similar H-charge combination along the second external direction and with (\ref{nsnsa3}) to give a Lorentz covariant expression
\be
{\cal A}_3+{\cal A}_4 = 2\pi\alpha' A_0 \frac{-2k_2\cdot k_3}{2 k_2 \cdot k_3 + k_3 \cdot k_3} \int \frac{dt}{t} Z(t) \;. \label{nsnsamp34}
\ee
Finally combining (\ref{nsnsamp34}) and (\ref{nsnsamp12}) gives the overall amplitude
\be
{\cal A} = 2\pi\alpha' A_0 \int \frac{dt}{t} \left( -Z(t) + t \frac{d}{dt} Z(t)\right) \;. \label{nsnsamp}
\ee
This amplitude contains the information we require to test the predictions of section \ref{sec:anommedsuper}.

Therefore the only contribution comes from the $N=2$ sector for which the CP trace in $A_0$ gives the correct group theory factors to match (\ref{nsanompredrun}) and (\ref{nsanompredanom}). This is all we need to extract from the prefactors of the amplitude (\ref{nsnsamp}) and now turn to looking at the $t$ integral
\be
\int_0^{\infty} \frac{dt}{t} \left( -Z(t) + t \frac{d}{dt} Z(t)\right) = \left[ Z\left(\infty\right) - Z\left(0\right) \right] - \int_0^{\infty} \frac{dt}{t} Z(t) \;. \label{anomrunfinal}
\ee
It is worth recalling the form of the internal classical open string partition function that appears in (\ref{nsnsamp})
\bea
Z(t) &=& {\cal Z}_{\mr{Int,Cl}}(t) = \sum_i \sum_{n,m} \delta_i e^{-\frac{t}{4\pi\alpha'}|\Delta X_i(m,n)|^2} \;, \nn \\
\Delta X_i(m,n) &\equiv& 2\pi \sqrt{\frac{\im{T}}{\im{U}}} \left( n + U m + X_i \right) \;.
\eea
Here $n$ and $m$ sum over the winding modes, $T$ and $U$ are the torus (along the untwisted direction) K\"ahler and complex-structure moduli. We have added a sum over $i$ which stands for the different brane stacks present in the construction so that $X_i$ denote the separation between the gaugino brane stack and another stack at the other end of the cylinder. We denote the gaugino brane stack $i=0$ and the sum over all the states includes the $X_{i=0}=0$ states and also the states which are strings stretching between the stack and other ones required for tadpole cancellation. The factor $\delta_i$ accounts for the Chan-Paton traces so that tadpole cancellation implies that in the UV $t \to 0$ the partition function vanishes, see section \ref{sec:themodel} for more detail, which guarantees finiteness and so\footnote{Since we have not introduced orientifolds strictly speaking the partition function does not vanish due to the $N=4$ untwisted tadpole. However the $N=2$ tadpoles are canceled and so do not contribute in the UV.}
\be
Z(0) = 0 \;.
\ee
In the IR $t \to \infty$ only the massless modes can contribute which gives
\be
Z(\infty) = 1 \;.
\ee
We see that the expression $\int dt \frac{dZ}{dt}$ has the key features of anomaly mediation: it arises in the ultraviolet,
but its value is determined purely by the infrared spectrum.
The last term of (\ref{anomrunfinal}) gives the field theory running as well as the threshold corrections. Below the winding modes scale $M_W$ we can write (\ref{anomrunfinal}) as
\be
Z(\infty) \left( 1 - \int_{\frac{1}{M_W^2}}^{\frac{1}{\mu^2}} \frac{dt}{t} = 1 - \ln\;\left(\frac{M_W^2}{\mu^2} \right) \right) \;,
\ee
where $\mu$ is an IR regulator. Using all the above we see that
(\ref{anomrunfinal}) precisely reproduces the field theory predictions of (\ref{nsanompredrun}) and (\ref{nsanompredanom}): we find an anomaly mediated contribution and a running mass of equal magnitude and opposite sign.

It is worth looking a little closer at the anomaly mediated contribution
\bea
\label{firstttt}
Z(0) - Z(\infty) & = & \underset{t \to 0}{\mr{lim}} \sum_{\Delta X_i \neq 0} \sum_{n,m} e^{-\frac{t}{4\pi\alpha'}|\Delta X_i(m,n)|^2} \;  \\
\label{secondttt}
& \equiv & \int_{0}^{\infty} dt \sum_i \sum_{n,m} \delta_i \frac{| \Delta X_i(m,n) |^2}{4 \pi \alpha'}
e^{-\frac{t}{4\pi\alpha'}|\Delta X_i(m,n)|^2}.
\eea
We see that although the UV part $Z(0)$ vanishes leaving just a contribution from the IR piece $Z(\infty)$, we can think of this as
 the contribution from all the heavy modes not present in the IR limit i.e. $Z(0)=0$ only if we also include the IR modes.
 Another way to see this is to note that in (\ref{secondttt}) all contributing terms must have $\Delta X_i \neq 0$ and so arise from
 a heavy string mode with non-zero winding. Furthermore since values of $t$
 larger than $|\Delta X_i|^{-2}$ give only an exponentially suppressed contribution, we see the
  dominant contribution arise from values of the loop parameter $t \sim |\Delta X_i|^{-2}$.
  In that sense the anomaly mediated contribution comes purely from the heavy (open string) modes in the theory that lead to the finite UV completion. This point of view is perhaps more appropriate in the closed string channel where these heavy open string modes become the light closed string modes that are interacting with the flux as in the supergravity analysis of section \ref{sec:anommedsuper}.

We present further evidence for our result in appendix \ref{sec:nsapp}. There we probe the anomaly mediated mass by studying a 4-point amplitude where one of the gaugini is decomposed into a scalar and fermion, see figure 4. The advantage of a 4-point calculation is that the open string momenta can all be taken on-shell from the start of the calculation.
From a formal perspective this is more robust than the 3-point calculation presented above in which the open-string momenta are taken on-shell only at the end, as it avoids the need to have off-shell intermediate steps. We find precisely the same result with the anomaly mediated mass arising in the same way.

\subsubsection*{Infrared properties}

We here briefly discuss some of the infrared properties of our expressions and connect to field theory discussions of anomaly mediation.

First note that the expressions (\ref{ir1}) and (\ref{ir2}) take the form where in the IR limit $t \to \infty$ they give a finite contribution. This is a field theory effect first pointed out in \cite{West,JackJonesWest,West:1991qt} and a stringy realisation of it for the case of Yukawa couplings was studied in \cite{Conlon:2010xb}. The schematic reason is that, in (\ref{ir1}) for example, if we take $z_2-w=ixt/2$ where $x$ is their relative separation then as $t \to \infty$ the expression (\ref{ir1}) tends to a constant factor times the momentum, which is Lorentz completed to $k_2 \cdot k_3$. Then if we also include the bosonic correlators the full amplitude takes the schematic form
\be
k^2 \int_0^{\infty} dt e^{-k^2 t} \;.
\ee
This integral evaluates to a finite constant value. To see this consider the different ranges of the values of $t$. For $t \ll \frac{1}{k^2}$ we have that the exponential is essentially 1 and therefore the contribution to the integral goes like $k^2 t$ which is very small. Near the range $t \sim \frac{1}{k^2}$ the integral gets an order 1 contribution. In the limit $t \gg \frac{1}{k^2}$ the integrand vanishes due to the exponential factor and so again there is no substantial contribution to the integral. Therefore we see that the contribution comes from modes around $t \sim \frac{1}{k^2}$. Since in the on-shell limit $k^2 \rightarrow 0$ this is a strict IR effect. This is a sign that this term, while real, should be associated with the 1PI action: it requires the existence of massless particles and comes from the $k \to 0$ limit of the loop integral.

The field theory interpretation of such terms is as $\frac{1}{\square}$ diagrams in the 1PI action.
In \cite{Dine:2007me} anomaly mediation is understood as an ultraviolet counter term to such infrared divergences: while the mass
term is generated in the ultraviolet, it appears as a necessary counterterm to cancel the effect of certain infrared loops.
This is a similar structure as appears in
the expressions in \cite{hepth9911029}, where the anomalous Lagrangian term is
\be
\frac{-g^2}{256 \pi^2} \int d^2 \Theta \, 2 \mc{E} W^{\alpha} W_{\alpha} \frac{1}{\square} \left( \bar{\mc{D}}^2 - 8 \mc{R} \right) \left[ 4 (T_R
-3T_G) R^{+} \right] \;.
\ee
It is interesting to note that our results are consistent with this discussion, as the anomaly-mediated mass term arises from a sum over (ultraviolet) winding modes while the infrared limit of the amplitude has $\frac{1}{\square}$ terms.
In principle we could hope to obtain a full match to the formalism of \cite{Dine:2007me} by computing the coefficient of all the
infrared $\frac{1}{\square}$ terms, however we leave this to future work.

%%%%%%%%%%%%%%%%%%%%%%%%%%%%%%%%%%%%%%%%%%%%%%%%%%%%%%%%%%%%%%%%%%%%%%%%%%%%%%%%%%%%%%%%%%%%%%%%%%%%%%%%%%%%%%%%%%%%%%%%%%%%%%%%
\subsection{RR Flux}
\label{sec:rrflux}
%%%%%%%%%%%%%%%%%%%%%%%%%%%%%%%%%%%%%%%%%%%%%%%%%%%%%%%%%%%%%%%%%%%%%%%%%%%%%%%%%%%%%%%%%%%%%%%%%%%%%%%%%%%%%%%%%%%%%%%%%%%%%%%%

We now describe the analogous computation for the case of RR flux. While to some extent the techniques are similar, overall this
computation is much more involved both technically and conceptually. We shall therefore simply present a summary of the calculation in this section
while giving the full details in appendix \ref{sec:rrapp}.

The first aspect which makes the RR computation more involved is that the comparison with supergravity expectations is not so simple.
As with the NSNS case, we expect to find both a running and an anomalous mass term in the string computation. With the NSNS case,
the presence of the pure $m_{3/2}$ term was equivalent to the presence of an anomalous mass term in the string computation, due to
a cancellation in the other contributions. In the RR case this is no longer the case, and a probe of the pure $m_{3/2}$ term
can only be done via both the relative sign and magnitude of the running and anomalous mass terms.

The vertex operator for the RR field strength is given by
\beq
V_F^{(-1/2,-1/2)} = N_F g_s e^{-\phi/2 - \tilde{\phi}/2} F_{mnp} \Theta (z) C \Gamma^{mnp} \tilde{\Theta} (\tilde{z}),
\eeq
where a factor of $g_s$ is included as we take the sphere to have a factor $g_s^{-2}$. The differing factors of $g_s$
between RR and NS-NS vertex operators relate to the
fact that the RR vertex operator directly involves the field strength $F_{mnp}$
whereas the NS-NS operator involves the 2-form potential.
$\Theta, \tilde{\Theta}$ are ten-dimensional spin fields and $C$ the charge conjugation operator, given by
\beq
C = \Gamma^0 \Gamma^3 \Gamma^5 \Gamma^7 \Gamma^9,
\eeq
and hence since $\Theta$ and $\tilde{\Theta}$ have the same chirality only odd forms are allowed, as we expect from IIB.
The form of the $C \Gamma^{mnp}$ structure determines the relative H-charges of the two spinors $\Theta$ and $\tilde{\Theta}$,
as it raises and lowers the H-charges of the individual operators.

In the on-shell zero-momentum limit the 3-form flux takes the simple form
$$
F_{mnp} = F_{345} \left( dX_3 \wedge dX_4 \wedge dX_5 + d\bar{X}_3 \wedge d\bar{X}_4 \wedge d\bar{X}_5 \right).
$$
However as we have already seen we must work at non-zero momentum and only take the $k \to 0$ limit at the end of the
computation. In this case the 3-form field strength we must use is given by
\be
\label{keats1}
F_3 = \left( dX_3 \wedge dX_4 \wedge dX_5 e^{ik \cdot X} + i X_5 \left(k \cdot dX\right) \wedge dX_3 \wedge dX_4 e^{ik \cdot X} + \hbox{c.c} \right).
\ee
The second term is necessary to satisfy the Bianchi identity $dF_3 = 0$. In contrast to the NS-NS computation, there are then
two separate flux contributions which must be individually computed and then summed.

For each of these options there are many picture-changing possibilities. In appendix \ref{sec:rrapp} we compute these, and obtain the
following results. The first flux term in eq. (\ref{keats1}) gives an overall amplitude
of (from (\ref{milton}))
\be
\label{homer}
\pi \alpha' \int \frac{dt}{t} \, \left[ (4 + 4t \frac{d}{dt}) Z(t) + \frac{k_2 \cdot k_3 + 2 k_2 \cdot k_2}{2 k_2 \cdot k_3 + k_3 \cdot k_3} Z(t) \right].
\ee
The second flux term in (\ref{keats1}) ends up giving (from (\ref{betjeman}))
\be
\label{betjeman15}
2 \pi \alpha' \int \frac{dt}{t} Z(t)
- 2 \pi \alpha' \int dt \frac{d}{dt} Z(t).
\ee
There are some points to draw from this expression. First, we see that we do find both anomalous and running mass terms present in this
amplitude, as we would expect based on the general supergravity structure.
However there is one crucial difference compared to the NS-NS amplitude. In the NS-NS amplitude, there was no ambiguity
in how to return from the off-shell limit to the on-shell theory (where the combination
of equations (\ref{nsnsamp34}) and (\ref{nsnsamp12}) led to a cancellation of all off-shell quantities). However in
(\ref{homer}) the off-shell terms do not cancel and so it is not clear how to extract a precise numerical answer.
Similar behaviour has been seen in the calculation of 3-point Yukawa couplings, where the off-shell ambiguity
can be resolved by going to a 4-point calculation where all particles can be put on shell \cite{Conlon:2010xb}.

In principle a 4-point calculation could also be performed for the case of RR flux, analogous to that
 presented in appendix C. However here given the calculational intricacy already present in
the 3-point calculation, without some improved technical methods the 4-point RR calculation seems prohibitive.

\subsection{2 gaugino picture changing}

At this point we also discuss a technical issue in the calculations. This occurs when we try and cancel the H-charges by
picture changing both gaugini fields into the $+1/2$ picture. If this is done, it turns out that there is no source of anomalous
mass terms for either the NS-NS or RR flux computations. This is not consistent with  (\ref{gmassanom2dam}) for any value
of $\delta_{AM}$ and so must be erroneous.

We do not fully understand the reason for this (although we note that the arguments for picture-changing independence in
\cite{Friedan:1985ge} apply directly to worldsheets without boundaries and do not seem to generalise straightforwardly to mixed open-closed
amplitudes). However we suspect the following.

The aim of the computation is to compute gaugino masses in a 3-form flux background. Properly, the computation of gaugino
masses in a flux background would require formulating the worldsheet CFT in the 3-form flux background. As described, we can
evade this by inserting the flux vertex operators on the worldsheet. However we should still be able to view this as
a leading term in an expansion in powers of flux of a worldsheet
action in which the flux vertex operator is exponentiated into the worldsheet. If it is possible to `integrate in' the flux
vertex operator, or view it as a perturbation on a flux-less background, then it would appear that the flux operator
should carry a net picture charge of zero. This would be necessary as, integrated in, there would be no flux vertex
operator and the gaugini would have to carry vanishing ghost charge by themselves. In this case an amplitude with both gaugini
picture changed (and so carrying non-zero ghost charge) would not be an appropriate picture to compute in.

%%%%%%%%%%%%%%%%%%%%%%%%%%%%%%%%%%%%%%%%%%%%%%%%%%%%%%%%%%%%%%%%%%%%%%%%%%%%%%%%%%%%%%%%%%
\section{Summary}
\label{sec:summary}
%%%%%%%%%%%%%%%%%%%%%%%%%%%%%%%%%%%%%%%%%%%%%%%%%%%%%%%%%%%%%%%%%%%%%%%%%%%%%%%%%%%%%%%%%%

In this paper we have studied supersymmetry breaking effects from a string theory perspective by calculating string scattering amplitudes on a $\mathbb{Z}_4$ orbifold with $D3$-branes on singularities. We have taken the approach of calculating predictions from supergravity formulae regarding gaugino masses and then checked these predictions by evaluating the appropriate correlators on the annulus. We could recreate brane-to-brane gravity mediated tree-level soft gaugino masses and showed that there are two contributions coming from the $N=4$ and $N=2$ open string winding modes which correspond to dilaton and blow-up mode mediation respectively.

The main result of this work concerns the supergravity formula for anomaly mediated gaugino masses
\be
\label{kilmarnock4}
m_{1/2} = -\frac{g^2}{16\pi^2}\left[\left(3T_G-T_R\right)m_{3/2} - \left(T_G-T_R\right)K_iF^i - \frac{2T_R}{d_R}F^i\partial_i\left(\mathrm{ln\;det\;}Z \right) + 2T_G F^I \partial_I \ln \left( \frac{1}{g_0^2}  \right) \right]\;.
\ee
We performed the first detailed check of this formula from a string theory approach. After noting the presence of the last term (which was not presented in \cite{hepth9911029}) we set out to probe the presence of the first term associated to the `superconformal anomaly' induced by supersymmetry breaking closed-string fluxes. For the case of NS H-flux the supergravity analysis predicts that if and only if the first term of (\ref{kilmarnock4}) is present should there be a 1-loop mass term for the gauginos. We probed this mass term by calculating the correlator $\left<H \lambda \lambda\right>$ and showed that the mass term is present.\footnote{We checked this further by calculating a 4-point amplitude with one of the gaugini decomposed into a scalar and a fermion.} Further, the string calculation also agreed with the predicted relative coefficient between this
mass term and the 1-loop running of the tree-level mass.

 Our results show how this mass term arises in string theory.
 We showed that it comes from a sum over UV open-string winding modes, which can be written as
 a total derivative of a partition function $Z_{Cl}$ with respect to the annular modular parameter. Integrated over all values of the modular parameter, the overall result is given purely by the difference
 $Z_{Cl}(\infty) - Z_{Cl}(0)$. The former is determined purely by the massless spectrum and gives the $\beta$-function, whereas the latter vanishes
 due to tadpole cancellation. In this way we obtain the fact that the anomaly-mediated mass is an ultraviolet effect determined purely
 by infrared physics.

For the case of RR-flux the results were not complete. The reason is that the supergravity prediction is less clean since a mass term arises with or without the first term of (\ref{kilmarnock4}) which means that the formula can only be fully checked by the relative coefficient between the 1-loop mass term and the 1-loop running of the tree-level mass term. This requires precise knowledge of all the numerical factors. However in calculating a 3-point amplitude there is sometimes an ambiguity in the off-shell formulation to do with the cancellation of the momentum factors in the on-shell limit (see \cite{Conlon:2010xb} for another example). This arose in the RR calculation which means that determining the precise numerical factors was not possible and would require a higher point computation. Nonetheless, although we could not reproduce their relative coefficient, we were able to recreate the presence of a mass term and a running term as predicted. It would be very interesting to try and resolve this issue with a higher point computation.

It would be interesting to see how these results are modified by the presence of orientifolds, where at 1-loop there are also the Mobius strip and Klein bottle diagrams. The inclusion of such diagrams would modify the calculation as in their presence the $N=1$ Chan-Paton trace is non-vanishing and is given by the orientifold charge. This means that there may be new contributions coming from $N=1$ string oscillator modes. There should also be field redefinitions for the twisted modes in going from the string to the supergravity calculation.

It would of course be very interesting to study the formula for soft scalar masses with a string calculation.
However since these come from the scalar potential they should involve two flux insertions which would make their calculation even
more difficult than the gaugino masses, in addition to the possible need to do a 2-loop computation.

\subsection*{Acknowledgments}

We thank Shanta de Alwis, Marcus Berg, Cliff Burgess, Kiwoon Choi, Michael Dine, Emilian Dudas, Jan Louis, Fernando Quevedo and Nathan Seiberg for helpful discussions.

The work of EP was supported in part by the European ERC Advanced Grant 226371 MassTeV, by the CNRS PICS no. 3059 and 4172,
by the grants ANR-05-BLAN-0079-02, the PITN contract PITN-GA-2009-237920 and the IFCPAR CEFIPRA programme 4104-2. MDG is supported by the German Science Foundation (DFB) under SFB 676. JC is supported by the Royal Society with a University Research Fellowship and Balliol College, Oxford.
JC thanks the KITP, Santa Barbara for hospitality while part of this work was carried out. We also thank for hospitality the Strings and GUTs workshop at the Max-Planck-Institut in Munich where this project was originated, and the String Phenomenology 2010 conference in Paris where some of this 
work has been presented.

\appendix

%%%%%%%%%%%%%%%%%%%%%%%%%%%%%%%%%%%%%%%%%%%%%%%%%%%%%%%%%%%%%%%%%%%%%%%%%%%%%%%%%%%%%%%%%%
\section{Gaugino threshold corrections}
\label{sec:thresholds}
%%%%%%%%%%%%%%%%%%%%%%%%%%%%%%%%%%%%%%%%%%%%%%%%%%%%%%%%%%%%%%%%%%%%%%%%%%%%%%%%%%%%%%%%%%

Here we calculate threshold corrections for gauginos. This provides
confidence in our ability to work with the gaugino vertex operators in a situation where the answer is already known; for supersymmetry should ensure that we obtain the same result as with gauge boson operators. This check is important as the gaugino running plays a role in the anomaly mediation supergravity predictions; in particular, the calculation implies that ``ghost derivative'' terms should be explicitly excluded from the off-shell extension (this point is moot for an on-shell calculation as their contributions then necessarily cancel). As a bonus, the calculation below can be easily extended to non-supersymmetric models where the gaugino threshold corrections are not identical to the gauge thresholds, for example when there are antibranes. This also applies to the calculation of kinetic mixing between gauginos located on different (anti)branes.
We proceed by computing the correlator $\langle \bar{\lambda} \lambda \rangle$.

The gaugino vertex operators are given in the (-1/2)-picture by
\begin{align}
V^{-1/2}_{\lambda} =& e^{-\phi/2} u^\alpha S_\alpha e^{ik \cdot X} \prod_{i=3}^5 e^{\frac{i}{2} H_i} \nonumber \\
V^{-1/2}_{\ov{\lambda}} =& e^{-\phi/2} \ov{u}^{\dot{\alpha}} S_{\dot{\alpha}} e^{ik \cdot X} \prod_{i=3}^5 e^{-\frac{i}{2} H_i}
\end{align}
$S^\alpha, S^{\dot{\alpha}}$ are chiral and anti-chiral  4d Weyl spinors $e^{\pm \frac{i}{2}(H_1 + H_2)}$, $e^{\pm \frac{i}{2}(H_1 - H_2)}$. The H-charges of these operators are $(+, + ,+, +, +)$ and $(+, -, -, -, -)$.
The annulus requires vanishing ghost charge and so we need to picture change one of the operators.

For the amplitude the only allowed Lorentz structure is $u \gamma^\mu \ov{u}$, since $u \gamma \gamma \gamma \ov{u}$ is equivalent.
The picture-changing operator $e^\phi \partial X_\mu \psi^\mu (w)$ may in general act on either internal or non-compact directions,
but in order to conserve $H$-charge must in this case act only on the non-compact directions.
In principle we could use the ready-made part of the $+1/2$ operator from Frieden, Martinec and Shenker \cite{Friedan:1985ge}:
\beq
V^{+1/2}_\lambda = u^\alpha e^{\phi/2}\bigg[\partial X^\mu + 2i \ap (k\cdot \psi) \psi^\mu \bigg] \gamma_{\mu \alpha \beta} S^\beta e^{ik \cdot X}.
\eeq
Note that this is four-component spinor notation, and raised indices are conjugate to lowered ones, so
\beq
S_\alpha S^\beta \sim (z-w)^{-1/2} \delta_\alpha^\beta.
\eeq
However, we will find that this is not the most useful approach, since it involves taking correlators with extra $\psi$ operators and then extracting the finite parts. The most straightforward approach is instead to calculate the full amplitude before taking the picture-changing
limit. However there are two crucial details. The first is that the most singular part of the amplitude must be excluded:
\beq
e^\phi \partial X^\mu \psi_\mu (z) V^{-1/2} (w) \sim 2\ap (z-w)^{-1} e^{\phi/2} u_\alpha \gamma^\mu_{\alpha \beta} k_\mu e^{ik \cdot X} + ... = 0 + ...
\eeq
This originates in the fact that picture-changing is a contour integral around a vertex operator, and the most singular terms (which
correspond to $\mc{O}(z-w)^{-2}$ double poles) should be excluded.

The second, directly related to the first, is that we immediately replace $e^{\phi(z)} e^{-\phi(w)/2}$ with $(z-w)^{1/2} e^{\phi (w)/2}$ and do not
 consider subleading terms from the ghost OPE. This is because ghost derivative terms are eliminated from the vertex operator by the equations of motion for the fermion; as we can see from the equation above there would be terms of the form
\beq
e^\phi \partial X^\mu \psi_\mu (z) V^{-1/2} (w) \supset e^{\phi/2} \frac{1}{2} \partial \phi (u_\alpha \gamma^\mu_{\alpha \beta} k_\mu) e^{ik \cdot X}
\eeq
which vanish by  $u_\alpha \gamma^\mu_{\alpha \beta} k_\mu=0$. One source of worry might be that since we are continuing the momenta off-shell we should in fact retain these terms. However, the reader can convince themself that this is not correct by performing a four-point calculation similar to that in \cite{Conlon:2010xb} or section (\ref{sec:nsapp}).

 However, we do need to include the subleading parts from the $\partial X$ and $\psi$ correlators,
  and in fact these will give us the result. Note that the above subtleties only arise
  because we are computing a two-point correlator; for four point amplitudes and above the on-shell conditions can be imposed more easily.
  It is however generally the case that for fermionic fields we should compute the full amplitude before performing the picture-changing; otherwise
contributing terms are missed.

We now evaluate the amplitude, placing the gaugino vertex operators at $z_1$ and $z_2$ and the picture-changing operator at $w$.
We find (continued off-shell)
\begin{align}
\langle V^{-1/2}_{\ov{\lambda}} V^{+1/2}_{\lambda} \rangle =&  u \gamma^\mu \ov{u} k_\mu \lim_{w \rightarrow z_2} (w-z_2)^{1/2} \bra e^{-\phi/2 (z_1)} e^{\phi/2 (z_2)}\ket \bra e^{ik\cdot X (z_1) } \partial X(w) e^{-ik \cdot X(z_2)} \ket  \nonumber \\
&\times \bra e^{\frac{i}{2} H (z_1)} e^{-\frac{i}{2} H (z_2)}\ket \bra e^{-\frac{i}{2} H (z_1)} e^{i H (w) } e^{\frac{-i}{2} H (z_2)}\ket \prod_{i=1}^3  \bra e^{-\frac{i}{2} H_i (z_1)} e^{\frac{i}{2} H_i (z_2)} \ket.
\end{align}

It is straightforward to write down the correlators using (\ref{hchargecorr})
\begin{align}
\bra e^{-\phi/2 (z_1)} e^{\phi/2 (z_2)}\ket =& \left(\frac{\theta_1 (z_1 - z_2)}{\theta_1^\prime (0)}\right)^{1/4}
\frac{1}{\theta_\alpha (\frac{z_2 - z_1}{2})},  \nonumber\\
\bra e^{\frac{i}{2} H (z_1)} e^{-\frac{i}{2} H (z_2)}\ket =& \left(\frac{\theta_1 (z_1 - z_2)}{\theta_1^\prime (0)}\right)^{-1/4} \theta_\alpha (\frac{z_1 - z_2}{2}),  \nonumber\\
\bra e^{ik\cdot X (z_1) } \partial X(w) e^{-ik \cdot X(z_2)} \ket =& i k_\nu \bigg(\partial_w G^{\mu \nu} (z_1 - w) - \partial_w G^{\mu\nu} (w-z_2)\bigg) \exp [ G^{\rho \kappa} (z_1 - z_2) k_\rho k_\kappa], \nonumber \\
\bra e^{-\frac{i}{2} H (z_1)} e^{i H (w) } e^{\frac{-i}{2} H (z_2)}\ket =& \left(\frac{\theta_1 (z_1 - z_2)}{\theta_1^\prime (0)}\right)^{1/4}\left(\frac{\theta_1 (z_1 - w)}{\theta_1^\prime (0)}\right)^{-1/2}\left(\frac{\theta_1 (w - z_2)}{\theta_1^\prime (0)}\right)^{-1/2}  \theta_\alpha (w - \frac{z_1 + z_2}{2} ).
\end{align}
Now, when performing the contour integration we require the fact that
\beq
\partial_w G^{\mu\nu} (w-z_2) \sim - \eta^{\mu \nu} \frac{2\ap}{w-z_2} + \mathcal{O} (w-z_2),
\eeq
whereas
\begin{align}
\bra e^{-\frac{i}{2} H (z_1)} e^{i H (w) } e^{\frac{-i}{2} H (z_2)}\ket \sim& (w-z_2)^{-1/2}\left(\frac{\theta_1 (z_1 - z_2)}{\theta_1^\prime (0)}\right)^{-1/4} \bigg[ 1 + \frac{1}{2} (w -z_2) \frac{\theta_1^\prime (z_1 - z_2)}{\theta_1 (z_1 - z_2)} \bigg] \nonumber \\
& \times \bigg[ \frac{\theta_\alpha (\frac{z_2- z_1}{2} )}{\theta_\alpha (0)} + (w-z_2) \frac{\theta_\alpha^\prime (\frac{z_2- z_1}{2} )}{\theta_\alpha (0)} \bigg].
\label{jabba}
\end{align}

Thus the whole amplitude is
\begin{align}
\bra V^{-1/2}_{\ov{\lambda}} V^{+1/2}_{\lambda}\ket =&  i u \gamma^\mu \ov{u}  \left(\frac{\theta_1 (z_1 - z_2)}{\theta_1^\prime (0)}\right)^{-1/4}  \exp [ G^{\rho \kappa} (z_1 - z_2) k_\rho k_\kappa] \nonumber \\
& \times \sum_\alpha \delta_\alpha \bigg[ \bigg\{ \ap k_\mu \frac{\theta_1^\prime (z_1 - z_2)}{\theta_1 (z_1 - z_2)} + \partial_{z_2} k^\nu G_{\mu \nu} (z_1 - z_2) \bigg\} \theta_\alpha (\frac{z_2- z_1}{2} ) \nonumber \\
&\qquad + 2\ap k_\mu \theta_\alpha^\prime (\frac{z_2- z_1}{2} ) \bigg] \frac{1}{\eta^3 (it)} \frac{1}{(4\pi^2 \ap t)^2} \prod_{i=1}^3  \bra e^{-\frac{i}{2} H_i (z_1)} e^{\frac{i}{2} H_i (z_2)} \ket_\alpha.
\end{align}
Clearly the last line has a structure familiar from supersymmetric gauge threshold corrections, while the middle line is different and thus should vanish in the supersymmetric limit. We have also indulged in a change of sign in the $\theta_\alpha$ functions; as the $\theta$-functions are either
even or odd this is consistent provided the numbers of such changes is zero modulo two.

Let us now evaluate this. For an $N=1$ supersymmetric orbifold without $N=2$ sectors the amplitude then simplifies to
\begin{align}
\bra V^{-1/2}_{\ov{\lambda}}(z_1) V^{+1/2}_{\lambda}(z_2) \ket =&  i u \gamma^\mu \ov{u}  \exp [ G^{\rho \kappa} (z_1 - z_2) k_\rho k_\kappa] \frac{1}{(4\pi^2 \ap t)^2}\nonumber \\
& \times \bigg[ \bigg\{ 3\ap k_\mu \frac{\theta_1^\prime (z_1 - z_2)}{\theta_1 (z_1 - z_2)} + \partial_{z_2} k^\nu G_{\mu \nu} (z_1 - z_2) \bigg\} + 2\ap k_\mu \sum_{i=1}^3 \frac{\vt^\prime (\theta_i) }{\vt(\theta_i)} \bigg].
\label{generalgaugino}\end{align}
To compute the whole amplitude, we must integrate over the modulus $t$ and the operator positions (fixing if desired one at the origin); we find that the piece in curly brackets vanishes on integration as an odd function. We are then left with
\begin{align}
\bra \bar{\lambda}\lambda\ket =& \int_0^\infty dt \frac{1}{(4\pi^2 \ap t)^2} \int_0^{it/2} dz_1 \bra V^{-1/2}_{\ov{\lambda}} (z_1,k) V^{+1/2}_{\lambda} \ket (0,-k)\nn\\
=&  i u \gamma^\mu \ov{u} k_\mu \int_0^\infty \frac{dt}{t} \frac{1}{(4\pi^2 \ap)^2} 4(\ap)^2 \sum_{i=1}^3 \frac{\vt^\prime (\theta_i) }{\vt(\theta_i)}.
\label{finalgaugino}
\end{align}
Including the Chan-Paton trace, we find this is
equal to the result for gauge thresholds \cite{09014350,Conlon:2010xb} on using the theta identity (\ref{extrathetaidentity}). In particular it is straightforward to repeat the last steps for $N=2$ sectors of orbifolds, and find that as expected \cite{09014350,Conlon:2010xb,Conlon:2009qa} the gauge couplings as measured by the gauginos run to the winding scale of the compactification. Note that if we were to erroneously include the ``ghost derivative'' terms then we would actually find that this term would be cancelled. Since this is the entire contribution for the supersymmetric case, we confirm our assertion that they should be excluded! Finally, for non-supersymmetric setups we will find contributions from all sectors, including the untwisted sector, where the result (\ref{generalgaugino}) is qualitatively the same as for the gauge bosons but clearly quantitatively different.

%%%%%%%%%%%%%%%%%%%%%%%%%%%%%%%%%%%%%%%%%%%%%%%%%%%%%%%%%%%%%%%%%%%%%%%%%%%%%%%%%%%%%%%%%%%
\section{Powers of coupling constants}
\label{sec:coupcon}
%%%%%%%%%%%%%%%%%%%%%%%%%%%%%%%%%%%%%%%%%%%%%%%%%%%%%%%%%%%%%%%%%%%%%%%%%%%%%%%%%%%%%%%%%%%

The gravity mediated contribution to the gaugini masses studied in section \ref{sec:gmgm} is computed on the annulus. However, from a supergravity perspective it is still a `tree-level' mass corresponding to $F^S \partial_S f_a$, rather than a 1-loop one. In this appendix we make this statement precise by studying the dependence on the string coupling $g_s$ of the amplitude.

We will work in conventions where no powers of $g_s$ are included in the vertex operators. This will mean that fields are in general
not canonically normalised, and this we will do at the end. The powers of $g_s$ then come from the diagram we are computing. These are
\bea
\hbox{Sphere} & & g_s^{-2} \\
\hbox{Disc} & & g_s^{-1} \\
\hbox{Annulus} & & g_s^{0}
\eea
Gauge and gaugino kinetic terms can be computed on the disk (tree-level) couplings and on the annulus (threshold corrections).
These generate amplitudes that look like (disk)
$$
\hbox{Gauge kinetic terms:   } \int d^4 x \frac{1}{g_s} F_{\mu \nu} F^{\mu \nu} \qquad
 \hbox{Gaugino kinetic terms:   }\int d^4 x \frac{1}{g_s} \bar{\lambda} \sigma^{\mu} \partial_{\mu} \lambda.
$$
and (annulus)
$$
\hbox{Gauge kinetic terms:   } \int d^4 x \beta \ln \left( \frac{\Lambda^2}{\mu^2} \right)  F_{\mu \nu} F^{\mu \nu} \qquad
\hbox{Gaugino kinetic terms:   } \int d^4 x \beta \ln \left( \frac{\Lambda^2}{\mu^2} \right) \bar{\lambda} \sigma^{\mu} \partial_{\mu} \lambda.
$$
Canonically normalised gauge and gaugino fields are then $\frac{A_{\mu}}{\sqrt{g_s}}$ and $\frac{\lambda}{\sqrt{g_s}}$.

The scalar and fermion kinetic terms are also generated on the disk (as we are dealing with
D3 branes and an underlying N=4 structure, these fields are just the extension of the above to include internal indices).
So these terms look like
\be
\frac{1}{g_s} \int d^4 x \partial_{\mu} \phi \partial^{\mu} \phi = \int d^4 x \partial_{\mu} \left( \frac{\phi}{\sqrt{g_s}} \right)
\partial^{\mu} \left( \frac{\phi}{\sqrt{g_s}} \right) \equiv \int d^4 x \partial_{\mu} \hat{\phi} \partial^{\mu} \hat{\phi}.
\ee
\be
\frac{1}{g_s} \int d^4 x \bar{\chi} \sigma^{\mu} \partial_{\mu} \chi = \int d^4 x \left( \frac{\bar{\chi}}{\sqrt{g_s}} \right) \sigma^{\mu} \partial_{\mu}
\left( \frac{\chi}{\sqrt{g_s}} \right) \equiv \int d^4 x \bar{\hat{\chi}} \sigma^{\mu} \partial_{\mu} \hat{\chi}.
\ee
Here $\hat{\phi}$ and $\hat{\chi}$ are canonically normalised fields.

The tree-level Yukawa coupling is generated on the disk and so takes the form
\be
\frac{1}{g_s} \int d^4 x \phi \chi \chi = \int d^4 x \left( \frac{\phi}{\sqrt{g_s}} \right) \left( \frac{\chi}{\sqrt{g_s}} \right)
\left( \frac{\chi}{\sqrt{g_s}} \right) \sqrt{g_s} \equiv \int d^4 x \hat{\phi} \hat{\chi} \hat{\chi} \sqrt{g_s} \;.
\ee
The one-loop Yukawa coupling from the annulus has the form
\bea
&&K \ln \left( \frac{M_W^2}{\mu^2} \right) \int d^4 x \phi \chi \chi =
K \ln \left( \frac{M_W^2}{\mu^2} \right) \int d^4 x \left( \frac{\phi}{\sqrt{g_s}} \right) \left( \frac{\chi}{\sqrt{g_s}} \right)
\left( \frac{\chi}{\sqrt{g_s}} \right) g_s^{3/2} \nn \\
&&\equiv K \ln \left( \frac{M_W^2}{\mu^2} \right) \int d^4 x \, \left( \hat{\phi} \hat{\chi} \hat{\chi} \right) \, g_s^{3/2} \;.
\eea
This is down by a factor of $g_s$ compared to the tree-level term.

The sphere diagram gives the graviton and gravitino kinetic terms. The sphere is normalised as $g_s^{-2}$, so we should get
schematically (linearising about flat space time to extract only the fluctuation of the graviton about Minkowski space, so indices
are raised and lowered using the Minkowski metric)
\bea
\hbox{ Graviton:  } & e^{-2 \phi} \int d^4 x \, \partial_{\mu} h \partial^{\mu} h & = \int d^4 x \partial_{\mu} \left( \frac{h}{g_s} \right)
\partial^{\mu} \left( \frac{h}{g_s} \right) \equiv \int d^4 x \, \partial_{\mu} \hat{h} \partial^{\mu} \hat{h} \nonumber \\
\hbox{ Gravitino: } & e^{-2 \phi} \int d^4 x \, \bar{\psi}_{3/2} \partial \, \psi_{3/2} & =
\int d^4 x \, \left( \frac{\bar{\psi}_{3/2}}{g_s} \right) \partial \left( \frac{\psi_{3/2}}{g_s} \right) \equiv
\int d^4 x \, \left( \hat{\bar{\psi}}_{3/2} \right) \partial \, \left( \hat{\psi}_{3/2}
\right) \;.
\eea
What is the gravitino mass? From a string perspective we would compute this by computing the 5-point amplitude
$\bra \phi_1 \phi_2 \phi_3 \psi_{3/2}  \psi_{3/2} \ket$ on the disk. We assume here that this amplitude is non-zero.
This corresponds to the gravitino mass induced by turning on the three boson fields.
There are volume factors that would require a more detailed computation but the overall prefactor must be $g_s^{-1}$ from the disk factor.
This should then give
\be
\frac{1}{g_s} \bra \phi_1 \phi_2 \phi_3 \psi_{3/2} \psi_{3/2} \ket = \lambda g_s^{5/2} \left( \frac{\phi_1}{\sqrt{g_s}} \right)
\left( \frac{\phi_2}{\sqrt{g_s}} \right) \left( \frac{\phi_3}{\sqrt{g_s}} \right) \left( \frac{\psi_{3/2}}{g_s} \right)
\left( \frac{\psi_{3/2}}{g_s} \right) \;.
\ee
As the fields in brackets are canonically normalised,
in terms of the string fields that enter the vertex operator, the gravitino mass is then given by
\be
m_{3/2} = \lambda g_s^{5/2} \left( \frac{\phi_1}{\sqrt{g_s}} \right)
\left( \frac{\phi_2}{\sqrt{g_s}} \right) \left( \frac{\phi_3}{\sqrt{g_s}} \right) \;,
\ee
where $\lambda$ is independent of the string coupling.
Now consider the amplitude we compute, which is $\bra \phi_1 \phi_2 \phi_3 \lambda_1 \lambda_2 \ket$ evaluated on the annulus.
This will give
\bea
\bra \phi_1 \phi_2 \phi_3 \lambda \lambda \ket & = & \kappa \phi_1 \phi_2 \phi_3 \lambda_1 \lambda_2 \\
& = & \kappa g_s^{5/2} \left( \frac{\phi_1}{\sqrt{g_s}} \right)
\left( \frac{\phi_2}{\sqrt{g_s}} \right) \left( \frac{\phi_3}{\sqrt{g_s}} \right)
\left( \frac{\lambda}{\sqrt{g_s}} \right) \left( \frac{\lambda}{\sqrt{g_s}} \right) \\
& \sim & m_{3/2} \left( \frac{\lambda}{\sqrt{g_s}} \right) \left( \frac{\lambda}{\sqrt{g_s}} \right).
\eea
This therefore gives a term in the Lagrangian which is a gaugino mass term directly proportional to the gravitino mass,
without any additional factors of the string coupling. This is consistent with the amplitude we have computed representing
a tree-level gaugino mass, despite the fact we are on the cylinder notwithstanding.

%%%%%%%%%%%%%%%%%%%%%%%%%%%%%%%%%%%%%%%%%%%%%%%%%%%%%%%%%%%%%%%%%%%%%%%%%%%%%%%%%%%%%%%%%%%%%%
\section{Higher-Point NS-NS Flux Computation}
\label{sec:nsapp}
%%%%%%%%%%%%%%%%%%%%%%%%%%%%%%%%%%%%%%%%%%%%%%%%%%%%%%%%%%%%%%%%%%%%%%%%%%%%%%%%%%%%%%%%%%%%%%

In this appendix we present a calculation of the NS H-flux anomaly mediated mass term using a 4-point amplitude. This is
formally more robust than the 3-point calculation presented in the main text
as the open string momenta can be taken on-shell from the start. As discussed in \cite{Conlon:2010xb}, the off-shell extension of a 3-point calculation can lead to ambiguous momentum prefactors (although the amplitude is unambiguous).\footnote{In \cite{Minahan}
it is claimed that there is a uniquely preferred way to go off-shell for 3-point closed string amplitudes. However the arguments
presented there involve modular invariance in a crucial way, and so do not seem to generalise to open string amplitudes
where modular invariance does not play a key role in the consistency of the theory.}
 For example we could obtain ratios $k_i^2/k_j^2$ which should become $1$ as we take the momenta to zero. In section \ref{sec:nsnsflux} we did not have such ambiguous ratios, but nonetheless performing the four-point check confirms that those results correct.

\begin{figure}\begin{center}
\epsfig{file=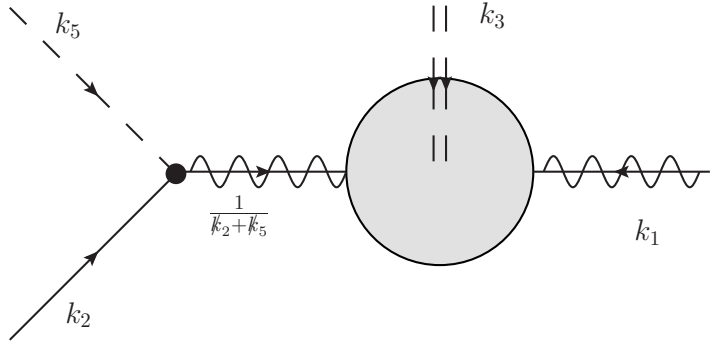,height=5cm}
\caption{Field theory limit of the four-point string diagram. The double scalar line denotes the flux insertion.}\end{center}
\label{4ptFluxfig}\end{figure}

We compute the higher-point amplitude and then factorise onto the relevant sub-diagram in order to obtain the relevant prefactors. Here we shall compute a four-point amplitude where a boson $\phi$ and fermion $\psi$ factorise in an appropriate limit onto one of the gaugini $\lambda$, as illustrated in figure \ref{4ptFluxfig}
\begin{align}
\mathcal{A}=\bra V_\lambda^{+1/2} (z_1, k_1) V_{\tilde{\psi}}^{-1/2} (z_2, k_2) V_{\tilde{\phi}}^0 (z_5,k_5) V_{NS-NS}^{0,0} (z_3, z_4, k_3) \ket
\end{align}
where $B =X^5 dX^3 \wedge dX^4 +  \ov{X}^5 d\ov{X}^3 \wedge d\ov{X}^4$; for convenience we write $z_3, z_4$ for the closed string coordinates with the understanding that $z_4 = - \bar{z}_3$; and
\begin{align}
V_{\tilde{\psi}}^{-1/2} =& e^{-\phi/2} e^{ik\cdot X} u_{\dot{\alpha}} \tilde{S}^{\dot{\alpha}} e^{-\frac{i}{2} H_3 + \frac{i}{2} H_4 + \frac{i}{2} H_5}, \nn \\
V_{\tilde{\phi}}^{-1} =& e^{-\phi}e^{ik\cdot X} \psi^3,  \nn \\
V_\lambda^{-1/2} =& e^{-\phi/2} e^{ik\cdot X}v_{\alpha} S^{\alpha} e^{\frac{i}{2} H_3 + \frac{i}{2} H_4 + \frac{i}{2} H_5}, \nn \\
V_{NS-NS}^{-1,-1} =& e^{-\phi - \tilde{\phi}} e^{ik\cdot X} \bigg( X^5 (\psi^3 \tilde{\psi}^4- \psi^4 \tilde{\psi}^3) + \ov{X}^5 (\ov{\psi}^3 \ov{\tilde{\psi}}^4- \ov{\psi}^4 \ov{\tilde{\psi}}^3)\bigg).
\end{align}
$X^5$ is the untwisted direction. $S^\alpha, S^{\dot{\alpha}}$ are chiral and anti-chiral  4d Weyl spinors $e^{\pm \frac{i}{2}(H_1 + H_2)}$, $e^{\pm \frac{i}{2}(H_1 - H_2)}$.

To reproduce the lower-point gaugino mass sub-diagram, we consider taking $z_5 \sim z_2$ and extract the pole in $k_2 \cdot k_5$ where
\begin{align}
V_{\tilde{\psi}^{\dot{\alpha}}}^{-1/2} (z_2) V_\phi^0 (z_5) =& e^{-\phi/2} e^{ik\cdot X} u_{\dot{\alpha}} \tilde{S}^{\dot{\alpha}} \Theta (z_2) V_\phi^0\nn\\
& \sim (z_2 - z_5)^{-1 + 2\ap k_2 \cdot k_5} k^5_\mu (\Gamma^\mu)_{\alpha \dot{\alpha}}V_{\lambda^\alpha}^{-1/2}.
\end{align}
Thus when we integrate over $z_5$, we we find a pole at $z_2$
\begin{align}
A \rightarrow& \frac{ k^5_\mu (\Gamma^\mu)_{\alpha \dot{\alpha}}}{2\ap k_2 \cdot k_5} u_{\dot{\alpha}} V_{\lambda^\alpha}^{-1/2}\nn\\
\rightarrow& \frac{1}{2\ap}\frac{1}{\slashed{k}_2 + \slashed{k}_5} u_{\dot{\alpha}} V_{\lambda^\alpha}^{-1/2}
\end{align}
since $\slashed{k}_2 u = 0$; this is what we expect from figure (\ref{4ptFluxfig}). So to extract the gaugino mass diagram we retain the piece proportional to $ k^5_\mu (\Gamma^\mu)_{\alpha \dot{\alpha}}/k_2 \cdot k_5$ from the amplitude and neglect the remaining terms.

We can clearly ignore the $X \psi \tilde{\psi}$ terms in the NS-NS operator due to $H$-charge -  we have only four PCOs - and so we retain only the second half of the flux operator. Moreover, we can readily see that we require either one or three PCOs to act on the external dimensions. In these case from considering the most general Lorentz structures possible the correlators must take the form
\begin{align}
\bra S_{\alpha} \tilde{S}_{\dot{\alpha}} \psi^\mu \ket =& A_1 (C\Gamma^\mu)_{\alpha \dot{\alpha}} \nn\\
\bra  S_{\alpha}\tilde{S}_{\dot{\alpha}} \psi^\mu \psi^\nu \psi^\rho \ket =& A_2 (C\Gamma^\mu)_{\alpha \dot{\alpha}} \eta^{\nu\rho} + B_2 (C\Gamma^\nu)_{\alpha \dot{\alpha}} \eta^{\mu\rho} + C_2 (C\Gamma^\rho)_{\alpha \dot{\alpha}} \eta^{\mu\nu} + D_2 (C\Gamma^\mu\Gamma^\nu \Gamma^\rho)_{\alpha \dot{\alpha}} ,
\end{align}
where $C$ is the charge-conjugation matrix, and the functions $A_1, B_2,C_2,D_2$ depend upon the PCO and vertex operator positions and in principle Lorentz-invariant functions of the momenta. Then in the first case we need only compute one amplitude to determine $A_1$, whereas for three external PCOs we must compute four amplitudes in order to completely determine all of the functions. Computing further amplitudes will yield no further information. In the former case, we can see that by choosing $S^\alpha = e^{-\frac{i}{2}H_1} e^{-\frac{i}{2}H_2} \equiv |--\ket, \tilde{S}^{\dot{\alpha}}  = e^{-\frac{i}{2}H_1} e^{+\frac{i}{2}H_2} \equiv |-+\ket, \mu=1+$ we find $\bra S^{--} \tilde{S}^{-+} \psi^{1+} \ket = A_1$.

To compute the three-PCO amplitudes, we have:
\begin{align}
\alpha = |--\ket, \dot{\alpha} = |+-\ket, \mu=1-, \nu=1+,\rho=2+ \rightarrow& -\frac{1}{2} C_2\nn\\
\alpha = |--\ket, \dot{\alpha} = |+-\ket, \mu=1-, \nu=2+,\rho=1+ \rightarrow& -\frac{1}{2} B_2\nn\\
\alpha = |--\ket, \dot{\alpha} = |+-\ket, \mu=2+, \nu=1-,\rho=1+ \rightarrow& -\frac{1}{2} A_2\nn\\
\alpha = |--\ket, \dot{\alpha} = |+-\ket, \mu=2+, \nu=1+,\rho=1- \rightarrow& -\frac{1}{2} A_2 - D_2\nn\\
\alpha = |--\ket, \dot{\alpha} = |+-\ket, \mu=1+, \nu=2+,\rho=1- \rightarrow& -\frac{1}{2} B_2 + D_2\nn\\
\alpha = |--\ket, \dot{\alpha} = |+-\ket, \mu=1+, \nu=1-,\rho=2+ \rightarrow& -\frac{1}{2} C_2 - D_2.
\end{align}
We need only compute four of these.

Once we have computed the fermionic part of the amplitudes, we recall that the PCOs contain a bosonic part, so total amplitude is given by the contraction of the Lorentz indices $\mu, \nu, \rho$ with $\bra \dot{X}_\mu \dot{X}_\nu \dot{X}_\rho \prod e^{ik\cdot X} \ket$ and then take the limit as the PCOs approach their vertex operators. When acting upon the NS-NS flux or the scalar, these yield simple factors of $k_3, k_5$ but for the gaugino there may be derivative terms.

In fact, since bosonic correlators in twisted directions vanish when integrated, we may only have the situation with three external PCOs; in one configuration the internal PCO acts on the gaugino, in another it acts on the NS-NS operator. So we must calculate $8$ amplitudes. However, we can in fact reduce these to the set which contain the appropriate pole between the boson and fermion; we can thus consider only four amplitudes. Choosing $\rho$ to be associated with the boson means that only $C_2$ and $D_2$ survive.

Now in  computing the four-point function the aim would be that we can take all vertex operators on shell, so we would think we could take for example $k_3^2=0$. However, once we have closed string operators on the annulus we find that there exist poles when the vertex operator approaches the boundary. If we were to go on shell then we would need to include explicit contact terms in the amplitude \cite{Green:1987qu}; we shall simply follow the standard procedure of analytically continuing the momenta, and take the limit $k_3^2 \rightarrow 0$ afterward. However, we are free to take the open string momenta on shell and use their equations of motion to simplify the calculation.

\subsection*{Internal PCO on gaugino}

Here we label $\rho\rightarrow \phi, (\mu,\nu) \rightarrow V_{NSNS}$ and so
\begin{align}
\mathcal{A} \supset -i  (k_3)_\mu (k_3)_\nu (k_5)_\rho    \bra  S_{\alpha} (z_1) \tilde{S}_{\dot{\alpha}} (z_2) \psi^\mu (z_3) \psi^\nu (z_4) \psi^\rho (z_5) \ket
\end{align}
multiplying the internal correlators. Here we can always use the four-theta identity, since in the $X^5$ direction the fermionic charges cancel the ghost charges.

\subsubsection*{$\mu=1-, \nu=1+,\rho=2+$}

We compute with the $H$-charges
\begin{align}
\psi^{+1/2}(z_1) =& \half(-,-,+,+,-) \;, \nn \\
\psi^{-1/2}(z_2) =& \half(+,-,-,+,+) \;, \nn \\
g(z_3) =& (-1,0,-1,0,0) \;,\nn\\
g(z_4) =& (+1,0,0,-1,0) \;,\nn\\
\phi^0 (z_5) =&(0,+1,+1,0,0) \:.
\end{align}

This yields $C_2$ with momentum $k_3^\mu k_3^\nu k_5^\rho$, giving $k_3^2 k_5^\rho C\Gamma_\rho$.

The amplitude gives
\begin{align}
2 (u_1 C \slashed{k}_5 u_2) k_3^2 \frac{\vt (-z_3+z_4 + \theta) \vt(z_2 - z_5 - \theta) }{\vt(z_2 -z_5) \vt(z_3-z_4) \vt(\theta) \vt(-\theta)} \bra \partial_n X^5 \ov{X}^5 \ket
\end{align}
This clearly contains the required pole in $z_2 - z_5$, which we can factorise onto to obtain
\begin{align}
2 \frac{(u_1 C \slashed{k}_5 u_2) k_3^2}{k_2 \cdot k_5} \frac{\vt (-z_3+z_4 + \theta) }{\vt(z_3-z_4)\vt(\theta)} \bra \partial_n X^5 \ov{X}^5 \ket.
\end{align}
This is the main expression for the anomaly-mediated mass in the main body of the paper, and shows that the prefactor of $k_3^2$ used there is correct. It is equivalent to computing with $H$-charges
\begin{align}
\psi^{+1/2}(z_1) =& \half(-,-,+,+,-) \;, \nn \\
\psi^{-1/2}(z_2) =& \half(+,+,+,+,+) \;, \nn \\
g(z_3) =& (-1,0,-1,0,0) \;,\nn\\
g(z_4) =& (+1,0,0,-1,0) \;.\nn\\
\end{align}

\subsubsection*{$\mu=2+, \nu=1+,\rho=1-$}

We compute with the charges
\begin{align}
\psi^{+1/2}(z_1) =& \half(-,-,+,+,-) \;, \nn \\
\psi^{-1/2}(z_2) =& \half(+,-,-,+,+) \;, \nn \\
g(z_3) =& (0,+1,-1,0,0) \;,\nn\\
g(z_4) =& (+1,0,0,-1,0) \;,\nn\\
\phi^0 (z_5) =&(-1,0,+1,0,0) \:.
\end{align}
This yields $D_2$ with momentum $k_3^\mu k_3^\nu k_5^\rho$, giving $k_3^\mu k_3^\nu k_5^\rho C\Gamma_\mu\Gamma_\nu \Gamma_\rho = k_3^2 k_5^\rho C\Gamma_\rho$.
However, this yields zero upon Riemann summation.

\subsection*{Internal PCO on NSNS}

Here $\mu$ becomes the PCO acting on the gaugino, $\nu$ on the LHS of the NSNS field, and $\rho$ still on the boson.
\begin{align}
\mathcal{A} \supset -i  (k_3)_\nu (k_5)_\rho  \lim_{u \rightarrow z_1} (u-z_1)^{1/2}\bigg[\bra \dot{X}_\mu (u)\prod e^{ik\cdot X} \ket  \bra  S_{\alpha} (z_1) \tilde{S}_{\dot{\alpha}} (z_2) \psi^\mu (u) \psi^\nu (z_4) \psi^\rho (z_5) \ket \bigg].
\end{align}

\subsubsection*{$\mu=1-, \nu=1+,\rho=2+$}

We compute with the charges
\begin{align}
\psi^{+1/2}(z_1) =& \half(-,---,+,+,+) \;, \nn \\
\psi^{-1/2}(z_2) =& \half(-,-,-,+,+) \;, \nn \\
g(z_3) =& (+1,0,-1,0,0) \;,\nn\\
g(z_4) =& (0,0,0,-1,-1) \;,\nn\\
\phi^0 (z_5) =&(0,+1,+1,0,0) \:.
\end{align}
This yields $C_2$ with momentum prefactor $k_1^\mu k_3^\nu k_5^\rho \rightarrow k_1 \cdot k_3 k_5^\rho C\Gamma_\rho $.  Extracting the pole piece we have the prefactor exactly as in the body of the paper (where we exchange $k_1 \leftrightarrow k_2$) and compute with the charges
\begin{align}
\psi^{+1/2}(z_1) =& \half(-,---,+,+,+) \;, \nn \\
\psi^{-1/2}(z_2) =& \half(-,+,-,+,+) \;, \nn \\
g(z_3) =& (+1,0,-1,0,0) \;,\nn\\
g(z_4) =& (0,0,0,-1,-1) \;
\end{align}
and the conjugate
\begin{align}
\psi^{+1/2}(z_1) =& \half(-,---,+,+,+) \;, \nn \\
\psi^{-1/2}(z_2) =& \half(-,+,-,+,+) \;, \nn \\
g(z_3) =& (+1,0,0,-1,0) \;,\nn\\
g(z_4) =& (0,0,-1,0,-1) \;.
\end{align}
These contributions can be calculated using the five-theta summation formula and are shown in section (\ref{sec:nsnsflux}) to yield running masses.

\subsubsection*{$\mu=2+, \nu=1+,\rho=1-$}

We compute with the charges
\begin{align}
PCO(u) =& (0,+1,0,0,0) \;,\nn\\
\psi^{+1/2}(z_1) =& \half(-,-,+,+,+) \;, \nn \\
\psi^{-1/2}(z_2) =& \half(+,-,-,+,+) \;, \nn \\
g(z_3) =& (+1,0,-1,0,0) \;,\nn\\
g(z_4) =& (0,0,0,-1,-1) \;,\nn\\
\phi^0 (z_5) =&(-1,0,+1,0,0) \:.
\end{align}
If we consider $D_2$ then we see that if it is contracted with $k_1$ then it vanishes by the equations of motion. However, this is exactly what we find preceding derivative terms. So we see that we must compute with the charges
\begin{align}
\psi^{+1/2}(z_1) =& \half(-,+,+,+,+) \;, \nn \\
\psi^{-1/2}(z_2) =& \half(+,-,-,+,+) \;, \nn \\
g(z_3) =& (+1,0,-1,0,0) \;,\nn\\
g(z_4) =& (0,0,0,-1,-1) \;,\nn\\
\phi^0 (z_5) =&(-1,0,+1,0,0) \:
\end{align}
multiplying
\begin{align}
 (ik_2^\mu \partial_{z_1} G_{12} + ik_3^\mu \partial_{z_1} G_{13} + ik_5^\mu \partial_{z_1} G_{15}) k_3^\nu k_5^\rho.
\end{align}
Extracting the pole term we can compute with the charges
\begin{align}
\psi^{+1/2}(z_1) =& \half(-,+,+,+,+) \;, \nn \\
\psi^{-1/2}(z_2) =& \half(-,-,+,+,+) \;, \nn \\
g(z_3) =& (+1,0,-1,0,0) \;,\nn\\
g(z_4) =& (0,0,0,-1,-1) \;,\nn\\
\end{align}
which, when we perform the Riemann summation, vanishes.

\subsection{Alternative Choice of Flux}

Since the previous calculation required us to take the limit as $k_3^2$ becomes on-shell, rather than insisting on this from the outset, here we demonstrate that the result is correct by calculating with a different choice of flux. Namely we take $\ov{X}^3 (\ov{\psi}^4 \ov{\tilde{\psi}}^5-\ov{\psi}^5 \ov{\tilde{\psi}}^4)$, equivalent to $B_2 = X_3 dX_4 \wedge dX_5 + c.c$.

The relevant term has both PCOs acting internally on the NSNS flux giving $\ov{\psi}^1 \ov{\psi}^2 \ov{\partial} \ov{X}^3$, and one acting internally on the gaugino in the $X^3$ direction. We have the charges
\begin{align}
\psi^{+1/2}(z_1) =& \half(-,-,+,+,-) \;, \nn \\
\psi^{-1/2}(z_2) =& \half(+,-,-,+,+) \;, \nn \\
g(z_3) =& (0,0,-1,-1,0) \;,\nn\\
g(z_4) =& (0,0,0,0,0) \;,\nn\\
\phi^0 (z_5) =&(0,+1,+1,0,0) \:.
\end{align}
The amplitude also contains the correlator $\bra \partial_n X^5 (z_1)\bar{\partial}\bar{X}^5 (z_4)\ket$.
When contracted we have the pole proportional to $C\slashed{k}_5/k_2 \cdot k_5$ as required, and can then compute with the charges
\begin{align}
\psi^{+1/2}(z_1) =& \half(-,-,+,+,-) \;, \nn \\
\psi^{-1/2}(z_2) =& \half(+,+,+,+,+) \;, \nn \\
g(z_3) =& (0,0,-1,-1,0) \;,\nn\\
g(z_4) =& (0,0,0,0,0) \;.
\end{align}
After we perform the Riemann summation we find that all dependence upon the position of the vertex operator positions drops out and we have
\begin{align}
A \supset \propto \frac{1}{k_2 \cdot k_5} (u_1 \slashed{k}_5 u_2) \int dt  \partial_t Z(t)
\end{align}
which gives us exactly an anomaly mediation mass term.

For completeness, here we can include the case where there are three PCOs acting externally.

\subsubsection*{$\mu=1-, \nu=1+,\rho=2+$}

We compute with the charges
\begin{align}
\psi^{+1/2}(z_1) =& \half(-,-,+,+,+) \;, \nn \\
\psi^{-1/2}(z_2) =& \half(+,-,-,+,+) \;, \nn \\
g(z_3) =& (-1,0,-1,-1,0) \;,\nn\\
g(z_4) =& (+1,0,0,0,-1) \;,\nn\\
\phi^0 (z_5) =&(0,1,+1,0,0) \:.
\end{align}
This multiplies $-2 k_3^2 (u_1 C \slashed{k}_5 u_2)$. We see that it may contain a pole in $k_3^2$ so we must treat it carefully. It reduces to the choice of charges
\begin{align}
\psi^{+1/2}(z_1) =& \half(-,-,+,+,+) \;, \nn \\
\psi^{-1/2}(z_2) =& \half(+,+,+,+,+) \;, \nn \\
g(z_3) =& (-1,0,-1,-1,0) \;,\nn\\
g(z_4) =& (+1,0,0,0,-1) \;.
\end{align}
Since this is not multiplied by any internal bosonic correlators this may not contribute to the anomaly mediated mass, since there is no derivative term for the partition function here. This may only contribute running mass terms.

\subsubsection*{$\mu=2+, \nu=1+,\rho=1-$}

We compute with the charges
\begin{align}
\psi^{+1/2}(z_1) =& \half(-,-,+,+,+) \;, \nn \\
\psi^{-1/2}(z_2) =& \half(+,-,-,+,+) \;, \nn \\
g(z_3) =& (0,+1,-1,-1,0) \;,\nn\\
g(z_4) =& (+1,0,0,0,-1) \;,\nn\\
\phi^0 (z_5) =&(-1,0,+1,0,0) \:.
\end{align}
This also multiplies $-k_3^2 (u_1 C \slashed{k}_5 u_2)$. It reduces to the choice of charges
\begin{align}
\psi^{+1/2}(z_1) =& \half(-,-,+,+,+) \;, \nn \\
\psi^{-1/2}(z_2) =& \half(-,-,+,+,+) \;, \nn \\
g(z_3) =& (0,+1,-1,-1,0) \;,\nn\\
g(z_4) =& (+1,0,0,0,-1) \;,
\end{align}
which vanishes by Riemann summation.

%%%%%%%%%%%%%%%%%%%%%%%%%%%%%%%%%%%%%%%%%%%%%%%%%%%%%%%%%%%%%%%%%%%%%%%%%%%%%%%%%%%%%%%%%%%%%%
\section{RR Flux computation}
\label{sec:rrapp}
%%%%%%%%%%%%%%%%%%%%%%%%%%%%%%%%%%%%%%%%%%%%%%%%%%%%%%%%%%%%%%%%%%%%%%%%%%%%%%%%%%%%%%%%%%%%%%

In this section we give the details of the RR flux computation that were omitted in the main text.
The relevant correlator involves 2 gauginos and one flux vertex operator $\left<\lambda\lambda F\right>$.  The vertex operator for the
RR field strength is given by
\beq
V_F^{(-1/2,-1/2)} = N_F g_s e^{-\phi/2 - \tilde{\phi}/2} F_{mnp} \Theta (z) C \Gamma^{mnp} \tilde{\Theta} (\tilde{z}),
\eeq
where a factor of $g_s$ is included as we take the sphere to have a factor $g_s^{-2}$. The differing factors of $g_s$
between RR and NS-NS vertex operators relate to the
fact that the RR vertex operator directly involves the field strength $F_{mnp}$
whereas the NS-NS operator involves the 2-form potential.
$N_F$ is a normalisation factor that is not important for our purposes. $\Theta, \tilde{\Theta}$ are ten-dimensional spin fields; $C$ is the charge conjugation operator, given by
\beq
C = \Gamma^0 \Gamma^3 \Gamma^5 \Gamma^7 \Gamma^9,
\eeq
and hence since $\Theta$ and $\tilde{\Theta}$ have the same chirality only odd forms are allowed, as we expect from IIB.
The form of the $C \Gamma^{mnp}$ structure then determines the
relative H-charges of the two spinors $\Theta$ and $\tilde{\Theta}$.

The fact that $F_{mnp}$ is itself the field strength implies that there are two distinct contributions to the RR vertex operator.
The physical state conditions are $dF_{mnp} = d*F_{mnp} = 0$.
As for the NS-NS 2-form potential, we take the RR 2-form potential to be
\be
C_2 = X_5 dX_3 \wedge dX_4 e^{ik \cdot X} + \hbox{c.c}.
\ee
This implies that the 3-form field strength has the form
\be
\label{keats}
F_3 = dC_2 = dX_3 \wedge dX_4 \wedge dX_5 e^{ik \cdot X} + i X_5 k_i dX_i \wedge dX_3 \wedge dX_4 e^{ik \cdot X} + \hbox{c.c}.
\ee
Although the second term naively vanishes as $k \to 0$, it is necessary to include it as momentum poles in
$1/k\cdot k$ can give a non-vanishing contribution in this limit. As long as we work at non-zero $k$
the second term is necessary to satisfy the Bianchi identity, and it is only at the end of the
computation that we can send $k \to 0$.
Both terms in (\ref{keats}) need to be considered separately, as they involve different H-charge structures among the
flux spinors. We shall consider the first (`standard') case to start with, and then subsequently consider the
second (`non-standard') possibility.

As described in appendix A, we do not include ghost derivative terms when evaluating the amplitudes.

\subsection{Standard Case}

As discussed in section \ref{sec:anommedsuper} we turn on RR flux of type $(3,0) + (0,3)$.
In this case only one choice of spinors is non-vanishing, since $\left(\Gamma^{345} \pm \Gamma^{\bar{3}\bar{4}\bar{5}} \right) |\pm \frac{1}{2},\pm\frac{1}{2},\pm\frac{1}{2} \ket$ is only non-vanishing for spinors that are either `all plus' or `all minus'.
As the gauginos have internal H-charges $| +\half, +\half, +\half \ket$, this implies the flux H-charges are
restricted to the case $|-\half, -\half, -\half \ket \otimes |-\half, -\half, -\half \ket$.

From the preceding discussion the `canonical picture' vertex operators then have charges
\begin{align}
g_1^{-1/2}(z_1) =& \half(-,-,+,+,+) \;, \nn \\
g_2^{-1/2}(z_2) =& \half(+,+,+,+,+) \;, \nn \\
\Theta^{-1/2} (w) =& \half(+,-,-,-,-) \;, \nn \\
\tilde{\Theta}^{-1/2} (\tilde{w}) =& \half(-,+,-,-,-) \;.
\end{align}
There is also a very similar case where we take $(+ -) \to (- +)$ and vice-versa. This gives identical expressions
but with $1$ and $2$ directions interchanged.
Canceling the ghost charge requires two PCOs which we take to act on $g_2(z_2)$ and $\tilde{\Theta}(\tilde{w})$.
There are a total of five separate cases which we evaluate in turn, with
the final amplitude given by the sum of all cases.

The computation is on an annulus with Dirichlet boundary conditions on the boundary for internal directions
and Neumann boundary conditions for external directions. As for the NSNS case, we can deal with these by replacing
 $\tilde{\Theta}(\tilde{w})$ by $-\Theta(-\bar{w})$ (Dirichlet) and
 $\tilde{\Theta}(\tilde{w})$ by $\Theta(-\bar{w})$ (Neumann).

\subsubsection*{Case 1 (Internal Picture Changing)}

The first amplitude corresponds to the PCOs acting on the internal direction. This gives the charges
\begin{align}
g_1^{-1/2}(z_1) =& \half(-,-,+,+,+) \;, \nn \\
g_2^{+1/2}(z_2) =& \half(+,+,+,+,-) \;, \nn \\
\Theta^{-1/2} (w) =& \half(+,-,-,-,-) \;, \nn \\
\tilde{\Theta}^{+1/2} (-\bar{w}) =& \half(-,+,-,-,+) \;.
\end{align}
There are two other cases trivially related to this by modifying the internal direction on which the picture changing acts.
The fermionic modes give a factor of\footnote{Throughout the calculations we often use the identity $\thba{\alpha}{\beta}{x}\thba{\alpha}{\beta}{y}=\thba{\alpha}{\beta}{-x}\thba{\alpha}{\beta}{-y}$
 to manipulate the amplitude to a form where the spin structure summation can be performed.}
\be
\frac{2\theta_1(\theta_1) \theta_1(\theta_2) \theta_1(-z_2 - \bar{w} + \theta_3)}{\theta_1(z_2 + \bar{w})} \;.
\ee
This vanishes unless $\theta_3 = 0$, and so the expression requires picture-changing in the
untwisted $N=2$ direction. In this case the fermionic part
reduces simply to a constant. Including the bosonic factors as in (\ref{expcorrreal}) then gives
\begin{align}
{\cal A}_1 =& A_0 \int \frac{dt}{t} \frac{1}{(t/2)^2} \int dz_1 dz_2 d^2 w \bra \partial_n X^5 (z_2) \bar{\partial} \overline{X}^5 (-\bar{w})\ket \nonumber\\
=& 2\pi\alpha' A_0  \int \frac{dt}{t}  ( 1 + t \frac{d}{dt} ) Z_{cl} (t) .
\end{align}
Note that the PCO applied to the RR field involves a $\bar{\partial}$ rather than a $\partial_n$, as the RR field is a bulk field
and the PCO only applies to the antiholomorphic component
Here $A_0$ is a normalisation constant and we have used (\ref{dirdercorrint}) and (\ref{classicalnormalderivative}) to evaluate the bosonic correlator.

\subsubsection*{Case 2A (External Picture Changing)}

We now consider picture changing applied to the external directions, for which there are four possibilities. In these cases derivative terms
arise and must be dealt with carefully. These occur when the OPE of the picture changing operator with the vertex operator has a pole
as its leading term, and it is necessary to consider higher order terms in the
OPE.\footnote{Specifically, these arise from the OPE of $e^{\phi} \partial X \cdot \psi (z)$ with
$e^{-\phi/2} e^{ik \cdot X} \bar{\psi}(w)$, which has as its leading term $\frac{ik}{z-w}$.}
The higher order terms can come from
contracting the derivative terms in the PCOs either with each other or with the momentum exponentials.
To take this into account, we treat the PCO itself as a vertex operator and take the corresponding contraction limit
of the PCO onto the vertex operator only at the end.\footnote{Based on the calculation of gaugino threshold corrections,
 we only include derivative terms from the bosonic and fermionic fields (not from the ghosts). We discuss this in more detail
 in appendix \ref{sec:thresholds}.}  The first case is given by the H-charges
\begin{align}
g_1^{-1/2}(z_1) =& \half(-,-,+,+,+) \;, \nn \\
PCO(x) =& (-1,0,0,0,0) \;, \nn \\
g_2^{+1/2}(z_2) =& \half(+,+,+,+,+) \;, \nn \\
\Theta^{-1/2} (w) =& \half(+,-,-,-,-) \;, \nn \\
PCO(y) =& (+1,0,0,0,0) \;, \nn \\
\tilde{\Theta}^{+1/2} (-\bar{w}) =& \half(-,+,-,-,-) \;.
\end{align}
For notational convenience we use $z_3 = w$ and $z_4 = - \bar{w}$ in order to make the pole structure clearer.
It is possible to perform the spin-structure sums directly, with the aid of the 4-theta Riemann identity.
Defining $\delta_2 \equiv x-z_2, \delta_4 \equiv -\bar{w} - y$, we have for an $\mc{N}=2$ sector
\begin{align}
\label{whwh}
&2\theta_1 (z_1 - z_3 + \frac{x+z_2 - y - z_4}{2}) \prod_{i=1}^3 \frac{\theta_1 (\frac{x-z_2 +z_4-y}{2} + \theta_i)}{\theta_1 (\theta_i)} \nonumber \\
=&-2\pi (\delta_2 + \delta_4) \theta_1 (z_1 - z_3 + z_2 - z_4)\bigg[1 + \frac{1}{2}(\delta_2 + \delta_4)  \frac{\theta_1^\prime (z_1 - z_3 + z_2 - z_4)}{\theta_1 (z_1 - z_3 + z_2 - z_4)} \bigg].
\end{align}
Now we turn to the spin-structure independent pieces from the fermionic correlators:
\begin{align}
&\theta_1(z_1-z_2)^{1/2}\theta_1(z_1 - z_3)^{-1} \theta_1(z_1-z_4)^{-1/2} \theta_1(z_1-x)^{1/2}\theta_1(z_1-y)^{-1/2} \nn \\
&\times \theta_1(z_2-z_3)^{-1/2}\theta_1(z_2 - z_4)^{-1}  \theta_1(x-z_2)^{-1/2} \theta_1(z_2-y)^{1/2}
\theta_1(x - y)^{-1}
\nn\\
&\times \theta_1 (z_3-z_4)^{1/2} \theta_1 (x-z_3)^{-1/2} \theta_1 (z_3-y)^{1/2} \theta_1 (x-z_4)^{1/2} \theta_1 (z_4-y)^{-1/2}\nn\\
=& \delta_2^{-1/2} \delta_4^{-1/2} \theta_1(z_1-z_2)\theta_1 (z_3-z_4)\theta_1(z_1 - z_3)^{-1} \theta_1(z_1-z_4)^{-1} \theta_1(z_2-z_3)^{-1}\theta_1(z_2 - z_4)^{-1}    \nn\\
&\times \bigg[ 1+\frac{1}{2} \delta_2 \bigg\{ -\frac{\theta_1^\prime (z_1 - z_2)}{\theta_1 (z_1 - z_2)} - \frac{\theta_1^\prime (z_2 - z_3)}{\theta_1 (z_2 - z_3)} - \frac{\theta_1^\prime (z_2 - z_4)}{\theta_1 (z_2 - z_4)}\bigg\} \nn\\
& + \frac{\delta_4}{2} \bigg\{ -\frac{\theta_1^\prime (z_1 - z_4)}{\theta_1 (z_1 - z_4)}-\frac{\theta_1^\prime (z_2 - z_4)}{\theta_1 (z_2 - z_4)} + \frac{\theta_1^\prime (z_3 - z_4)}{\theta_1 (z_3 - z_4)}\bigg\}\bigg].
\end{align}
The final part we need is the bosonic correlator, and the poles in $\delta_2$ and $\delta_4$ that it contains.
This is given by
\bea
\left<\partial_t X^1(x) \bar{\partial} \overline{X}^1(y) \prod_i e^{ik_i \cdot X(z_i)} \right> & = &
\left(\frac{\alpha'}{2}\right)^2 k_2^{1+}k_3^{1-} \frac{1}{\delta_2\delta_4} \nonumber \\
& & + \frac{\alpha'}{4}
\left[ \frac{k_2^{1+}}{\delta_2} \left( k_1^{1-} G^1_{\bar{\partial}}(-\bar{w},z_1) + k_2^{1-} G^1_{\bar{\partial}}(-\bar{w},z_2) + k_3^{1-} G^1_{\bar{\partial}}(-\bar{w},w) \right) \right. \nn \\
 & & - \left. \frac{k_3^{1-}}{\delta_4} \left( k_1^{1+} G^1(z_2,z_1) + k_3^{1+} G^1(z_2,w) + k_3^{1+} G^1(z_2,-\bar{w}) \right) \right]\;. \label{boscor2a}
\eea
Here we have defined
\bea
G^1\left(z_1,z_2\right) & \equiv & \left<\partial_tX^1(z_1)\overline{X}^1(z_2)\right> = -\frac{\alpha'}{2}\left[ \frac{\theta_1^{'}(z_1-z_2)}{\theta_1(z_1-z_2)}+\frac{\theta_1^{'}(z_1+\bar{z}_2)}{\theta_1(z_1+\bar{z}_2)} - \frac{\theta_1^{'}(\bar{z}_1-\bar{z}_2)}{\theta_1(\bar{z}_1-\bar{z}_2)} - \frac{\theta_1^{'}(\bar{z}_1+z_2)}{\theta_1(\bar{z}_1+z_2)} + ... \right] \;,  \nn \\
G^1_{\bar{\partial}} \left(z_1,z_2\right) & \equiv & \left<\bar{\partial} \overline{X}^1(z_1)X^1(z_2)\right> =
-\alpha'\left[  \frac{\theta_1^{'}(\bar{z}_1-\bar{z}_2)}{\theta_1(\bar{z}_1-\bar{z}_2)} + \frac{\theta_1^{'}(\bar{z}_1+z_2)}{\theta_1(\bar{z}_1+z_2)} + ... \right], \label{g1def}
\eea
where the ellipses denote non singular terms.

We now want to analyse the pole structure of these terms. In principle there can be terms involving zero, one or two derivatives
(equivalently, zero, one or two powers of $\delta_2$ and $\delta_4$). The zero derivative term clearly vanishes due to the powers of $\delta$ in (\ref{whwh}).
We now analyse the other derivative terms.

First consider the case of the `double pole' that comes from the term $\frac{1}{\delta_2 \delta_4}$. As $z_{3/4} \to z_1$ there is a momentum pole,
coming from terms of the form
\bea
& & \half \left( - \frac{\theta_1(z_{34})}{\theta_1(z_{14}) \theta_1(z_{13}) \theta_1(z_{14})} + \frac{\theta_1(z_{34})}
{\theta_1(z_{14}) \theta_1(z_{13}) \theta_1(z_{34})} \right) \\
& = & \half \left( \frac{-(\epsilon + \bar{\epsilon})}{\epsilon \bar{\epsilon} \bar{\epsilon}} + \frac{(\epsilon + \bar{\epsilon})}{
\epsilon \bar{\epsilon} (\epsilon + \bar{\epsilon})} \right) \to 0,
\eea
using the fact that $\int d\epsilon d \bar{\epsilon} \frac{1}{\epsilon^2} = 0$ due to the angular integral.
As $z_{3/4} \to z_2$ we get in the same fashion
\be
\int d\epsilon d \bar{\epsilon} \frac{(\epsilon + \bar{\epsilon})}{\epsilon \bar{\epsilon}} \left( \frac{1}{\epsilon} + \frac{2}{(-\bar{\epsilon})}
+ \frac{1}{(\epsilon + \bar{\epsilon})} \right) = 0.
\ee

There are also `single pole' terms involving only a single factor of $\delta_2^{-1}$ or $\delta_4^{-1}$. As $z_{3/4} \to z_1$
we have
\bea
\label{dryden}
& & \frac{(\epsilon + \bar{\epsilon})}{\epsilon \bar{\epsilon}} \left( 2 k_1^{1-} k_2^{1+} \frac{1}{\bar{\epsilon}} +
k_2^{1+} k_3^{1-} \frac{1}{\epsilon + \bar{\epsilon}} \right) \theta_1(z_{12}) \theta_1(z_{12}) \theta_1(z_{12})^{-1}
\theta_1(z_{12})^{-1} \\
& = & \frac{(2 k_1^{1-} k_2^{1+} + k_3^{1-} k_2^{1+})}{\epsilon \bar{\epsilon}}.
\eea
As $z_{3/4} \to z_2$ we have
\bea
\label{shelley}
& & \frac{\epsilon + \bar{\epsilon}}{\epsilon \bar{\epsilon}}
\left( \frac{k_3^{1-} k_3^{1+}}{2} \left( -\frac{1}{\epsilon} + \frac{1}{\bar{\epsilon}} \right)
+ 2 \frac{k_2^{1+} k_2^{1-} }{\bar{\epsilon}} + \frac{k_2^{1+} k_3^{1-}}{\epsilon + \bar{\epsilon}} \right) \theta_1(z_{21}) \theta_1(z_{12}) \theta_1(z_{12})^{-1}
\theta_1(z_{12})^{-1} \\
& = & - \left( \frac{2 k_2^{1+} k_2^{1-} + k_2^{1+} k_3^{1-}}{\epsilon \bar{\epsilon}} \right).
\eea
The differing factor of $\theta_1(z_{21})$ comes from $\theta_1(z_1 +z_2 - z_3 - z_4)$ as $z_{3/4} \to z_2$.
Both (\ref{dryden}) and (\ref{shelley}) are naively divergent, as $\int d \epsilon d \bar{\epsilon} \frac{1}{\epsilon \bar{\epsilon}} = \infty$.

As $z_{3/4} \to z_1$ the divergence is regulated by the term from the bosonic $\langle e^{ik \cdot X} \rangle$ correlator,
\be
\vert \epsilon \vert^{2 k_1 \cdot k_3 + k_3 \cdot k_3}.
\ee
As $z_{3/4} \to z_2$ the regularisation is
\be
\vert \epsilon \vert^{2 k_2 \cdot k_3 + k_3 \cdot k_3},
\ee
giving overall
\be
\frac{(2 k_1^{1-} \cdot k_2^{1+} + k_2^{1+} \cdot k_3^{1-})}{(2 k_1 \cdot k_3 + k_3 \cdot k_3)}
- \frac{2 k_2^{1+} \cdot k_2^{1-} + k_2^{1+} \cdot k_3^{1-}}{ 2 k_2 \cdot k_3 + k_3 \cdot k_3} = 0.
\ee
once we use $k_1 + k_2 + k_3 = 0$.

We then conclude that this method of PCOs does not contribute to the amplitude.

\subsubsection*{Case 2B (External Picture Changing)}

In this case the H-charges are given by
\begin{align}
g_1^{-1/2}(z_1) =& \half(-,-,+,+,+) \;, \nn \\
g_2^{+1/2}(z_2) =& \half(+++,+,+,+,+) \;, \nn \\
\Theta^{-1/2} (w) =& \half(+,-,-,-,-) \;, \nn \\
\tilde{\Theta}^{+1/2} (-\bar{w}) =& \half(---,+,-,-,-) \;.
\end{align}
There are no derivative terms since the PCO derivatives must be contracted with the exponentials as the H-charges of the PCOs and the vertex operators are the same sign. In this case the Riemann summation of the spin structure dependent parts gives
\be
2\theta_1(z_1 - w) \prod_{i=1}^3 \theta_1(z_2 + \bar{w} - \theta_i) \;.
\ee
The spin structure independent parts give
\be
\theta_1(z_1-w)^{-1} \theta_1(z_2+\bar{w})^{-3} \;.
\ee
Including the bosonic factors, we find overall for the $N=2$ sector
$$
\frac{2\theta_1(z_2 + \bar{w} + \theta)\theta_1(z_2 + \bar{w} - \theta)}{\theta_1(z_2+\bar{w})^{2}\theta_1(\theta)\theta_1(-\theta)} \;.
$$
This expression can be simplified using (\ref{theta1identity}). Alternatively,
the amplitude has an overall prefactor of $k_2 \cdot k_3$ and to get a non-zero answer we need to identify a pole as
$w\to z_2$. However, the angular integral vanishes
\be
\int d^2 w \frac{1}{(w-z_2)^2} \sim \int dr \frac{1}{r} \left[ e^{-2i \theta} \right]_0^{\pi} =0 \;.
\ee
In consequence we conclude that this possibility of PCOs also does not contribute to the amplitude.

\subsubsection*{Case 2C (External Picture Changing)}

In this case the H-charges are given by
\bea
g_1^{-1/2}(z_1) &=& \half(-,-,+,+,+) \;, \nn \\
PCO(u) &=& (0,-1,0,0,0) \;, \nn \\
g_2^{+1/2}(z_2) &=& \half(+,+,+,+,+) \;, \nn \\
\Theta^{-1/2} (z_3) &=& \half(+,-,-,-,-) \;, \nn \\
\tilde{\Theta}^{+1/2} (z_4) &=& \half(-,+++,-,-,-) \;.
\eea
This has an overall momentum factor of $k_3^{2-}$ from picture changing $\tilde{\Theta}$.
There are derivative terms we need to take into account.
The spin structure summation can be evaluated using the identity (\ref{derivthetaiden}),
which allows it to be converted into a form which can be evaluated using the 4-theta Riemann identity.
If we denote $\delta = z_2 - u$, then after a bit of manipulation the spin-structure dependent part becomes
\bea
& & \frac{\delta \theta_1(z_2 - z_1) \theta_1(z_1 - z_4) \theta_1^{'}(0) \theta_1(\frac{1}{2}\left(z_2 + z_1 - z_3 - z_4\right))
\theta_1(-z_3 + z_4 + \theta) \theta_1(z_3 - z_4 + \theta)}{\theta_1(z_2 - z_1 -z_3 + z_4) \theta_1(z_1 - z_3)}  \\
& & + \frac{2 \theta_1(z_2 - z_3) \theta_1(z_3 - z_4)
\theta_1(z_1 - z_3 + \delta/2) \theta_1(z_2 - z_4 - \delta/2) \theta_1(z_1 - z_4 - \delta/2 - \theta)
\theta_1(z_1 - z_4 - \delta/2 + \theta)}{\theta_1(z_2 - z_1 -z_3 + z_4) \theta_1(z_1 - z_3)}. \nonumber
\eea
We can write this as
\bea
& & 2 \frac{ \theta_1(z_2 - z_3) \theta_1(z_3 - z_4) \theta_1(z_2 - z_4) \theta_1(z_1 - z_4 - \theta)
\theta_1(z_1 - z_4 + \theta)}{\theta_1(z_2 - z_1 -z_3 + z_4)} \left[ 1 + \right. \\
& & \left. \frac{\delta}{2} \left(
\frac{\theta_1^{'}(z_1 - z_3)}{\theta_1(z_1 - z_3)} - \frac{\theta_1^{'}(z_2 - z_4)}{\theta_1(z_2 - z_4)}
- \frac{\theta_1^{'}(z_1 - z_4 - \theta)}{\theta_1(z_1 - z_4 - \theta)} -
\frac{\theta_1^{'}(z_1 - z_4 + \theta)}{\theta_1(z_1 - z_4 + \theta)} \right) \right] \nonumber \\
& & - \frac{\delta \theta_1(z_2 - z_1) \theta_1(z_1 - z_4) \theta_1^{'}(0) \theta_1(\frac{1}{2}\left(z_2 + z_1 - z_3 - z_4\right))
\theta_1(z_3 - z_4 - \theta) \theta_1(z_3 - z_4 + \theta)}{\theta_1(z_2 - z_1 -z_3 + z_4) \theta_1(z_1 - z_3)}. \nn
\eea
The spin structure independent part gives
\bea
& & \theta_1(z_2 - z_1) \theta_1(z_2 - z_4)^{-2} \theta_1(z_1 - z_3)^{-1} \theta_1(z_1 - z_4)^{-1} \nn \\
& &\times\left[ 1 + \frac{\delta}{2} \left( \frac{\theta_1^{'}(z_2 - z_1)}{\theta_1(z_2 - z_1)}
+ \frac{\theta_1^{'}(z_2 - z_3)}{\theta_1(z_2 - z_3)} - 3 \frac{\theta_1^{'}(z_2 - z_4)}{\theta_1(z_2 - z_4)} \right)
\right].
\eea
The bosonic correlator is
\be
\label{abccbd}
\left<\partial_t X^{2+} (x)e^{ik\cdot X}\right> = -\frac{\alpha'}{2}
\left[ k_1^{2+}G^1(x,z_1) + k_2^{2+}G^1(x,z_2) + k_3^{2+}G^1(x,z_3) \right],
\ee
where the correlator $G^1(z_1,z_2)$ is defined in (\ref{g1def}).

For non-derivative terms, the combination of the fermionic parts gives
\be
\frac{2 \theta_1(z_2 - z_3) \theta_1(z_3 - z_4) \theta_1(z_2 - z_1) \theta_1(z_1 - z_4 - \theta) \theta_1(z_1 - z_4 + \theta)}
{\theta_1(z_2 - z_1 - z_3 +z_4) \theta_1(z_1 - z_3) \theta_1(z_1 - z_4) \theta_1(z_2 - z_4)}.
\ee
It is not difficult to see that no pole can be
obtained from this expression.

To get a non-zero answer we must go the
derivative terms that depend on $\delta$. In this case we can obtain a pole using the
second contraction in (\ref{abccbd}), in the limit where $z_{3/4} \to z_1$.
Evaluating this, we find a contribution of
\be
\pi \alpha' \int \frac{dt}{t} \frac{k_3^{2-} k_2^{2+}}{2 k_2 \cdot k_3  + k_3 \cdot k_3} Z(t).
\ee

\subsubsection*{Case 2D (External Picture Changing)}

The final consistent way of picture changing external coordinates is given by
\bea
g_1^{-1/2}(z_1) &=& \half(-,-,+,+,+) \;, \nn \\
g_2^{+1/2}(z_2) &=& \half(+,+++,+,+,+) \;, \nn \\
\Theta^{-1/2} (z_3) &=& \half(+,-,-,-,-) \;, \nn \\
PCO(x) &=& (0,-1,0,0,0) \;,\nn \\
\tilde{\Theta}^{+1/2} (z_4) &=& \half(-,+,-,-,-) \;.
\eea
At the end we can replace $z_3$ and $z_4$ by $w$ and $-\bar{w}$.
Let us denote $\delta = z_4 - x$. Then again using the identity (\ref{derivthetaiden})
the spin-structure dependent part for the $N=2$ sector can be written as
\bea
& & -2 \frac{ \theta_1(z_2 - z_1) \theta_1(z_1 - z_4) \theta_1(z_2 - z_4) \theta_1(z_2 - z_3 - \theta)
\theta_1(z_2 - z_3 + \theta)}{\theta_1(z_2 - z_1 -z_3 + z_4)} \left[ 1 + \right. \nonumber \\
& & \left. \frac{\delta}{2} \left(
\frac{\theta_1^{'}(z_2 - z_4)}{\theta_1(z_2 - z_4)} - \frac{\theta_1^{'}(z_1 - z_3)}{\theta_1(z_1 - z_3)}
+ \frac{\theta_1^{'}(z_2 - z_3 - \theta)}{\theta_1(z_2 - z_3 - \theta)} +
\frac{\theta_1^{'}(z_2 - z_3 + \theta)}{\theta_1(z_2 - z_3 + \theta)} \right) \right] \nonumber \\
& & - \frac{\delta \theta_1(z_2 - z_3) \theta_1(z_3 - z_4) \theta_1^{'}(0) \theta_1(z_2 + z_1 - z_3 - z_4)
\theta_1(z_2 - z_1 + \theta) \theta_1(z_2 - z_1 - \theta)}{\theta_1(z_2 - z_1 -z_3 + z_4) \theta_1(z_1 - z_3)}.
\eea
The spin structure independent part gives
\bea
& & \theta_1(z_2 - z_4)^{-2} \theta_1(z_3 - z_4) \theta_1(z_2 - z_3)^{-1} \theta_1(z_1 - z_3)^{-1} \nn \\
& &\times\left[ 1 + \frac{\delta}{2} \left( \frac{\theta_1^{'}(z_1 - z_4)}{\theta_1(z_1 - z_4)}
+ \frac{\theta_1^{'}(z_3 - z_4)}{\theta_1(z_3 - z_4)} - 3 \frac{\theta_1^{'}(z_2 - z_4)}{\theta_1(z_2 - z_4)} \right)
\right]. \nonumber
\eea
The bosonic correlator is
$
\langle \bar{\partial} X^{2+} (x) \prod_i e^{ik \cdot X(z_i)} \rangle,
$
which gives
\be
\label{abccba}
k_2^{2-}\left<\bar{\partial} X^{2+}(x)e^{ik\cdot X}\right> = -\frac{\alpha'}{2} k_2^{2-}
\left[ k_1^{2+}G_{\bar{\partial}}^1(x,z_1) + k_2^{2+}G_{\bar{\partial}}^1(x,z_2) + k_3^{2+}G_{\bar{\partial}}^1(x,z_3) \right].
\ee
The correlator $G_{\bar{\partial}}^1(z_1,z_2)$ is defined in (\ref{g1def}).

Without the inclusion of derivative terms, the fermionic terms give
\be
\frac{2 \theta_1(z_2 - z_1) \theta_1(z_1 - z_4) \theta_1(z_3 - z_4) \theta_1(z_2 - z_3 - \theta) \theta_1(z_2 - z_3 + \theta)}
{\theta_1(z_2 - z_1 - z_3 +z_4) \theta_1(z_2 - z_3) \theta_1(z_2 - z_4) \theta_1(z_1 - z_3)}.
\ee
This has a potential pole as $z_{3/4} \to z_2$. We can obtain this pole in two ways, using either the first or third
 term of (\ref{abccba}). These give a momentum structure of the form (using $2k_2 \cdot k_3 + k_3 \cdot k_3 = -2 k_1
 \cdot k_3 - k_3 \cdot k_3 $)
\be
\label{ya2}
\pi \alpha' \int \frac{dt}{t} 2 \frac{\left( k_3^{2+} k_2^{2-} + 2k_2^{2+} k_2^{2-} \right)}{2 k_2 \cdot k_3 + k_3 \cdot k_3}.
\ee
We also need to include the derivative terms of $\mc{O}(\delta)$. Evaluating these we find that there is again a pole
as $z_{3/4} \to z_2$. Evaluating this we obtain
\be
\label{ya1}
\pi \alpha' \int \frac{dt}{t} - \frac{k_2^{2-} k_3^{2+}}{2 k_2 \cdot k_3 + k_3 \cdot k_3} Z(t).
\ee
Combining (\ref{ya1}) and (\ref{ya2}) we get
\be
\pi \alpha' \int \frac{dt}{t} \, \frac{k_3^{2+} k_2^{2-} + 2 (k_2^{2+} k_2^{2-} + k_2^{2-} k_2^{2+})}{2 k_2 \cdot k_3 + k_3 \cdot k_3} Z(t).
\ee

If we now combine all the standard cases, we have
\be
\pi \alpha' \int \frac{dt}{t} \, (2 + 2 t \frac{d}{dt}) Z(t) + \frac{k_3^{2+} k_2^{2-} + k_3^{2-} k_2^{2+} + 2 (k_2^{2+} k_2^{2-} + k_2^{2-} k_2^{2+})}{2 k_2 \cdot k_3 + k_3 \cdot k_3} Z(t).
\ee
This term should be Lorentz completed by including the case where the external H-charges of the flux spinors $\Theta$ and $\tilde{\Theta}$
are exchanged. This will give an identical contribution except with $1$ and $2$ directions interchanged.
We then finally emerge with
\be
\label{milton}
\pi \alpha' \int \frac{dt}{t} \, \left[ (4 + 4t \frac{d}{dt}) Z(t) + \frac{k_2 \cdot k_3 + 2 k_2 \cdot k_2}{2 k_2 \cdot k_3 + k_3 \cdot k_3} Z(t) \right].
\ee

\subsection{Non-standard Case}

We now consider contributions from the non-standard case. There are again two basic options for the flux H-charges in the canonical
picture, coming from $k_3^{1+} dX^{1-}$ and $k_3^{2+} dX^{2-}$. We consider these separately.

\subsubsection*{Case 3A}

In this case we start with H-charges
\bea
g^{-1/2}_1(z_1) &=& \half(-,-,+,+,+) \;, [-1/2] \nn \\
g^{-1/2}_2(z_2) &=& \half(+,+,+,+,+) \;, [-1/2] \nn \\
\Theta^{-1/2}(z_3) & = & \half(-,-,-,-,+) \;, [-1/2] \nn \\
\tilde{\Theta}^{-1/2}(z_4) & = & \half(-,+,-,-,-) \;, [-1/2]
\eea
There is a bosonic factor of $k_3^{1+} X_5(z, \bar{z})$, which pairs with $\Gamma^{1+}$.\footnote{This arises as the flux term is
$k_3^{1+} dX^{1-} \wedge dX_3 \wedge dX_4$, and then the form of the RR vertex operator means this flux is paired with a $\Gamma^{1+}$.}
The factor of $\Gamma^{1+}$ then fixes the H-charges of the RR vertex operator due to the structure
$\Theta C \Gamma^{\alpha \beta \gamma} \tilde{\Theta} e_{\alpha \beta \gamma}$. $\Gamma^{1+}$ raises
the $-1/2$ component of $\tilde{\Theta}$ before it is lowered again by $C$.

We will apply PCOs to operators 2 and 4. There are two basic options. In the first we picture change
\bea
g^{-1/2}_2(z_2) = \half(+,+,+,+,+) \; [-1/2] & \to & \half(+++,+,+,+,+) \; [+1/2], \nn \\
\tilde{\Theta}^{-1/2}(z_4) = \half(-,+,-,-,-) \; [-1/2] & \to & \half(-,+,-,-,---), [+1/2] \nn
\eea
In the second we picture change
\bea
g^{-1/2}_2(z_2) = \half(+,+,+,+,+) \; [-1/2] & \to & \half(+,+,+,+,-) \; [+1/2], \nn \\
\tilde{\Theta}^{-1/2}(z_4) = \half(-,+,-,-,-) \; [-1/2] & \to & \half(+,+,-,-,-) [+1/2]. \nn
\eea

We start with the first case, where the H-charges behave as
\bea
g^{-1/2}_1(z_1) = \half(-,-,+,+,+) \;, [-1/2] & \to & \half(-,-,+,+,+) \; [-1/2],  \\
g^{-1/2}_2(z_2) = \half(+,+,+,+,+) \; [-1/2] & \to & \half(+++,+,+,+,+) \; [+1/2], \nn \\
\Theta^{-1/2}(z_3) = \half(-,-,-,-,+) \;, [-1/2] & \to & \half(-,-,-,-,+) \; [-1/2], \nn \\
\tilde{\Theta}^{-1/2}(z_4) = \half(-,+,-,-,-) \; [-1/2] & \to & \half(-,+,-,-,---) [+1/2]. \nn
\eea
In this case there are no derivative terms. The spin structure dependent part can be evaluated easily using the
4-$\theta$ Riemann identity, and gives
\be
\theta_1(z_1 - z_4) \theta_1(-z_2 + z_4 + \theta) \theta_1(-z_2 + z_4 - \theta) \theta_1(-z_2 + z_3).
\ee
The spin structure independent parts give
\be
\theta_1(z_1 - z_4)^{-1} \theta_1(z_2 - z_3)^{-1} \theta_1(z_2 - z_4)^{-2},
\ee
giving overall
\be
\frac{\theta_1(-z_2 + z_4 + \theta) \theta_1(-z_2 + z_4 - \theta)}{\theta_1(z_2 - z_4)^2 \theta_1(\theta) \theta_1(-\theta)}.
\ee
This has no pole as the angular integral cancels as $z_4 \to z_2$.

We now consider the second case, where the H-charges behave as
\bea
g^{-1/2}_1(z_1) & = & \half(-,-,+,+,+) \;, [-1/2] \to \half(-,-,+,+,+) \;, [-1/2]  \\
g^{-1/2}_2(z_2) & = & \half(+,+,+,+,+) \; [-1/2] \to \half(+,+,+,+,-) \; [+1/2], \nn \\
\Theta^{-1/2}(z_3) & = & \half(-,-,-,-,+) \;, [-1/2] \to \half(-,-,-,-,+) \; [-1/2], \nn \\
G(u) & \to & (++,0,0,0,0) \nn \\
\tilde{\Theta}^{-1/2}(z_4) & = & \half(-,+,-,-,-) \; [-1/2] \to \half(-,+,-,-,-) [+1/2]. \nn
\eea
As stated previously we only include derivative terms coming from
the fermion and boson fields and do not include ghost derivatives. We first evaluate the ghost + fermion contributions.

The spin structure dependent parts give
\bea
&&\theta_{\alpha \beta}\left(\frac{-z_1 + z_2 - z_3 + z_4}{2} + (u-z_4)\right)\theta_{\alpha \beta}\left(\frac{-z_1 + z_2 - z_3 + z_4}{2}\right)
\theta_{\alpha \beta}\left(\frac{z_1 + z_2 - z_3 - z_4}{2} + \theta\right) \nn \\
&&\times\theta_{\alpha \beta}\left(\frac{z_1 + z_2 - z_3 - z_4}{2} - \theta\right).
\eea
This gives
\be
-\theta_1(z_2 - z_3) \theta_1(z_1 - z_4) \theta_1(\theta) \theta_1(-\theta)
\left[ 1 + \left( \frac{u-z_4}{2} \right) \left( \frac{\theta_1^{'}(z_2 - z_3)}{\theta_1(z_2 - z_3)}
+ \frac{\theta_1^{'}(z_1 - z_4)}{\theta_1(z_1 - z_4)} \right) \right].
\ee
The spin structure independent parts give
\be
\theta_1(z_1 - z_4)^{-1} \theta_1(z_2 - z_3)^{-1} \left[
1 + \frac{(u-z_4)}{2} \left( \frac{\theta_1^{'}(z_1 - z_4)}{\theta_1(z_1 - z_4)}
- \frac{\theta_1^{'}(z_2 - z_4)}{\theta_1(z_2 - z_4)} + \frac{\theta_1^{'}(z_3 - z_4)}{\theta_1(z_3 - z_4)}
\right) \right].
\ee
So combined, the fermionic and ghost parts give
\be
- \left( 1 + \frac{(u-z_4)}{2} \left[ \frac{\theta_1^{'}(z_2 - z_3)}{\theta_1(z_2 - z_3)}
- \frac{\theta_1^{'}(z_2 - z_4)}{\theta_1(z_2 - z_4)} + \frac{2 \theta_1^{'}(z_1 - z_4)}{\theta_1(z_1 - z_4)}
+ \frac{\theta_1^{'}(z_3 - z_4)}{\theta_1(z_3 - z_4)} \right] \right)
\ee

Now lets consider the bosonic part. The basic bosonic correlator we want is
\be
k_3^{1+} \langle (\partial + \bar{\partial}) \bar{X}^5(z_2) X^5(z_3, \bar{z}_3) \bar{\partial} X^{1-}(u)
e^{ik \cdot X(z_1)} e^{ik \cdot X(z_2)} e^{ik \cdot X(z_3)}
\ee
The $X^5$ correlator is in the Dirichlet direction. This is given by
\be
\langle \bar{X}^3(z_2) X^3(z_3, \bar{z}_3) \rangle = - \alpha' \left[ \ln \theta_1(z_2 - z_3)
+ \ln \theta_1(\bar{z}_2 - \bar{z}_3) - \ln \theta_1(\bar{z}_2 + z_3) - \ln \theta_1(z_2 + \bar{z}_3) \right].
\ee
Therefore, using $z_2 = - \bar{z}_2$ we have
\be
\langle (\partial + \bar{\partial}) \bar{X}^5(z_2) X^5(z_3, \bar{z}_3) \rangle_{D,qu}
= - 2 \alpha' \left[ \frac{\theta_1^{'}(z_2 - z_3)}{\theta_1(z_2 - z_3)} - \frac{\theta_1^{'}(z_2 + \bar{z}_3)}{\theta_1(z_2 + \bar{z}_3)} \right].
\ee
Note this vanishes when $z_3$ is on the boundary.

Meanwhile, we also need the external correlator
$
\langle \bar{\partial} X^{1-}(u) e^{ik \cdot X(z_1)} e^{ik \cdot X(z_2)} e^{ik \cdot X(z_3)} \rangle.
$
The Neumann correlator is
\be
\langle X(u) \bar{X}(z) \rangle = - \alpha' \left[ \ln \theta_1(u-z) + \ln \theta_1(\bar{u} - \bar{z})
+ \ln \theta_1(u + \bar{z}) + \ln \theta_1(\bar{u} + \bar{z}) \right]
\ee
and so
\be
\langle \bar{\partial} X(u) \bar{X}(z) \rangle = - \alpha' \left[ \frac{\theta_1^{'}(\bar{u} - \bar{z})}{\theta_1(\bar{u} - \bar{z})}
+ \frac{\theta_1^{'}(\bar{u} + z)}{\theta_1(\bar{u} + z)} \right].
\ee
 Then
\bea
\langle \bar{\partial} X^{1-}(u) e^{ik \cdot X(z_1)} e^{ik \cdot X(z_2)} e^{ik \cdot X(z_3)} \rangle
& = & -\alpha' k_3^{1-} \left[ \frac{\theta_1^{'}(\bar{u} - \bar{z}_3)}{\theta_1(\bar{u} - \bar{z}_3)}
+ \frac{\theta_1^{'}(z_3 + \bar{z}_3)}{\theta_1(z_3 + \bar{z}_3)} \right] \nonumber \\
& & -2 \alpha' k_2^{1-} \frac{\theta_1^{'}(z_2 + \bar{z}_3)}{\theta_1^{'}(z_2 + \bar{z}_3)}
-2 \alpha' k_1^{1-} \frac{\theta_1^{'}(z_1 + \bar{z}_3)}{\theta_1(z_1 + \bar{z}_3)}.
\eea

Now let's combine the fermionic and bosonic parts. For non-derivative terms we have
\be
\langle (\partial + \bar{\partial}) \bar{X}^5(z_2) X^5(z_3, \bar{z}_3) \rangle
\ti (- \alpha' k_3^{1+} k_3^{1-} \frac{\theta_1^{'}(z_3 + \bar{z}_3)}{\theta_1(z_3 + \bar{z}_3)})
\ee
This gives a contribution
\be
\label{sappho1}
\langle (\partial + \bar{\partial}) \bar{X}^5(z_2) \Delta X^5(z_3, \bar{z}_3) \rangle \ti \frac{\alpha' k_3^{1+} k_3^{1-}}{2 k_3 \cdot k_3},
\ee
which is an anomalous mass term. The other contribution is
\be
4 \alpha'^2 k_3^{1+} k_2^{1-} \frac{\theta_1^{'}(z_2 + \bar{z}_3)}{\theta_1^{'}(z_2 + \bar{z}_3)}
\frac{\theta_1^{'}(z_2 - z_3)}{\theta_1^{'}(z_2 - z_3)},
\ee
which then gives a contribution to a running mass of
\be
\label{sappho2}
4 \pi \alpha'^2 \frac{k_3^{1+} k_2^{1-}}{2 k_2 \cdot k_3 + k_3 \cdot k_3} \int \frac{dt}{t}.
\ee

Now consider derivative terms. Here we have
\be
\alpha' k_3^{1-} k_3^{1+} \langle (\partial + \bar{\partial}) \bar{X}^5(z_2) X^5(z_3, \bar{z}_3) \rangle
\half \left( \frac{\theta_1^{'}(z_2 - z_3)}{\theta_1(z_2 - z_3)} - \frac{\theta_1^{'}(z_2 - z_4)}{\theta_1(z_2 - z_4)}
+ \frac{2 \theta_1^{'}(z_1 - z_4)}{\theta_1(z_1 - z_4)} + \frac{\theta_1^{'}(z_3 - z_4)}{\theta_1(z_3 - z_4)} \right).
\ee
As $z_3 \to z_4$, we have
\be
\label{sappho3}
- \frac{\alpha' k_3^{1-} k_3^{1+}}{4 k_3 \cdot k_3} \langle (\partial + \bar{\partial}) \bar{X}^5(z_2) \Delta X^5(z_3, \bar{z}_3) \rangle.
\ee
This gives a mass term of opposite sign to the previous mass term.

As $z_{3/4} \to z_2$, we have
\be
\label{sappho4}
\alpha' k_3^{1-} k_3^{1+} \frac{-2 \alpha'}{2} \left( \frac{1}{z_2 - z_3} - \frac{1}{z_2 + \bar{z}_3} \right)
\left( \frac{1}{z_2 - z_3} - \frac{1}{z_2 + \bar{z}_3} \right) = \frac{2 \pi \alpha'^2 k_3^{1-} k_3^{1+}}{2 k_2 \cdot k_3 + k_3 \cdot k_3}
\int \frac{dt}{t} Z(t).
\ee

If we now combine equations (\ref{sappho1}), (\ref{sappho2}), (\ref{sappho3}) and (\ref{sappho4}) we get
\be
\label{qq}
2 \pi \alpha' \frac{2 k_3^{1+} k_2^{1-} + k_3^{1-} k_3^{1+}}{2 k_2 \cdot k_3 + k_3 \cdot k_3} \int \frac{dt}{t} Z(t)
- \frac{2 \pi \alpha' k_3^{1-} k_3^{1+}}{k_3 \cdot k_3} \int dt \frac{d}{dt} Z(t).
\ee

\subsubsection*{Case 3B}

We now consider the other contribution coming from the other set of H-charges. In this case we start with H-charges
\bea
g^{-1/2}_1(z_1) &=& \half(-,-,+,+,+) \;, [-1/2] \nn \\
g^{-1/2}_2(z_2) &=& \half(+,+,+,+,+) \;, [-1/2] \nn \\
\Theta^{-1/2}(z_3) & = & \half(+,+,-,-,+) \;, [-1/2] \nn \\
\tilde{\Theta}^{-1/2}(z_4) & = & \half(-,+,-,-,-) \;, [-1/2]
\eea
The bosonic factor is now $k_3^{2-} X^5(z, \bar{z})$. There are two basic options for the picture changing. The first is to
picture change to
\bea
g^{-1/2}_1(z_1) &=& \half(-,-,+,+,+) \;, [-1/2] \nn \\
g^{-1/2}_2(z_2) &=& \half(+,+,+,+,-) \;, [+1/2] \nn \\
\Theta^{-1/2}(z_3) & = & \half(+,+,-,-,+) \;, [-1/2] \nn \\
PCO(u) & = & \half(0,--,0,0,0) \;, \nn \\
\tilde{\Theta}^{-1/2}(z_4) & = & \half(-,+,-,-,-) \;, [+1/2]
\eea
The spin-structure dependent part of this amplitude is
\be
\theta_1 \left(z_2 - z_4 + \frac{(z_4 - u)}{2} \right)
\theta_1 \left(-z_1 + z_3 + \frac{(z_4 - u)}{2} \right) \theta_1(\theta) \theta_1(-\theta),
\ee
which evaluates to
\be
\theta_1(z_2 - z_4) \theta_1(-z_1 + z_3) \theta(\theta) \theta(-\theta)
\left[ 1 + \frac{(z_4 - u)}{2} \left[ \frac{\theta_1^{'}(z_2 - z_4)}{\theta_1(z_2 - z_4)}
+ \frac{\theta_1^{'}(-z_1 + z_3)}{\theta_1(-z_1 +z_3)} \right] \right].
\ee
In a similar way the spin structure independent part evaluates to
\be
\theta_1(z_1 - z_3)^{-1} \theta_1(z_2 - z_4)^{-1} \left[ 1 - \frac{(u-z_4)}{2} \left(
\frac{\theta_1^{'}(z_1 - z_4)}{\theta_1(z_1 - z_4)} + \frac{\theta_1^{'}(z_2 - z_4)}{\theta_1(z_2 - z_4)}
- \frac{\theta_1^{'}(z_3 - z_4)}{\theta_1(z_3 - z_4)} \right) \right].
\ee
Combining these we have for the fermionic and ghost part
\be
1 - \frac{(u-z_4)}{2} \left[ 2 \frac{\theta_1^{'}(z_1 - z_4)}{\theta_1(z_2 - z_4)}
- \frac{\theta_1^{'}(z_1 - z_3)}{\theta_1(z_1 - z_3)}
+ \frac{\theta_1^{'}(z_1 - z_4)}{\theta_1(z_1 - z_4)}
- \frac{\theta_1^{'}(z_3 - z_4)}{\theta_1(z_3 - z_4)} \right].
\ee
The bosonic correlator we want is
$
k_3^{2-} \langle \partial \bar{X}^5(z_2) X^5(z_3, \bar{z}_3) \bar{\partial} X_2(u) \prod e^{ik \cdot X} \rangle.
$
There are two basic contributing parts here. First we have
\be
\langle \partial \bar{X}^5(z_2) X^5(z, \bar{z}) \rangle_{Qu} = -2 \alpha' \left[ \frac{\theta_1^{'}(z_2 - z_3)}{\theta_1(z_2 - z_3)}
- \frac{\theta_1^{'}(z_2 + \bar{z}_3)}{\theta_1(z_2 + \bar{z}_3)} \right].
\ee
We also have
\bea
\langle \bar{\partial} X_2(u) \prod e^{ik \cdot X(z)} \rangle & = & - \alpha' k_3^{2+} \left[ \frac{\theta_1^{'}(\bar{u}-\bar{z}_3)}
{\theta_1(\bar{u}-\bar{z}_3)} + \frac{\theta_1^{'}(z_3 + \bar{z}_3)}{\theta_1(z_3 + \bar{z}_3)} \right] \nonumber \\
& & -2 \alpha' k_2^{2+} \frac{\theta_1^{'}(z_2 + \bar{z}_3)}{\theta_1(z_2 + \bar{z}_3)} - 2 \alpha' k_1^{2+}
\frac{\theta_1^{'}(z_1 + \bar{z}_3)}{\theta_1(z_1 + \bar{z}_3)}.
\eea
We now want to combine the bosonic and fermionic correlators. The non-derivative terms give a mass term, coming from
\be
\langle (\partial + \bar{\partial}) \bar{X}^5(z_2) X^5(z_3, \bar{z}_3) \rangle_{Cl} \ti -\alpha' k_3^{2-} k_3^{2+}
\frac{\theta_1^{'}(z_3 + \bar{z}_3)}{\theta_1(z_3 + \bar{z}_3)}.
\ee
Factorising $z_3$ onto the boundary then gives an anomalous mass term coming from
\be
\label{arnold1}
\langle (\partial + \bar{\partial}) \bar{X}^5 (z_2) \Delta X^5(z_3, \bar{z}_3) \rangle_{Cl} \ti \frac{\alpha' k_3^{2-} k_3^{2+}}{2 k_3 \cdot k_3}.
\ee
Here $\Delta X^5(z_3, \bar{z}_3) = X^5(1/2+iZ_3) - X^5(iZ_3)$. The classical correlator then gives an anomalous mass term.

There is also a running mass term which comes from directly factorising $z_{3/4} \to z_2$. The magnitude of this is given by
\be
\label{arnold2}
\frac{4 \pi \alpha' k_2^{2+} k_3^{2-}}{2 k_2 \cdot k_3 + k_3 \cdot k_3} \int \frac{dt}{t} Z(t).
\ee

We also need to consider the derivative terms. These give
$$
-\alpha' k_3^{2-} k_3^{2+} \langle \left( \partial + \bar{\partial} \right) \bar{X}^5(z_2) X^5(z_3, \bar{z}_3) \rangle
\half \left( \frac{2 \theta_1^{'}(z_2 - z_4)}{\theta_1(z_2 - z_4)} - \frac{\theta_1^{'}(z_1 - z_3)}{\theta_1(z_1 - z_3)}
+ \frac{\theta_1^{'}(z_1 - z_4)}{\theta_1(z_1 - z_4)} - \frac{\theta_1^{'}(z_3 - z_4)}{\theta_1(z_3 - z_4)} \right).
$$
As $z_3 \to z_4$ we get an anomalous mass term of the form
\be
\label{arnold3}
- \frac{\alpha' k_3^{2-} k_3^{2+}}{4 k_3 \cdot k_3} \langle (\partial + \bar{\partial}) \bar{X}^5(z_2) \Delta X^5(z_3, \bar{z}_3) \rangle_{Cl}.
\ee
As $z_{3/4} \to z_2$ we obtain a running mass term of the form
\be
\label{arnold4}
\frac{2 \pi \alpha'^2 k_3^{2+} k_3^{2-}}{2 k_2 \cdot k_3 + k_3 \cdot k_3} \int \frac{dt}{t} Z(t).
\ee
Equations (\ref{arnold1}), (\ref{arnold2}), (\ref{arnold3}) and (\ref{arnold4}) combine to give
\be
\label{qqq}
2 \pi \alpha' \frac{2 k_2^{2+} k_3^{2-} + k_3^{2+} k_3^{2-}}{2 k_2 \cdot k_3 + k_3 \cdot k_3} \int \frac{dt}{t} Z(t)
- \frac{2 \pi \alpha' k_3^{2-} k_3^{2+}}{k_3 \cdot k_3} \int dt \frac{d}{dt} Z(t).
\ee

Finally we need to consider the other picture changing option, which is
\bea
g^{-1/2}_1(z_1) &=& \half(-,-,+,+,+) \;, [-1/2] \nn \\
g^{-1/2}_2(z_2) &=& \half(+,+,+,+,+) \;, [+1/2] \nn \\
PC(u) & = & \half(0,--,0,0,0) \;, \nn \\
\Theta^{-1/2}(z_3) & = & \half(+,+,-,-,+) \;, [-1/2] \nn \\
\tilde{\Theta}^{-1/2}(z_4) & = & \half(-,+,-,-,---) \;, [+1/2]
\eea
Here the bosonic factors are $k_3^{2-} \partial X^{2+}(u)$, with us being interested in the limit $u \to z_2$.
The spin structure independent part is
\bea
&&\theta_1(z_1 - z_2) \theta_1(z_1 - z_3)^{-1} \theta_1(z_1 - z_4)^{-1} \theta_1(z_2 - z_4)^{-1} \nn \\
&& \times \left[ 1 - \half (u-z_2) \left[ \frac{\theta_1^{'}(z_1 - z_2)}{\theta_1(z_1 - z_2)} +
\frac{\theta_1^{'}(z_2 - z_3)}{\theta_1(z_2 - z_3)} + \frac{\theta_1^{'}(z_2 - z_4)}{\theta_1(z_2 - z_4)} \right] \right].
\eea
The spin structure dependent part is
\bea
&&\frac{
\theta_{\alpha \beta}(\frac{-z_1 + z_2 + z_3 - z_4}{2})
\theta_{\alpha \beta}(\frac{-z_1 - z_2 + z_3 + z_4}{2} + (z_2 - u))}
{\theta_{\alpha \beta}(\frac{z_1 - z_2 + z_3 - z_4}{2})} \nn \\
&& \times \theta_{\alpha \beta}(\frac{z_1 + z_2 - z_3 - z_4}{2} + \theta)
\theta_{\alpha \beta}(\frac{z_1 + z_2 - z_3 - z_4}{2} - \theta)
\theta_{\alpha \beta}(\frac{z_1 + z_2 + z_3 - 3z_4}{2}) \;.
\eea
This is a little tricky as there no direct way of doing the full sum. However we can still extract the result by
being careful. First, let's evaluate this without the derivative term (i.e. with $u = z_2$). In this case we can
rewrite the amplitude as
\bea
&&\frac{
\theta_{\alpha \beta}(\frac{-z_1 + z_2 + z_3 - z_4}{2})
\theta_{\alpha \beta}(\frac{-z_1 - z_2 + z_3 + z_4}{2} + (z_2 - u))}
{\theta_{\alpha \beta}(\frac{-z_1 + z_2 - z_3 + z_4}{2})}  \\
&& \times \theta_{\alpha \beta}(\frac{z_1 + z_2 - z_3 - z_4}{2} + \theta)
\theta_{\alpha \beta}(\frac{z_1 + z_2 - z_3 - z_4}{2} - \theta)
\theta_{\alpha \beta}(\frac{-z_1 - z_2 - z_3 + 3z_4}{2})\;, \nn
\eea
which evaluates to
\be
\frac{\theta_1(z_2 - z_4)\theta_1(-z_1 + z_4+\theta) \theta_1(-z_1 + z_4 - \theta) \theta_1(z_2 - z_3)
\theta_1(-z_3 + z_4)}{\theta_1(-z_1 + z_2 -z_3 + z_4)}.
\ee
The presence of the $\theta_1(z_3 - z_4)$ zero means it is not possible to obtain a momentum pole from the non-derivative
terms, as when combined with the spin-structure independent terms and the bosonic correlator $\langle \partial X e^{ik \cdot X} \rangle$
we see that we can get at most a single pole as we factorise $z_{3/4}$ onto $z_1$ or $z_2$.

We therefore need to consider derivative terms, for which the bosonic amplitude is simply $k_3^{2-} k_2^{2+}$. The spin-structure independent
fermionic terms are
\be
\theta_1(z_1 - z_2) \theta_1(z_1 - z_3)^{-1} \theta_1(z_1 - z_4)^{-1} \theta_1(z_2 - z_4)^{-1}
\left[ \frac{\theta_1^{'}(z_1 - z_2)}{\theta_1(z_1 - z_2)} +
\frac{\theta_1^{'}(z_2 - z_3)}{\theta_1(z_2 - z_3)} + \frac{\theta_1^{'}(z_2 - z_4)}{\theta_1(z_2 - z_4)} \right].
\ee
This shows potential poles as $z_{3/4} \to z_1$ and as $z_{3/4} \to z_2$. Note that a single zero is sufficient to eliminate these poles.
In this limit, $z_3 = z_4 + \mc{O}(\epsilon)$. In the spin-structure dependent sum we can therefore put $z_3 = z_4$, as any error in this
approximation cannot contribute to a pole. Doing so, we can simplify the spin structure dependent terms to
$$
\theta_{\alpha \beta}(\frac{-z_1 - z_2 + 2z_3}{2} + (z_2 - u))
\theta_{\alpha \beta}(\frac{z_1 + z_2 - 2 z_3}{2} + \theta)
\theta_{\alpha \beta}(\frac{z_1 + z_2 - 2 z_3}{2} - \theta)
\theta_{\alpha \beta}(\frac{-z_1 - z_2 + 2 z_3}{2}),
$$
for which we can perform the Riemann summation to obtain
$$
\theta_1 \left(\frac{z_2 - u}{2} \right) \theta_1(\theta) \theta_1(-\theta) \theta_1(-z_1 - z_2 + 2 z_3).
$$
The zero as $u \to z_2$ cancels off against the pole in the derivative term, and we can now evaluate the contributions.
There are two separate contributions, one from the limit $z_{3/4} \to z_2$, and another from the limit $z_{3/4} \to z_1$.
The limit $z_{3/4} \to z_2$ has a prefactor
$$
\frac{\theta_1(z_1 - z_2) \theta(z_1 - z_2)}{\theta_1(z_2 - z_1) \theta_1(z_2 - z_1)} \to 1,
$$
and the limit $z_{3/4} \to z_1$ has a prefactor
$$
\frac{\theta_1(z_1 - z_2) \theta(z_1 - z_2)}{\theta_1(z_1 - z_2) \theta_1(z_1 - z_2)} \to 1.
$$
Overall we then get
$$
k_3^{2+}k_3^{2-} \left( \frac{1}{2k_2 \cdot k_3 + k_3 \cdot k_3} + \frac{1}{2 k_1 \cdot k_3 + k_3 \cdot k_3} \right) = 0
$$
using $k_1 + k_2 + k_3 = 0$. So then there is no contribution from this choice of picture changing,

We then combine (\ref{qq}) and (\ref{qqq}), together with the analogous terms that come from exchanging $(+-) \to (-+)$
H-charges for the fluxes which complete the Lorentz structure. These then give the final contribution from the 'non-standard'
vertex operator terms,
\be
\label{betjeman}
2 \pi \alpha' \int \frac{dt}{t} Z(t)
- 2 \pi \alpha' \int dt \frac{d}{dt} Z(t).
\ee
with a contribution from both running and anomalous mass terms.

\subsection{Summary}

Overall we then have from the standard case
\be
\label{milton2}
\pi \alpha' \int \frac{dt}{t} \, \left[ (4 + 4t \frac{d}{dt}) Z(t) + \frac{k_2 \cdot k_3 + 2 k_2 \cdot k_2}{2 k_2 \cdot k_3 + k_3 \cdot k_3} Z(t) \right].
\ee
and from the non-standard case
\be
\label{betjeman2}
2 \pi \alpha' \int \frac{dt}{t} Z(t)
- 2 \pi \alpha' \int dt \frac{d}{dt} Z(t).
\ee
Let us make the important points about these expressions. First, we obtain both anomalous and running contributions to
gaugino masses. This is an important check on the general structure of 1-loop gaugino masses in flux backgrounds and in conjunction
with the NSNS computation shows how anomalous gaugino masses arise in string theory. However
the expression is not unambiguous; in contrast to the NSNS case the numerical result depends on the details of the off-shell
prescription. This means that it is not possible to extract the numerical relationship between the anomalous and running mass
which is necessary to test the formula (\ref{kilmarnock}) fully. This inherent conceptual ambiguity sits on top of the
non-trivial task of ensuring that given the calculational complexity
no errant signs or factors of 2 are present in the expressions (\ref{milton2}) and (\ref{betjeman2}).
In principle the conceptual ambiguity could be resolved by going to a higher point amplitude where all fields can be put
on-shell, but this appears calculationally prohibitive.

%%%%%%%%%%%%%%%%%%%%%%%%%%%%%%%%%%%%%%%%%%%%%%%%%%%%%%%%%%%%%%%%%%%%%%%%%%%%%%%%%%%%%%%%%%%%%%
\section{Useful identities}
\label{sec:useiden}
%%%%%%%%%%%%%%%%%%%%%%%%%%%%%%%%%%%%%%%%%%%%%%%%%%%%%%%%%%%%%%%%%%%%%%%%%%%%%%%%%%%%%%%%%%%%%%

The standard notation for the Jacobi Theta functions is:
\begin{equation}
\theta \left[ \begin{array}{c} a \\ b \end{array} \right] (z;\tau) = \sum_{n = -\infty}^{\infty} \mathrm{exp} \bigg[ \pi i (n+a)^2 \tau + 2 \pi i (n+a)(z+b) \bigg]
\end{equation}
A common definition is $\theta_{\alpha \beta} \equiv \theta \left[ \begin{array}{c} \alpha/2 \\ \beta/2 \end{array} \right]$, and
\begin{eqnarray}
\theta_1 \equiv \theta_{11} & \theta_2 \equiv \theta_{10} \nonumber \\
\theta_3 \equiv \theta_{00} & \theta_4 \equiv \theta_{01}.
\end{eqnarray}

Expansions of the functions for $q=e^{\pi i \tau}$ are
\begin{eqnarray}
\theta_{00} (z, \tau) & = \theta_3 =& 1 + 2\sum_{n=1}^{\infty} q^{n^2} \cos 2\pi n z \nonumber \\
\theta_{01} (z, \tau) & = \theta_4 =& 1 + 2\sum_{n=1}^{\infty} (-1)^n q^{n^2} \cos 2\pi n z \nonumber \\
\theta_{10} (z, \tau) & = \theta_2 =& 2q^{1/4}\sum_{n=0}^{\infty} q^{n(n+1)} \cos \pi (2n+1) z \nonumber \\
\theta_{11} (z, \tau) & = \pm \theta_1 =& 2q^{1/4}\sum_{n=0}^{\infty} (-1)^n q^{n(n+1)} \sin \pi (2n+1) z \nonumber \\
\end{eqnarray}

The Dedekind $\eta$ function is defined as
\begin{eqnarray}
\eta (\tau) &=& q^{1/12} \prod_{m=1}^{\infty} (1-q^{2m}) \\
&=& \left[ \frac{\theta_1^{\prime} (0,\tau)}{-2\pi} \right]^{1/3}.
\end{eqnarray}

The generalised Riemann summation formula is
\begin{align}
&\sum_{\alpha, \beta} (-1)^{\alpha + \beta + \alpha \beta}\prod_{i=1}^{4} \tab{\alpha/2+c_i}{\beta/2+d_i}{z_i}{\tau}=2\tab{1/2}{1/2}{\sum_i z_i/2}{\tau} \tab{1/2 +c_2 }{1/2 + d_2 }{\frac{z_1 + z_2 - z_3-z_4}{2} }{\tau}\nonumber \\
&\times \tab{1/2 +c_3 }{1/2 + d_3 }{\frac{z_1 - z_2 + z_3-z_4}{2} }{\tau}\tab{1/2 +c_4 }{1/2 + d_4 }{\frac{z_1 - z_2 - z_3+z_4}{2} }{\tau} \label{rieiden1}.
\end{align}
An identity useful in simplifying amplitudes after Riemann summation is
\begin{align}
\frac{\vt(x+y)\vt(x-y)(\vt^\prime (0))^2}{\vt(x)^2 \vt(y)\vt(-y)} =&
\partial_y^2 \log \vt(y) - \partial_x^2 \log\vt(x).
\end{align}
In particular,
\begin{align}
(2\alpha^\prime) \frac{\theta_1(-z_1 + z_2 + \theta) \theta_1(-z_1 + z_2
- \theta)\vt^\prime(0)^2}{\theta_1(z_{12})^2\vt(\theta)\vt(-\theta)}
=&(2\alpha^\prime) \partial_{\theta}^2\log \vt (\theta) +
\partial_{z_1}^2 G_{12} + \frac{8\pi \alpha^\prime}{t}.
\label{theta1identity}\end{align}
We also have the five-theta identity \cite{Atick:1986rs}:
\begin{align}
&\sum_\nu \delta_\nu \theta_\nu (z_1) \theta_\nu (z_2) \theta_\nu (z_3) \theta_\nu (z_4) \theta_\nu (z_5) \theta_\nu^{-1} (z_1 + z_2 + z_3 + z_4 + z_5) \nonumber \\
&=-2 \theta_1 (z_1 + z_2 + z_3 + z_4) \theta_1 (z_2 + z_3 + z_4 + z_5) \theta_1(z_1 + z_3 + z_4 + z_5) \nonumber \\
&\times \theta_1 (z_1 + z_2 + z_4 + z_5)\theta_1 (z_1 + z_2 + z_3 + z_5) \theta_1^{-1} (2[z_1 + z_2 + z_3 + z_4 + z_5]).
\label{FiveTheta}\end{align}
Another identity useful for threshold corrections is
\be
\sum_{\alpha} \frac{\theta_{\alpha}^{''}(0)}{\eta^3} \prod_i \frac{\theta_{\alpha}(u_I)}{\theta_1(u_I)} =
- 2 \pi \sum_{i=1}^3 \frac{\theta^\prime \left[ \begin{array}{c} 1/2 \\ 1/2 \end{array}\right](\theta_i,it)}{\tab{1/2}{1/2}{\theta_i}{it}}.
\label{extrathetaidentity}
\ee
Yet another identity which is useful taken from \cite{Atick:1986rs} is
\bea
&&\theta_{\alpha\beta}\left(\half(-z_1+z_2+z_3-z_4)\right)\theta_{\alpha\beta}\left(\half(z_1+z_2-z_3-z_4)\right) = \frac{\theta_{\alpha\beta}\left(\half(z_1-z_2+z_3-z_4)\right)}{\theta_1(-z_1+z_2-z_3+z_4)\theta_1(z_1-z_3)} \nn \\
&&\left[\theta_{\alpha\beta}\left(\half(-z_1-z_2+3z_3-z_4)\right)\theta_1(-z_1+z_2)\theta_1(z_1-z_4) - \right.\nn \\ &&\left. \theta_{\alpha\beta}\left(\half(3z_1-z_2-z_3-z_4)\right)\theta_1(z_2-z_3)\theta_1(z_3-z_4) \right] \;. \label{derivthetaiden}
\eea
An equivalent way of re-expresing this (by relabelling the $z_i$) is
\begin{align}
\vartheta_\nu (\frac{z_1 - z_2 + z_3 - z_4}{2}) \vartheta_\nu (\frac{z_1 + z_2 - z_3 - z_4}{2}) =& \frac{\vartheta_\nu (\frac{-z_1 + z_2 + z_3 - z_4}{2})}{\vt(z_1 - z_2 - z_3 + z_4) \vt(z_2 - z_3)} \\
&\times \bigg[ \vartheta_\nu (\frac{-z_1 - z_2 + 3 z_3 - z_4}{2}) \vt(z_1 - z_2) \vt(z_2 - z_4) \nn\\
&\qquad - \vartheta_\nu (\frac{-z_1 + 3z_2 - z_3 - z_4}{2}) \vt(z_1 - z_3) \vt(z_3 - z_4) \bigg]. \nn
\end{align}

The general Poisson resummation formula is
\be
\label{poisson}
\sum_{n_i} \exp (-\pi t n_i A_{ij} n_j) = \frac{1}{t^{\frac{\hbox{dim}(A)}{2}} (\hbox{det} A)^{\half}}
\sum_{m_i} \exp (-\frac{\pi}{t} m_i A^{-1}_{ij} m_j) \;.
\ee
A useful simple form is
\be
\sum_{n} \exp (-\pi t A (n+c)^2 ) = \frac{1}{\sqrt{At}} \sum_{m} \exp (-\frac{\pi}{At} m^2 - 2\pi i m c) \;.
\label{poissonsimple}
\ee

\end{document}